\documentclass[a4paper,11pt]{article}
\pdfoutput=1 %

\usepackage{jheppub} %

\usepackage{silence}
\usepackage{dcolumn}
\usepackage{bm}

\usepackage{float}
\usepackage{graphicx}
\usepackage{breqn}
\usepackage{mathtools}
\usepackage{listings}
\usepackage{comment}
\usepackage[bottom]{footmisc} 
\usepackage{cleveref}
\usepackage{booktabs}
\usepackage{tabularx}
\usepackage{makecell}
\usepackage[utf8]{inputenc}
\usepackage{siunitx}

\usepackage{xcolor}
\usepackage[normalem]{ulem}

\usepackage{placeins}

\newcommand{\infinity}{\infty}

\begin{document}

\interfootnotelinepenalty=10000 %

\crefname{figure}{Fig.}{Figs.}
\Crefname{figure}{Figure}{Figures}
\crefname{table}{Tab.}{Tabs.}
\Crefname{table}{Table}{Tables}
\crefname{equation}{Eq.}{Eqs.}
\Crefname{equation}{Equation}{Equations}
\crefname{section}{Sec.}{Secs.}
\Crefname{section}{Section}{Sections}
\Crefname{appendix}{Appendix}{Appendices}
\crefname{appendix}{Appendix}{Appendices}
\creflabelformat{equation}{#2\textup{#1}#3}
\renewcommand{\crefrangeconjunction}{--}

\newcommand{\reftexts}{Ref.~}
\newcommand{\lips}[1]{\tilde{d} #1 \;}
\newcommand{\vecIV}[1]{#1} %
\newcommand{\vecIII}[1]{\vec{#1}} %
\newcommand{\vecII}[1]{\boldsymbol{#1}} %

\newcommand{\cole}[1]{%
{#1}%
}

\newcommand{\coleTwo}[1]{%
{#1}%
}

\newcommand{\coleThree}[1]{%
{#1}%
}

\newcommand{\coleFour}[1]{%
{#1}%
}

\newcommand{\coleFive}[1]{%
{#1}%
}

\newcommand{\rev}[1]{%
{#1}%
}

\newcommand{\revTwo}[1]{%
{#1}%
}

\newcommand{\revThree}[1]{%
{#1}%
}

\makeatletter

\newcommand{\coll}[2]{%
	$\coll@process{#1} + \coll@process{#2}$%
}

\newcommand{\collThree}[3]{%
	$\coll@process{#1} / \coll@process{#2} + \coll@process{#3}$%
}

\newcommand{\collFour}[4]{%
  $\coll@process{#1} / \coll@process{#2} / \coll@process{#3} + \coll@process{#4}$%
}

\newcommand{\coll@process}[1]{%
  \ifx#1A%
    #1%
  \else\ifx#1B%
    #1%
  \else\ifx#1x%
    #1%
  \else\ifx#1p%
    #1%
  \else\ifx#1d%
    #1%
  \else
    \coll@checkHeThree{#1}%
  \fi\fi\fi\fi\fi%
}

\newcommand{\coll@checkHeThree}[1]{%
  \ifnum\pdfstrcmp{#1}{He3}=0 %
    {}^3\mathrm{He}%
  \else
    \mathrm{#1}%
  \fi
}

\makeatother

\title{Statistical analysis of pQCD energy loss across system size, flavor, $\sqrt{s_{NN}}$, and $p_T$ }

\author[a]{Coleridge Faraday}
\author[a,b]{and W.\ A.\ Horowitz}
\emailAdd{frdcol002@myuct.ac.za}
\emailAdd{wa.horowitz@uct.ac.za}
\affiliation[a]{Department of Physics\char`,{} University of Cape Town\char`,{} Private Bag X3\char`,{} Rondebosch 7701\char`,{} South Africa}
\affiliation[b]{Department of Physics\char`,{} New Mexico State University\char`,{} Las Cruces\char`,{} New Mexico\char`,{} 88003\char`,{} USA}

\date{\today}

\abstract{
	We present suppression predictions from our pQCD-based energy loss model, which receives small system size corrections, for high-$p_T$ $\pi$, $D$ and $B$ meson $R_{AB}$ as a function of centrality, flavor, $\sqrt{s_{NN}}$, and $p_T$ from large to small collision systems at RHIC and LHC. 
	A statistical analysis is used to constrain the effective strong coupling in our model to available high-$p_T$ suppression data from central heavy-ion collisions at RHIC and LHC, yielding good agreement with all available data. We estimate two important theoretical uncertainties in our model, stemming from: the transition between vacuum and hard thermal loop propagators in the collisional energy loss, and from the angular cutoff on the radiated gluon momentum. 
	The model uncertainties lead to significant uncertainties in the extracted $\alpha_s^{\text{eff.}}$ of $\mathcal{O}(30\%)$, much larger than the uncertainties associated with the extraction procedure; however, the final uncertainty on the constrained $R_{AB}$ and on extrapolations of $R_{AB}$ to regions where it was not constrained, $\lesssim 20\%$, is significantly smaller than one might naively expect.
	We find the best fit extracted $\alpha_s^{\text{eff.}} = \num{0.41(14:10)}$ at RHIC and $\alpha_s^{\text{eff.}} = \num{0.37(11:8)}$ at LHC.
	When applying the statistical extraction of $\alpha_s$ to different subsets of experimental data, we find, consistently, that the extracted $\alpha_s$ remains relatively unchanged across heavy- and light-flavor final states and across central, semi-central, and peripheral collisions.
We make predictions from our large-system-constrained model for small systems and find good agreement with the photon-normalized $R^{\pi^0}_{d \text{Au}} \simeq 0.75 $ in $0\text{--}5\%$ centrality \coll{d}{Au} collisions by PHENIX. 
However, we find strong disagreement with the measured $R^{h^{\pm}}_{p \text{Pb}} \gtrsim 1$ in $0\text{--}5\%$ centrality \coll{p}{Pb} collisions by ALICE and ATLAS; we argue that this disagreement is due, in large part, to centrality bias. %
We make predictions for the ratio of suppression in \coll{He3}{Au} and \coll{p}{Au} collisions, which may in the future be used to disentangle final- from initial-state suppression in small systems.
We then compare our results to various subsets of data, which allows us to estimate the preferred: low-$p_T$ scale at which non-perturbative processes become important, scales at which the strong coupling runs, and scale at which vacuum propagators transition to thermally modified propagators in collisional energy loss.
}

\maketitle

\flushbottom

\section{Introduction}\label{sec:introduction}

The quark-gluon plasma (QGP) is a state of matter that is formed at extreme temperatures and densities in the ultra-relativistic collision of heavy-ions \cite{Shuryak:2014zxa,Busza:2018rrf,ALICE:2022wpn}. Due to the short lifetime of the QGP, one must infer its existence and properties from particles that are formed during the collision. A number of observables have been proposed that are associated with the production of a QGP including multi particle correlations \cite{STAR:2000ekf, PHENIX:2003qra,ALICE:2010suc}, strangeness enhancement \cite{Rafelski:1982pu,STAR:2003jis,ALICE:2013xmt}, and quarkonia suppression \cite{Matsui:1986dk,STAR:2009irl,ALICE:2012jsl,CMS:2012bms}. These observables show that a QGP is formed in central heavy-ion collisions at RHIC and LHC. 
Surprisingly, these observables appear to be insensitive to the size of the colliding systems, but rather smoothly turn on as a function of multiplicity \cite{ATLAS:2015hzw, ALICE:2023ulm, ATLAS:2013jmi, ALICE:2014dwt, CMS:2015yux, PHENIX:2013ktj, PHENIX:2014fnc, PHENIX:2015idk, PHENIX:2016cfs, PHENIX:2017xrm, ALICE:2016sdt, ALICE:2015mpp, ALICE:2013wgn, Grosse-Oetringhaus:2024bwr, ALICE:2024vzv}.

High-$p_T$ particle suppression, where a parton loses energy as it traverses through the medium, is another observable related to QGP formation \cite{Gyulassy:2004zy,Wiedemann:2009sh,Majumder:2010qh,Connors:2017ptx}. %
High-$p_T$ particle suppression in central \coll{A}{A} collisions at RHIC \cite{PHENIX:2001hpc,STAR:2003pjh,STAR:2003fka,PHENIX:2006ujp,PHENIX:2008saf,STAR:2009fqa,PHENIX:2012jha,STAR:2018zdy} and LHC \cite{ALICE:2010yje,ALICE:2012ab,CMS:2012aa,CMS:2017qjw, CMS:2017uoy,ALICE:2018vuu, ALICE:2018lyv,ALICE:2019hno, ATLAS:2022kqu} has been unambiguously linked to partonic energy loss in a QGP.
In high multiplicity small systems, however, some measurements indicate a high-$p_T$ particle suppression \cite{PHENIX:2021dod,PHENIX:2023dxl}, while other measurements show an enhancement \cite{ALICE:2014xsp,ALICE:2019fhe, ATLAS:2022kqu}. %
From the theoretical side, it is not obvious whether high multiplicity small collision systems generate a droplet of QGP of great enough size and temperature to yield measurable high-$p_T$ particle suppression.
There is thus a need for quantitative predictions of high-$p_T$ suppression from careful theoretical calculations to be compared with experimental data. %
Generating these quantitative predictions of high-$p_T$ suppression in small systems, constrained by a rigorous statistical analysis of central \coll{A}{A} suppression data, is the main purpose of this work.

In order to provide such quantitative theoretical predictions, one needs several quantities under reasonably precise theoretical control: 1) the production spectra of high-$p_T$ partons, 2) the energy lost by these high-$p_T$ partons as they travel through the QGP, 3) a model for the soft medium dynamics through which the high-$p_T$ parton propagates, and 4) a hadronization mechanism. The spectral input is under good theoretical control from pQCD calculations \cite{Cacciari:1998it,Eskola:2002kv}. 
We will use the well-calibrated hydrodynamics calculations as the model for the medium evolution through which our partons propagate \cite{Schenke:2020mbo}.
At low-moderate momenta $\lesssim 10 ~\mathrm{GeV}$  hadronization processes appear to be modified by the presence of a hot, dense, medium; however, above these momenta vacuum processes appear to describe the hadronization physics well \cite{ALICE:2021bib,ALICE:2022exq,CMS:2023frs,LHCb:2018weo,ALICE:2017thy,PHENIX:2003tvk}.

For the energy loss processes, we will use well-controlled perturbative QCD calculations of collisional \cite{Wicks:2008zz,Braaten:1991jj,Braaten:1991we} and radiative energy loss \cite{Gyulassy:2000er,Djordjevic:2003zk}, both of which include small system size corrections \cite{Kolbe:2015rvk,Kolbe:2015suq,Wicks:2008zz}. 
Previous works systematically investigated some of the theoretical uncertainties associated with these energy loss processes \cite{Faraday:2023mmx,Faraday:2024gzx}.
In this work, we investigate the uncertainties associated with 1) restricting the emitted gluon’s phase space in order to make the calculation self-consistent with the large formation time approximation \cite{Kolbe:2015rvk,Faraday:2023mmx,Faraday:2023uay} and 2) the transition between vacuum and hard thermal loop (HTL) \cite{Braaten:1989mz, Klimov:1982bv, Pisarski:1988vd, Weldon:1982aq, Weldon:1982bn} propagators in the collisional energy loss derivation \cite{Romatschke:2004au,Gossiaux:2008jv,Wicks:2008zz}.  We focus on these two uncertainties for the following reasons.  The phase space restriction was a recent proposal \cite{Faraday:2023mmx} for enforcing self-consistency of the numerical evaluation of the radiative energy loss with the large formation time approximation, and the consequences of this proposed restriction on the phase space have not yet been explored.  We choose to explore the consequences of the systematic uncertainty associated with the propagators used in the collisional energy loss derivation as this is a major source of uncertainty that is not currently under theoretical control \cite{Romatschke:2004au, Wicks:2008zz,Faraday:2024gzx}.

So as to make quantitative predictions for small systems, we calibrate our energy loss model on central large systems and confirm its consistency across known experimental results. We perform this calibration via a $\chi^2$ comparison of our energy loss model predictions to central heavy-ion suppression data by extracting the single parameter for our model, the effective strong coupling constant $\alpha_s^{\text{eff.}}$. We check for consistency by confirming that this $\alpha_s^{\text{eff.}}$ calibration is consistent across centrality and flavor, where running coupling is not expected to be important. 
Previous works \cite{Armesto:2009zi,JET:2013cls,Andres:2016iys,Casalderrey-Solana:2018wrw,Xie:2020zdb,JETSCAPE:2021ehl,Xie:2022ght,JETSCAPE:2024cqe,Karmakar:2024jak} have made sophisticated statistical comparisons between their energy loss models and experimental data to characterize properties of the QGP. Additionally, other numerous works have considered energy loss in small systems \cite{Kolbe:2015rvk,Park:2016jap,Huss:2020whe,Huss:2020dwe,Ke:2022gkq,Zakharov:2021uza,Zakharov:2023vee}. However, to our knowledge, no other work has performed a sophisticated statistical analysis of large-system collisions that provides fully constrained, zero-parameter predictions for peripheral heavy-ion collisions and high-multiplicity small-system collisions, enabling a rigorous test of the consistency of QGP formation in these small systems. That is the primary purpose of this work.

One difficulty associated with making predictions for centrality cut observables are various selection biases called \emph{centrality bias}, which may correlate the hard and soft modes in the collision \cite{ALICE:2014xsp, ATLAS:2016xpn, ALICE:2018ekf, ATLAS:2023zfx,ATLAS:2024qsm}. One response to these small system selection biases is to consider only minimum bias collisions \cite{Huss:2020whe,Huss:2020dwe}; however, in such collisions the expected energy loss signal is reduced. 
In \cite{Huss:2020whe,Huss:2020dwe,Brewer:2021kiv}, it was suggested that \coll{O}{O} collisions would be an ideal candidate for such a minimum bias measurement.
The question of whether a QGP forms in minimum bias light-ion collisions is distinct from, though related to, the question of whether it forms in central \collFour{p}{d}{He3}{A} collisions. Both are central to this work. However, when comparing to centrality-cut small-system data, care must be taken in interpreting the results due to potential centrality biases.
What we will show is that there is a clear consistency between our energy loss model predictions and peripheral \coll{A}{A} measurements at RHIC \cite{PHENIX:2008saf,PHENIX:2012jha} and LHC \cite{ATLAS:2022kqu,ALICE:2018vuu,CMS:2016xef} and central \coll{d}{Au} measurements by PHENIX \cite{PHENIX:2023dxl}.
We will also show that our results are in stark contrast with $R_{pA}$ measurements at LHC \cite{ALICE:2014xsp,ALICE:2019fhe,ATLAS:2022kqu}, although we believe this discrepancy is due to the much larger centrality biases associated with this measurement.
Considering the former question, as to whether there is measurable suppression in light-ion collisions, we also provide predictions for neutral pions produced in central and minimum bias \coll{O}{O} and \coll{Ne}{Ne} collisions, where we find $\sim 30\%$ suppression in minimum bias \coll{O}{O} and \coll{Ne}{Ne} collisions.

An alternative interpretation of the PHENIX measured $R_{d \mathrm{Au}} < 1$ and the ATLAS $R_{p \mathrm{Pb}} > 1$ is provided by the color fluctuation model \cite{Alvioli:2014eda, Alvioli:2017wou}, which attributes the nuclear modification in small systems to initial-state effects \cite{Perepelitsa:2024eik,Perepelitsa:2014yta}. 
In this work we do not include initial-state effects. Instead, we construct an energy-loss-only baseline prediction for small systems, derived from a large-system $\chi^2$-constrained energy loss model with explicit small-system size corrections. 
We make predictions for the ratio of $R_{AB}$ in \coll{He3}{Au} collisions to that in \coll{p}{Au} collisions as a function of centrality and $p_T$. 
In the scenario where final state energy loss leads to suppression, this ratio is expected to be less than one due to the path-length dependence of energy loss. Conversely, initial-state effects from, for example, the color fluctuation model decrease with system size, presumably leading to this ratio being greater than one \cite{Perepelitsa:2024eik}.
A future measurement of this ratio at RHIC, normalized by prompt photons \cite{PHENIX:2023dxl}, could help disentangle final- and initial-state contributions to suppression in small systems.

Having set up the statistical extraction machinery, we explore the physics conclusions we may reach by examining different subsets of experimental data.
We first consider hadronization, 
a non-perturbative phenomenon that is not under good theoretical control.
There are experimental and theoretical reasons to believe that the presence of a dense QCD medium will modify the vacuum hadronization process \cite{PHENIX:2003tvk,ALICE:2017thy,ALICE:2021bib}. 
Our model does not include any medium modification to hadronization.
We will compute the $p$-value of our model predictions as a function of the minimum $p_T$ allowed in the range of central \coll{A}{A} collision data considered and 
attribute statistically significant deviations of our model from data to non-perturbative contributions to the suppression pattern, such as a medium modification to the hadronization process.
We then consider the running of the strong coupling.
LHC charged hadron data now exist over many decades of transverse momenta \cite{ATLAS:2022kqu,CMS:2016xef}. One naturally expects that over such a large $p_T$ range, the strong coupling will run appreciably.  Several derivations consider the next-to-leading order, logarithmically enhanced contributions to the transport coefficient $\hat{q}$ \cite{Blaizot:2014bha,Arnold:2021pin,Muller:2021wri,Ghiglieri:2022gyv} that is used in many energy loss models. Additionally, several energy loss models have implemented reasonable choices for the scales at which the strong coupling runs \cite{Xu:2014ica, Xu:2015bbz, Peshier:2006ah,JETSCAPE:2024cqe,Djordjevic:2013xoa}; however, so far, there has not been a first-principles derivation of these scales. There is also currently a significant tension between the highest $p_T$ charged hadron suppression measured by ATLAS \cite{ATLAS:2022kqu} and CMS \cite{CMS:2016xef}.  
Currently, our energy loss model neglects running coupling effects.  By dividing up the $p_T$ range of data and comparing to our energy loss model, we are able to show that while the CMS data \cite{CMS:2016xef} naturally requires a monotonic decrease of the strong coupling with the measured particle transverse momentum, the ATLAS data \cite{CMS:2016xef} requires a non-monotonic dependence of the coupling on $p_T$.  We are additionally able to examine the effect of the change in medium temperature from RHIC to LHC to find that the data favors the couplings to run as $\alpha_s^2(2 \pi T)\alpha_s(E)$ and the collisional energy loss to be given by hard thermal loops \cite{Braaten:1989mz, Klimov:1982bv, Pisarski:1988vd, Weldon:1982aq, Weldon:1982bn}.

This paper is organized as follows. In \cref{sec:model} we describe in detail our energy loss model; in \cref{sec:statistical_analysis} we present the details of our statistical analysis; in \cref{sec:global_extraction} we present the results of our global extraction of $\alpha_s^{\text{eff.}}$; in \cref{sec:model_robustness} we discuss the self consistency of our model across centrality and flavor; in \cref{sec:small_systems} we extrapolate our model to peripheral and small systems; in \cref{sec:missing_physics} we perform a statistical analysis on various subsets of data to qualitatively investigate medium modifications to hadronization, running coupling effects, and the transition between HTL and vacuum propagators in collisional energy loss; in \cref{sec:jet_transport_coefficient_and_comparison_with_other_work} we convert our extraction of $\alpha_s^{\text{eff.}}$ to an extraction of $\hat{q} / T^3$ and compare our results with previous extractions; and in \cref{sec:conclusion} we conclude.

\section{Physics model}
\label{sec:model}

This work utilizes the energy loss model introduced and described in our previous work \cite{Faraday:2023mmx,Faraday:2024gzx}, which itself is based on the Wicks-Horowitz-Djordjevic-Gyulassy (WHDG) energy loss model \cite{Wicks:2005gt}. Our energy loss model includes small system size corrections \cite{Kolbe:2015rvk,Wicks:2008zz} to both the pQCD radiative and collisional energy loss, uses fluctuating initial conditions \cite{Schenke:2020mbo}, and takes into account realistic production spectra \cite{Cacciari:1998it} and fragmentation functions \cite{Cacciari:2005uk,deFlorian:2007aj}. 
In this section we will briefly review the energy loss model.

\subsection{Radiative energy loss}
\label{sec:radiative_energy_loss}

\subsubsection{DGLV radiative energy loss}
\label{sec:dglv_radiative_energy_loss}

The Djordjevic-Gyulassy-Levai-Vitev (DGLV) opacity expansion \cite{Djordjevic:2003zk, Gyulassy:1999zd} gives the inclusive differential distribution of medium-induced gluon radiation from a high-$p_T$ parent parton moving through a smooth brick of QGP. Interactions with the medium are modeled as elastic scatters with infinitely massive scattering centers using the Gyulassy-Wang potential \cite{Gyulassy:1993hr}, which then induce gluon radiation.
The expansion is in the expected number of scatterings or the \textit{opacity} $L / \lambda_g$, where $L$ is the length of the QGP brick and $\lambda_g$ is the mean free path of a gluon in the QGP. The DGLV single inclusive gluon radiation spectrum is, to first order in opacity, \cite{Gyulassy:2000er,Djordjevic:2003zk}
	\begin{gather}
		\frac{\mathrm{d} N^g_{\text{DGLV}}}{\mathrm{d} x}=  \frac{C_R \alpha_s^{\text{eff.}}}{\pi \lambda_g} \frac{1}{x} \int \frac{\mathrm{d}^2 \mathbf{q}_1}{\pi} \frac{\mu^2}{\left(\mu^2+\mathbf{q}_1^2\right)^2} \int \frac{\mathrm{d}^2 \mathbf{k}}{\pi} \int \mathrm{d} \Delta z \, \bar{\rho}(\Delta z) \nonumber\\
 -\frac{2\left\{1-\cos \left[\left(\omega_1+\tilde{\omega}_m\right) \Delta z\right]\right\}}{\left(\mathbf{k}-\mathbf{q}_1\right)^2+m_g^2+x^2 M^2}\left[\frac{\left(\mathbf{k}-\mathbf{q}_1\right) \cdot \mathbf{k}}{\mathbf{k}^2+m_g^2+x^2 M^2}-\frac{\left(\mathbf{k}-\mathbf{q}_1\right)^2}{\left(\mathbf{k}-\mathbf{q}_1\right)^2+m_g^2+x^2 M^2}\right].
 \label{eqn:DGLV_dndx}
\end{gather}
In the above, we have used the notation $\omega \equiv x E^+ / 2,~\omega_0 \equiv \mathbf{k}^2 / 2 \omega,~\omega_i \equiv (\mathbf{k} - \mathbf{q}_i)^2 / 2 \omega$, $\mu_i \equiv \sqrt{\mu^2 + \mathbf{q}_i^2}$, and $\tilde{\omega}_m \equiv (m_g^2 + M^2 x^2) / 2 \omega$ following \cite{Djordjevic:2003zk, Kolbe:2015rvk}. Additionally, $\mathbf{q}_i$ is the transverse momentum of the $i^{\mathrm{th}}$ gluon exchanged with the medium, $\mathbf{k}$ is the transverse momentum of the radiated gluon, $\Delta z$ is the distance between production of the hard parton, and scattering, $C_R$ is the quadratic Casimir of the hard parton representation ($C_F = 4 / 3$ for quarks and $C_A = 3$ for gluons), $x$ is the fraction of parent parton's plus momentum carried away by the radiated gluon, $M$ is the mass of the incident parton, $m_g \approx \mu / \sqrt{2}$ is the mass of the radiated gluon \cite{Wicks:2005gt}, and $\alpha_s^{\text{eff.}}$ is the strong coupling. 

 The quantity $\bar{\rho}(\Delta z)$ is the \emph{distribution of scattering centers} in $\Delta z$ and is defined in terms of the density of scattering centers $\rho(\Delta z)$ in a static brick,
\begin{equation}
	 \bar{\rho}(\Delta z) \equiv \frac{A_{\perp}}{N} \rho(\Delta z),
  \label{eqn:density_scattering_centers}
\end{equation}
where $\Delta z$ is in the direction of propagation, $N$ is the number of scattering centers, $A_{\perp}$ is the perpendicular area of the brick, and $\int \mathrm{d}z \; \bar{\rho}(\Delta z) = 1$. 
We will assume an exponential form $\bar{\rho}(\Delta z) = (2/L) \exp [-(2 \Delta z)/L]$ which allows the $\Delta z$ integral to be carried out analytically \cite{Djordjevic:2003zk} and thereby simplifies the numerics. 
Note that this assumption means that we do not have to calculate either $N$ or $A_{\perp}$ in \cref{eqn:density_scattering_centers}.
Future work will consider a more realistic distribution of scattering centers determined by the full hydrodynamic evolution of the initial-state \cite{Bert:2024}. 

\subsubsection{Short path length correction to DGLV radiative energy loss}
\label{sec:radiative_energy_loss_correction}

The derivation of the modification to the radiative energy loss in the DGLV \cite{Vitev:2002pf,Djordjevic:2003zk} opacity expansion approach with the relaxation of the large path length assumption $L \gg \mu^{-1}$ was considered in \cite{Kolbe:2015rvk,Kolbe:2015suq}. In the derivation of this short path length correction (SPLC), all assumptions and approximations made in the original GLV and DGLV derivations were kept, except that the short path length approximation $L\gg\mu^{-1}$ was relaxed. The single inclusive radiative gluon distribution, including both the original DGLV contribution as well as the short path length correction, is
\small
	\begin{gather}
    \frac{\mathrm{d} N^g_{\text{DGLV+SPLC}}}{\mathrm{d} x}=  \frac{C_R \alpha_s L}{\pi \lambda_g} \frac{1}{x} \int \frac{\mathrm{d}^2 \mathbf{q}_1}{\pi} \frac{\mu^2}{\left(\mu^2+\mathbf{q}_1^2\right)^2} \int \frac{\mathrm{d}^2 \mathbf{k}}{\pi} \int \mathrm{d} \Delta z \, \bar{\rho}(\Delta z) \nonumber\\
   \times\left[-\frac{2\left\{1-\cos \left[\left(\omega_1+\tilde{\omega}_m\right) \Delta z\right]\right\}}{\left(\mathbf{k}-\mathbf{q}_1\right)^2+m_g^2+x^2 M^2}\left[\frac{\left(\mathbf{k}-\mathbf{q}_1\right) \cdot \mathbf{k}}{\mathbf{k}^2+m_g^2+x^2 M^2}-\frac{\left(\mathbf{k}-\mathbf{q}_1\right)^2}{\left(\mathbf{k}-\mathbf{q}_1\right)^2+m_g^2+x^2 M^2}\right] \right. \\
   +\frac{1}{2} e^{-\mu_1 \Delta z}\left(\left(\frac{\mathbf{k}}{\mathbf{k}^2+m_g^2+x^2 M^2}\right)^2\left(1-\frac{2 C_R}{C_A}\right)\left\{1-\cos \left[\left(\omega_0+\tilde{\omega}_m\right) \Delta z\right]\right\}\right. \nonumber\\
   \left.\left.+\frac{\mathbf{k} \cdot\left(\mathbf{k}-\mathbf{q}_1\right)}{\left(\mathbf{k}^2+m_g^2+x^2 M^2\right)\left(\left(\mathbf{k}-\mathbf{q}_1\right)^2+m_g^2+x^2 M^2\right)}\left\{\cos \left[\left(\omega_0+\tilde{\omega}_m\right) \Delta z\right]-\cos \left[\left(\omega_0-\omega_1\right) \Delta z\right]\right\}\right)\right],\nonumber
   \label{eqn:full_dndx}
\end{gather}
\normalsize
\noindent where the first two lines of the above equation are the original DGLV result \cite{Gyulassy:2000er,Djordjevic:2003zk}, \cref{eqn:DGLV_dndx}, while the last two lines are the short path length correction.
Of particular importance is the large formation time assumption, which allows one to systematically neglect a significant number of diagrams in both the original DGLV derivation \cite{Vitev:2002pf,Djordjevic:2003zk} and in the short path length correction \cite{Kolbe:2015rvk,Kolbe:2015suq}. In our previous work \cite{Faraday:2023mmx}, we found that the large formation time approximation was not satisfied self-consistently by \emph{either} the unmodified DGLV result or the DGLV result which receives a short path length correction.

\subsubsection{Large formation time approximation}
\label{sec:large_formation_time_approximation}

In our previous work \cite{Faraday:2023mmx}, we found that the faster growth in $p_T$ of the short path length correction compared to standard DGLV in conjunction with the much larger size of the short path length correction for gluons compared to quarks led to $R_{AA} > 1$ for $p_T \gtrsim 100 ~\mathrm{GeV}$ pions produced in $0\text{--}5\%$ centrality \coll{Pb}{Pb} collisions. We argued that the large effect of the SPLC was due to a breakdown of a particular approximation---the large formation time (LFT) approximation---used in the derivation of both DGLV and DGLV + SPLC. The LFT approximation assumes that
\begin{equation}
	\frac{(\mathbf{k} - \mathbf{q}_1)^2}{2 x E} \ll \sqrt{\mu^2 + \mathbf{q}_1^2} \quad \iff \quad \frac{(\mathbf{k} - \mathbf{q}_1)^2}{2 x E \sqrt{\mu^2 + \mathbf{q}_1^2}} \ll  1.
	\label{eqn:lft_assumption}
\end{equation}
The LFT approximation is particularly important in the derivation of the short path length correction (SPLC) to DGLV \cite{Kolbe:2015suq,Kolbe:2015rvk}, where it is used to set the new contributions from 9 out of the 11 diagrams to zero. %
We previously calculated \cite{Faraday:2023mmx} the expectation value of dimensionless ratios $R$ that were assumed small under the various approximations and assumptions in the calculation, including the collinear, soft, and large formation time approximations. Explicitly, the procedure to calculate the expectation value of a function $R(\{ X_i \})$, depending on the set of variables $\{X _i\}$, is
\begin{equation}
  \langle R \rangle \equiv \frac{\int \mathrm{d} \{X_i\} ~ R(\{X_i\}) \; \left | \frac{\mathrm{d} E}{\mathrm{d} \{ X_i \}} \right |}{\int \mathrm{d} \{X_i\}~ \left | \frac{\mathrm{d} E}{\mathrm{d} \{X_i\}} \right |},
    \label{eqn:SPL_expected_value}
\end{equation}
where $\{X_i\}$ can be any of $\{\mathbf{k}, \mathbf{q}, x, \Delta z\}$ and $\mathrm{d} \{ X_i \} \equiv \prod_i \mathrm{d} X_i$. Also note that $R$ can depend on quantities that are not integrated over, such as $\{L, E, \mu\}$. This expectation value is calculated with respect to the differential energy loss $d E/d \{X_i\}$, since the energy loss is closely related to the $R_{AB}$, the observable of interest in this work. While we self-consistently found that $\left\langle R \right\rangle \ll 1$ in the collinear and soft approximations, we observed $\left\langle R \right\rangle \gg 1$ at high $p_T$ for the LFT approximation in \emph{both} DGLV and DGLV + SPLC.

\begin{figure}[!b]
	\centering
	\includegraphics[width=\linewidth]{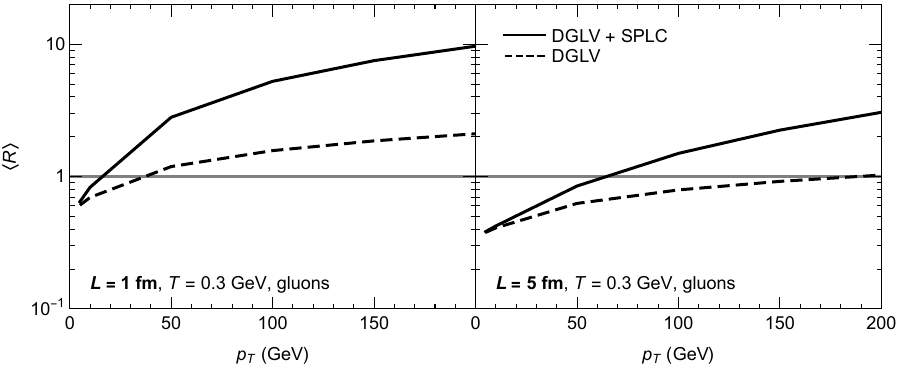}
	\caption{Plot of the expectation value $\left\langle R \right\rangle$ of the ratio $R \equiv (\mathbf{k} - \mathbf{q})^2 / (2 x E \sqrt{\mu^2 + \mathbf{q}^2})$ as a function of incident momentum $p_T$. Theoretical radiative energy loss curves are shown for DGLV (solid) and DGLV + SPLC (dashed). The large formation time approximation used in the derivation of DGLV and DGLV + SPLC assumes $R \ll 1$. Theoretical curves are produced for gluons moving through a brick of QGP at constant temperature of $T = 0.3 ~\mathrm{GeV}$ with a constant path length of $L = 1 ~\mathrm{fm}$ (left) and $L = 5 ~\mathrm{fm}$ (right). }
	\label{fig:lft_consistency_old_bounds}
\end{figure}

\Cref{fig:lft_consistency_old_bounds} shows the expectation of $\left\langle R \right\rangle$ where $R \equiv ((\mathbf{k} - \mathbf{q}_1)^2)/(2 x E \sqrt{\mu^2 + \mathbf{q}_1^2})$ as a function of the incident energy $E$ for gluons at constant length $L = 1 ~\mathrm{fm}$ (left) and $L = 5 ~\mathrm{fm}$ (right), both at constant temperature $T = 0.3 ~\mathrm{GeV}$. 
These values of length and temperature are similar to the average values for central \coll{p}{Pb} collisions ($L \sim 1 ~\mathrm{fm}$) and \coll{Pb}{Pb} collisions ($L \sim 5 ~\mathrm{fm}$). We only plot the results for gluons; light, charm, and bottom quarks all have significantly smaller values of $\left\langle R \right\rangle$.
	We observe in the right panel of the figure that for $L = 5 ~\mathrm{fm}$ the LFT approximation is violated for $p_T \gtrsim 200 ~\mathrm{GeV}$ in the DGLV result and for $p_T \gtrsim 80 ~\mathrm{GeV}$ when the SPLC is included. In the left panel, we observe that for $L = 1 ~\mathrm{fm}$ the LFT approximation is violated for $p_T \gtrsim 50 ~\mathrm{GeV}$ for the DGLV result and for $p_T \gtrsim 10 ~\mathrm{GeV}$ when the SPLC is included. Based on these results, we concluded in \cite{Faraday:2023mmx} that the breakdown of the large formation time approximation likely explains the large $\mathcal{O}(200\%)$ short path length correction to energy loss at high $p_T$.

In \cite{Faraday:2023mmx,Faraday:2023uay}, we suggested that one could restrict the phase space of radiated gluon emission to regions where the large formation time approximation is valid, similar to the standard procedure performed using the collinear approximation \cite{Armesto:2011ht}. For the collinear approximation, $k^+ \gg k^- \implies 2 xE \gg |\mathbf{k}|$ and $p^+ \gg p^- \implies |\mathbf{k}| \ll 2 (1-x)E$. 
In our previous works \cite{Faraday:2023mmx,Faraday:2024gzx}, we adopted the prescription \cite{Wicks:2005gt,Armesto:2011ht,Horowitz:2009eb}
\begin{equation}
	|\mathbf{k}|_{\text{max}}  =  2 x E(1-x).
	\label{eqn:coll_upper_bound}
\end{equation}

We may follow a similar procedure for the LFT approximation, which requires that \cref{eqn:lft_assumption} holds. We may proceed by assuming that $|\mathbf{k}| \gg |\mathbf{q}|$, which we validated by
numerically checking that the typical $|\mathbf{q}|$ which contributes to energy loss is $\mathcal{O}(20\text{--}30)\%$ of the typical $|\mathbf{k}|$ for $p_T \gtrsim 50 ~\mathrm{GeV}$ for phenomenologically applicable path lengths and temperatures. Moreover, the typical $|\mathbf{k} - \mathbf{q}|^2$ is within $5\%$ of the typical $\mathbf{k}^2$, weighted by their contribution to energy loss, indicating that $\mathbf{k} \gg \mathbf{q}$ is a good approximation. Under this assumption, we obtain
\begin{equation}
	|\mathbf{k}|_{\text{max}}^{\text{LFT}}  =  \sqrt{2 x E} \left(\mu^2 + \mathbf{q}^2\right)^{1 /4 }.
	\label{eqn:LFT_upper_bound}
\end{equation}

\begin{figure}[!t]
	\centering
	\includegraphics[width=\linewidth]{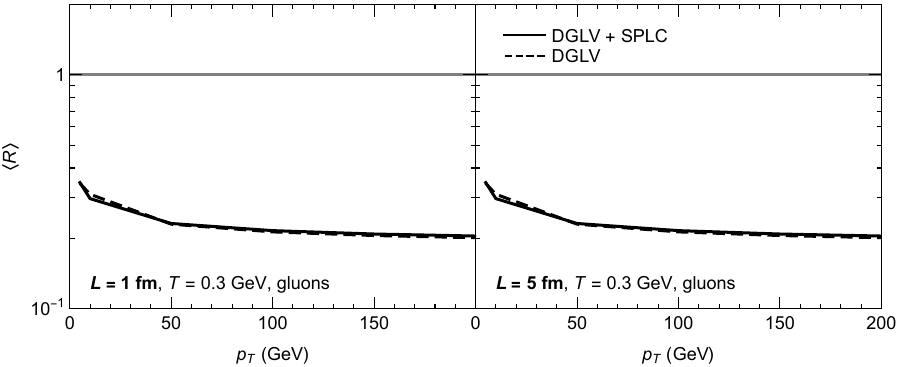}
	\caption{Plot of the expectation value $\left\langle R \right\rangle$ of the ratio $R \equiv \mathbf{k}^2 / (2 x E \sqrt{\mu^2 + \mathbf{q}^2})$ as a function of $p_T$. Theoretical radiative energy loss curves are shown for DGLV (dashed) and DGLV + SPLC (solid) with the upper bound on the transverse radiated gluon momentum from the collinear + large formation time approximation in \cref{eqn:LFT_coll_upper_bound}. The large formation time approximation used in the derivation of DGLV and DGLV + SPLC assumes $R \ll 1$.} 
	\label{fig:lft_consistency}
\end{figure}

To ensure that both the LFT and collinear approximations are respected, we impose the combined constraint
\begin{equation}
	|\mathbf{k}|_{\text{max}}^{\text{LFT + coll.}}  =  \min\left[2 x E (1-x), \sqrt{2 x E} \left(\mu^2 + \mathbf{q}^2\right)^{1 /4 }\right].
	\label{eqn:LFT_coll_upper_bound}
\end{equation}

We now return to the computation of $\left\langle R \right\rangle$ where $R = \mathbf{k}^2 / (2 x E \sqrt{\mu^2 + \mathbf{q}^2})$, however, with the new bound from \cref{eqn:LFT_coll_upper_bound} that ensures both the large formation time and collinear approximations are respected. \Cref{fig:lft_consistency} plots $\left\langle R \right\rangle$ as a function of incident energy $E$ for the new upper bound in \cref{eqn:LFT_coll_upper_bound} for both DGLV (dashed) and DGLV + SPLC (solid).
From the figure, we observe that our new cutoff on the transverse radiated gluon momentum, given by \cref{eqn:LFT_coll_upper_bound}, self-consistently results in $\langle R \rangle  \sim 0.2 \ll 1$  for both the DGLV and DGLV + SPLC results.

\begin{figure}[!tpb]
	\centering
	\includegraphics[width=\linewidth]{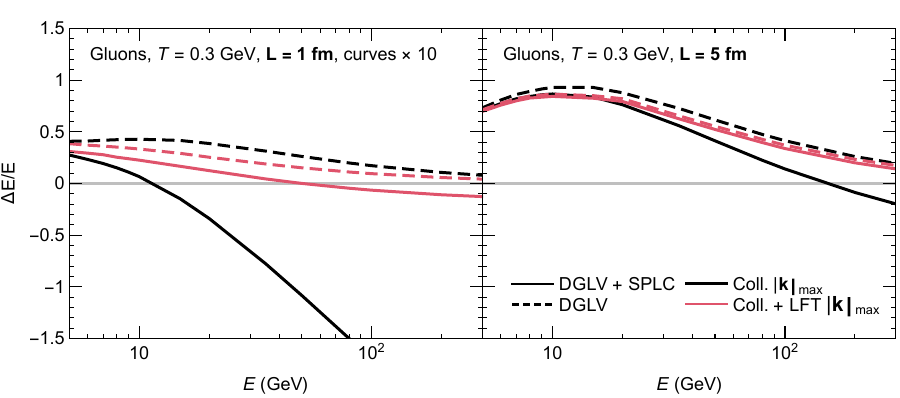}
	\caption{$\Delta E / E$ as a function of $E$ for incident gluons moving through a brick of QGP with constant temperature $T = 0.3 ~\mathrm{GeV}$ and constant length $L = 1 ~\mathrm{fm}$ (left) and $L = 5 ~\mathrm{fm}$ (right). Theoretical radiative energy loss curves are shown for DGLV (solid) and DGLV + SPLC (dashed) in conjunction with the upper bound on the transverse radiated gluon momentum from the collinear approximation only (black) and from the collinear + large formation time approximation (red). All curves shown in the left panel are multiplied by 10 for purposes of illustration.}
	\label{fig:deltaEoverE_small_large_LFT_kmax}
\end{figure}

Now that we have shown that our new large formation time + collinear (LFT + coll.) upper bound ensures that the theory self-consistently respects the large formation time approximation, we turn our attention to the effects of this new upper bound on the energy loss. \Cref{fig:deltaEoverE_small_large_LFT_kmax} plots the fractional energy loss $\Delta E / E$ as a function of the incident energy $E$ for gluons moving through a brick of QGP with constant temperature $T = 0.3 ~\mathrm{GeV}$ and constant length $L = 1 ~\mathrm{fm}$ (left) and $L = 5 ~\mathrm{fm}$ (right). Theoretical energy loss curves are shown for standard DGLV energy loss from \cref{eqn:DGLV_dndx} (solid) and for DGLV + SPLC \cite{Kolbe:2015rvk,Kolbe:2015suq} (dashed) in combination with the bound on $|\mathbf{k}|_{\text{max}}$ determined according to the collinear approximation from \cref{eqn:coll_upper_bound} (black) as well as the LFT + coll.\ bound from \cref{eqn:LFT_coll_upper_bound} (red).

In the figure, we observe that the SPLC is dramatically reduced when using the new LFT + coll.\ bound. We note that while at very large energies and very small path lengths, there is a small amount of negative energy loss (energy gain) present when the SPLC is included, this effect is unlikely to be phenomenologically relevant since the absolute magnitude of the energy gain is small. 
Since the LFT + coll.\ kinematic cutoff scales as $\sim \sqrt{E}$ while the collinear kinematic cutoff scales as $\sim E$, we expect that at large energies the LFT + coll.\ kinematic cutoff restricts the phase space for gluon emission more than the collinear kinematic cutoff. 
We see numerically in \cref{fig:deltaEoverE_small_large_LFT_kmax} that the LFT + coll.\ upper bound reduces the energy loss in both DGLV and DGLV + SPLC by $\sim \! 20\%$ %
for $p_T \gtrsim 50 ~\mathrm{GeV}$.
This reduction of energy loss may, at least in part, describe the fast rise of the charged hadron $R_{AA}$ in $p_T$, as opposed to, for instance, running coupling effects. %

\begin{figure}[!b]
	\centering
	\includegraphics[width=\linewidth]{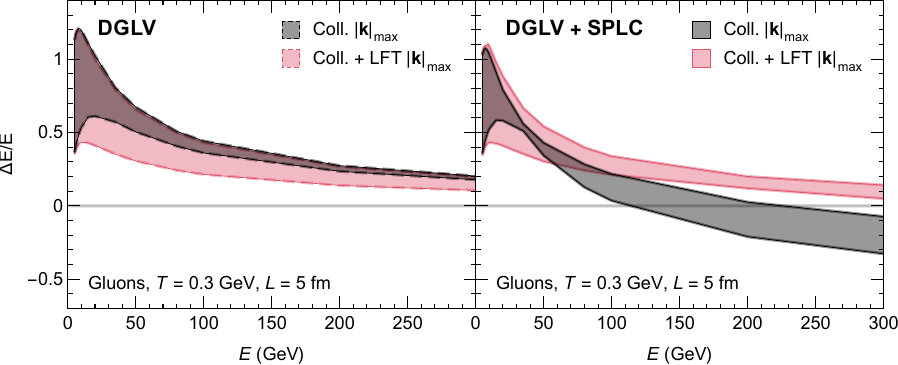}
	\caption{$\Delta E / E$ as a function of incident parton energy $E$ for gluons traversing a brick of QGP with $L = 5 ~\mathrm{fm}$ and $T = 0.3 ~\mathrm{GeV}$. Band widths are generated by varying the upper bound $|\mathbf{k}|_{\text{max}}$ on the transverse radiated gluon momentum $\mathbf{k}$ by factors of two up and down. Results are shown for $|\mathbf{k}|_{\text{max}}$ determined by ensuring that the collinear approximation is self-consistently satisfied (black) and by ensuring that both the collinear and large formation time approximations are self-consistently satisfied (red). The left figure shows results for DGLV energy loss and the right for DGLV + SPLC energy loss.}
	\label{fig:deltaE_collinear_LFT_varying_kmax}
\end{figure}

	The $\sqrt{E}$ vs $E$ scalings in the $|\mathbf{k}|_{\text{max}}$ upper bound likely make our results more sensitive to the exact value of the kinematic cutoff chosen when the LFT + coll.\ $|\mathbf{k}|_{\text{max}}$ is used. To investigate this sensitivity to the exact cutoff value, we consider varying the upper bound by a factor, which we will refer to as the ``$|\mathbf{k}|_{\text{max}}$ multiplier", which varies in the range $[0.5, 2]$. This procedure is similar to that followed in \cite{Horowitz:2009eb}, where the authors found that the energy loss has significant sensitivity to the exact angle used to cutoff the radiation according to the collinear approximation. \Cref{fig:deltaE_collinear_LFT_varying_kmax} plots the fractional radiative energy loss $\Delta E / E$ as a function of the incident energy $E$ for LFT + coll.\ $|\mathbf{k}|_{\text{max}}$ (red) and collinear $|\mathbf{k}|_{\text{max}}$ (black). The left panel shows the DGLV results and the right panel shows the DGLV + SPLC results. We see in the left panel of the figure that the DGLV with the collinear $|\mathbf{k}|_{\text{max}}$ has smaller sensitivity to the value of the $|\mathbf{k}|_{\text{max}}$ multiplier than the DGLV result with the LFT + coll.\ $|\mathbf{k}|_{\text{max}}$. One may understand this as the $\sqrt{E}$ scaling of the LFT + coll.\ $|\mathbf{k}|_{\text{max}}$ cutting off the phase space earlier than the $E$ scaling of the collinear $|\mathbf{k}|_{\text{max}}$. From \cref{eqn:DGLV_dndx}, we see that the spectrum goes like $d N / d \mathbf{k} \sim 1 / \mathbf{k}$ at high $p_T$, and so cutting the spectrum off earlier in $\mathbf{k}$ leads to a larger sensitivity to the exact value of the cutoff.
	In the right panel we observe a reduced sensitivity to the exact value of the cutoff when the LFT + coll.\ $|\mathbf{k}|_{\text{max}}$ is used compared to the collinear $|\mathbf{k}|_{\text{max}}$ for DGLV + SPLC. 
	Numerical investigation indicates that SPLC does not fall off as quickly in $\mathbf{k}$ compared to DGLV, leading to an increased sensitivity to the $|\mathbf{k}|_{\text{max}}$ multiplier when only the coll.\ $|\mathbf{k}|_{\text{max}}$ is used. Comparing the LFT + coll.\ bands for DGLV (left) and DGLV + SPLC (right), we observe that the sensitivity is slightly reduced for DGLV + SPLC compared to DGLV over the full $p_T$ range. This reduced sensitivity is because an increase in the $|\mathbf{k}|_{\text{max}}$ upper bound typically leads to a slight reduction in the energy loss for the DGLV + SPLC result, which stems from the fact that the DGLV + SPLC energy loss is negative at large $\mathbf{k}$ while DGLV alone is positive at large $\mathbf{k}$.
	The sensitivity to the value of the $|\mathbf{k}|_{\text{max}}$ multiplier used will be propagated through to the $R_{AB}$ in this analysis and treated as a fundamental theoretical uncertainty, which is further discussed in \cref{sec:theoretical_unc}.

\subsection{Collisional energy loss}
\label{sec:collisional_energy_loss}

In this work, we present results with two different implementations of the collisional energy loss, which serves to estimate the theoretical uncertainty present in the collisional energy loss sector. The two implementations differ in how they transition between hard thermal loop (HTL) \cite{Braaten:1989mz, Klimov:1982bv, Pisarski:1988vd, Weldon:1982aq, Weldon:1982bn} propagators, which are valid for small momentum transfers, and vacuum propagators, which are valid at large momentum transfers. One approach to treat this theoretical uncertainty is to minimize the dependence on the transition momentum scale $q^*$ where one switches between the propagators \cite{Gossiaux:2008jv}; however, in this work we rather consider the transition scale a fundamental theoretical uncertainty. We have discussed this uncertainty at length in our previous work \cite{Faraday:2024gzx}.

In order to estimate the effect of this transition scale uncertainty on the predictions, we present results with two separate collisional energy loss implementations: Braaten and Thoma (BT) \cite{Braaten:1991we,Braaten:1991jj} and HTL-only \cite{Wicks:2008zz}. The BT collisional energy loss calculation \cite{Braaten:1991jj, Braaten:1991we} uses HTL propagators at small momentum transfers and vacuum propagators at large momentum transfers. Additionally, the calculation is performed in the small-coupling and high-temperature regime where the dependence on $q^*$ falls away. The HTL-only \cite{Wicks:2008zz} collisional energy loss calculation uses only HTL propagators for all momentum transfers. Additionally, the full kinematics for hard momentum transfers are kept.

\subsubsection{Braaten and Thoma collisional energy loss}
\label{sec:braaten_thoma_elastic}

The BT collisional energy loss \cite{Braaten:1991jj, Braaten:1991we} of a quark is calculated under two approximations: $E \ll M^2 / T$ and $E \gg M^2 / T$, where $M$ is the mass of the incident quark, and $T$ is the medium temperature. For $E \ll M^2 / T$ the energy loss $dE$ differential in the traversed distance $z$ is
\begin{equation}
\frac{\mathrm{d} E}{\mathrm{d} z} = 2 C_R \pi \alpha_s^2 T^2 \left[\frac{1}{v}- \frac{1-v^2}{2 v^2}\ln \frac{1+v}{1-v}\right] \ln \left(2^{\frac{n_f}{6+n_f}} B(v) \frac{E T}{m_g M}\right)\left(1+\frac{n_f}{6}\right),
\label{eqn:elastic_energy_loss_low}
\end{equation}
where $B(v)$ is a smooth function satisfying constraints listed in \cite{Braaten:1991we} which simplifies the formula, $v$ is the incident velocity, and $n_f$ is the number of active quark flavors in the plasma. For $E \gg M^2/T$ one finds
\begin{equation}
    \frac{\mathrm{d} E}{\mathrm{d} z} = 2 C_R \pi \alpha_s^2 T^2 \left(1 + \frac{n_f}{6}\right)  \ln \left(2^{\frac{n_f}{2(6+n_f)}} \, 0.92 \frac{\sqrt{E T}}{m_g}\right).
    \label{eqn:elastic_energy_loss_high}
\end{equation}
The energy loss at arbitrary incident energy is taken to be the connection of these two asymptotic results such that $\mathrm{d} E / \mathrm{d} z$ is continuous. 

\subsubsection{HTL-only collisional energy loss}
\label{sec:pure_htl_elastic_energy_loss}

The HTL-only collisional energy loss is calculated using only HTL propagators for all momentum transfers but with the full kinematics of the hard momentum transfer retained. In this section, we present the results needed to replicate the energy loss in our calculation; more detailed steps were produced in \cite{Wicks:2008zz} and our previous work \cite{Faraday:2024gzx}. We use the ``HTL-X1" procedure from \cite{Wicks:2008zz}.

The number of elastic scatters $N$ differential in the path length $z$, $dN / d z$, may be written as \cite{Wicks:2008zz}
\begin{equation}
		\frac{dN}{d z} = \frac{4 C_R \alpha_s^2}{\pi} \int_q \frac{n_B(\omega)}{q} \left( 1 - \frac{\omega^2}{q^2} \right)^2
		\sum_m\left[C_{L L}^{m}\left|\Delta_L\right|^2+2 C_{L T}^{m} \operatorname{Re}\left(\Delta_L \Delta_T\right)+C_{T T}^m\left|\Delta_T\right|^2\right],
	\label{eqn:scattering_rate_final}
\end{equation}
where
\begin{equation}
	\int_q \equiv \int \frac{d^3 q d \omega}{2 \pi} \frac{E}{E^{\prime}} \delta\left(\omega +E-E^{\prime}\right),
	\label{eqn:integral_q}
\end{equation}
$E$ is the energy of the incident parton, the index $m$ sums over all of the medium partons ($N_c^2 - 1$ gluons and $4 N_c n_f$ quarks), $\omega$ is the transferred energy, and $n_B$ is a Bose distribution. For energy loss, it is most convenient to integrate the angular dependence, assuming the system is isotropic, leaving the integral in $\omega$ and in $q \equiv | \vec{q}|$. We also perform a change of variables $\epsilon \equiv \omega / E$. 
That is, 
\begin{equation}
	\int_q = \int d q d \epsilon \; \frac{E}{E^{\prime}} \delta\left(\epsilon E +E-E^{\prime}\right).
	\label{eqn:integral_q_new}
\end{equation}

The coefficients $C_{L L}^{m}$, $C_{L T}^{m}$, and $C_{T T}^m$ are given in terms of various thermal integrals \cite{Wicks:2008zz}
\begin{equation}
	C^{m} \equiv \int_{\frac{1}{2}(\omega+q)}^{\infty} k^0 k d k\left(n_{m}\left(k^0-\omega\right)-n_{m}\left(k^0\right)\right) c^{\prime},
\end{equation}
where $n_m$ is the distribution function for the medium particle $m$ (a Bose distribution for gluons and a Fermi distribution for quarks) and the coefficients $c^\prime$ are
\begin{align}
	c_{L L}^{\prime} \equiv  & \left(\left(1+\frac{\omega}{2 E}\right)^2-\frac{q^2}{4 E^2}\right)\left(\left(1-\frac{\omega}{2 k^0}\right)^2-\frac{q^2}{4\left(k^0\right)^2}\right), \\
	c_{L T}^{\prime} \equiv  & 0, \\
	c_{T T}^{\prime} \equiv  & \frac{1}{2}\left(\left(1+\frac{\omega}{2 E}\right)^2+\frac{q^2}{4 E^2}-\frac{1-v^2}{1-\frac{\omega^2}{q^2}}\right) \left(\left(1+\frac{\omega}{2 k^0}\right)^2+\frac{q^2}{4\left(k^0\right)^2}-\frac{1-v_k^2}{1-\frac{\omega^2}{q^2}}\right).
\end{align}
In the above $v_k (v)$ is the velocity of the medium parton (incident parton), and the $c^\prime$ are independent of whether the medium parton is a quark or a gluon since we have assumed that the quarks and gluons differ only by the statistical distribution and the Casimir. One may then calculate the $C$ coefficients in terms of various thermal integrals, which one may perform analytically,
\begin{equation}
	I_n^m \equiv \int_{\frac{1}{2} (\omega + q)}^\infty dk \; (n_m (k-\omega) - n_m(k)) k^n.
	\label{eqn:general_thermal_integral}
\end{equation}
We provide the results of these integrals for bosons ($+$) and fermions ($-$),
\begin{align*}
I_0^\pm &= \pm T \ln\left(\frac{1 \mp e^{-\kappa_+}}{1 \mp e^{-\kappa_-}}\right) \\
I_1^\pm &= \pm T^2 \left[\operatorname{Li}_2\left(\pm e^{-\kappa-}\right) - \operatorname{Li}_2\left(\pm e^{-\kappa+}\right)\right] + \kappa_+ T I_0^\pm \\
I_2^\pm &= \pm 2 T^3 \left[\operatorname{Li}_3\left(\pm e^{-\kappa-}\right) - \operatorname{Li}_3\left(\pm e^{-\kappa+}\right)\right] \\
&\quad+ 2 \kappa_+ T I_1^\pm - \kappa_+^2 T^2 I_0^\pm,
\end{align*}
where $\kappa_{\pm} = (\omega \pm q)/2 T$ and $\operatorname{Li}_n$ is the polylogarithm function.

Finally, we integrate out the $z$ dependence to calculate the number of elastic scatterings $N$, differential in the lost energy fraction $\epsilon$,
\begin{align}
	\frac{dN}{d \epsilon}  &= \frac{d\omega}{d \epsilon} \frac{d}{d \omega} \int d z \frac{dN}{d z} \\
	&\simeq L E \frac{dN}{d z \; d\omega}.
	\label{eqn:dndx_pure_htl}
\end{align}
In the last step, we have assumed that $d N /dz$ is independent of $z$, which is true in a static brick. In future work \cite{Bert:2024}, we will consider a more realistic model for the medium, which includes the $z$ dependence in the temperature $T(z)$.

\subsection{Fluctuations and total energy loss}
\label{sec:total_energy_loss}

The total number of gluons that are radiated $\left\langle N^g \right\rangle$  and total number of elastic scatters $\left\langle N^s \right\rangle$ that occur are determined by the theoretical calculations in \cref{sec:radiative_energy_loss,,sec:collisional_energy_loss}, respectively. However, fluctuations in both quantities are important to consider, as energy loss is far less efficient than energy gain for the nuclear modification factor  $R_{AB}$  due to the steeply falling production spectrum \cite{Gyulassy:2001kr,Faraday:2024gzx}. Following \cite{Gyulassy:2001kr}, we proceed by assuming that gluon radiation and elastic scatters are both individually independent, and thereby individually follow Poisson distributions. 
Assuming the independence of the collisional and radiative energy loss kernels is reasonable because the radiative energy loss kernel has no collisional component: the derivation of the radiative energy loss kernel assumes the scattering potentials are infinitely massive \cite{Gyulassy:1993hr}. 
We take the probability for a parton to lose a fraction $x$ of its energy to radiative (collisional) processes given that a single gluon radiation (elastic scatter) has occurred to be $P_{\text{rad}}^{(1)}(x) = \left\langle N^g \right\rangle^{-1} dN^g / dx$ ($P_{\text{coll}}^{(1)}(x) = \left\langle N^s \right\rangle^{-1} dN^s / dx$). Then, the probability for a parton to lose a fraction $\epsilon$ of its energy through radiative processes is~\cite{Gyulassy:2001nm}
\begin{multline}
	P_{\text{rad}}(\epsilon) = e^{-N^g} \delta(x)  +\\ \sum_{n = 1}^{\infty} \frac{e^{- N^g} \left(N^g\right)^n}{n!} \int d x_1 \cdots d x_n \; P_{\text{rad}}^{(1)}(x_1)  \cdots  P_{\text{rad}}^{(1)}(x_n) \delta\left[\epsilon - (x_1  + \cdots + x_n)\right],
\end{multline}
and similarly for collisional energy loss. %

This procedure is used for the radiative energy loss and the HTL-only collisional energy loss; however, this procedure cannot be used for the BT collisional energy loss implementation because the number of elastic scatters is not a finite quantity in the BT calculation. 
We therefore model the distribution of BT collisional energy loss with a Gaussian whose mean is given by the BT calculation and whose width is given by the fluctuations dissipation theorem \cite{Moore:2004tg}.
In our previous work \cite{Faraday:2024gzx}, we compared the effects of treating the HTL-only energy loss as Gaussian vs.\ Poisson and found the difference to be $\mathcal{O}(5\text{--}15\%)$ at mid- to high-$p_T$. We therefore expect that the modeling of BT collisional energy loss as a Gaussian distribution does not significantly affect our results.

In our model, the contributions from the radiative and collisional energy loss are calculated separately. 
We proceed by assuming that the amount of collisional energy loss is independent of the amount of radiative energy loss, which yields the total probability distribution
\begin{align}
  P_{\text{tot}}(\epsilon) = \int d x \, P_{\text{coll}}(x)P_{\text{rad}}(\epsilon-x).
\end{align}

\subsection{Nuclear modification factor and spectra}
\label{sec:nuclear_modification_factor}

The \emph{leading hadron nuclear modification factor} $R^h_{AB}(p_T)$, for the collision \coll{A}{B} and final state hadron $h$, is defined as
\begin{equation}
    R^h_{AB}(p_T) \equiv \frac{\mathrm{d} N^{AB \to h + X} / \mathrm{d} p_T}{\langle N_{\text{coll}} \rangle \mathrm{d} N^{pp \to h + X} / \mathrm{d} p_T},
    \label{eqn:nuclear_modification_factor}
\end{equation}
where $\mathrm{d} N^{AB / pp \to h + X} / \mathrm{d} p_T$ is the inclusive differential number of measured $h$ hadrons in \coll{A}{B} / \coll{p}{p} collisions, and $\langle N_{\text{coll}} \rangle$ is the expected number of binary collisions usually calculated according to the Glauber model \cite{Glauber:1970jm,Miller:2007ri}. In this work, we only compare to the leading hadron nuclear modification factor. However, future extensions of our model will consider the azimuthal anisotropy through multi-particle \cite{Bert:2024} as well as dihadron correlations.

We make several assumptions about the underlying collision and partons to access the $R_{AB}$ theoretically from within our energy loss model. We first assume, following \cite{Wicks:2005gt,Horowitz:2010dm}, that the spectrum of produced partons $q$ in the initial-state of the plasma formed by the collisions \coll{A}{B} (before energy loss) is $d N^{q}_{AB\text{, i}} / d p_i = N_{\text{coll}} \times dN^q_{pp} / dp_i$, where $d N^q_{pp} / d p_i$ is the parton production spectrum in \coll{p}{p} collisions. 
This assumption is equivalent to neglecting initial-state effects, for instance initial-state energy loss and nuclear parton distribution functions (nPDFs), which is justified in heavy-ion collisions by the measured $R_{AA}$ consistent with unity for probes that do not interact strongly with the QGP \cite{CMS:2012oiv, CMS:2011zfr}. Quantitatively, nPDF-only calculations of the nuclear modification factor in \coll{Pb}{Pb} collisions predict a $\sim 10\%$ enhancement for $p_T \sim 10\text{--}200 ~\mathrm{GeV}$ jets \cite{Helenius:2012wd}, significantly smaller than the $\sim 80\%$ suppression due to energy loss.
In smaller systems such as \collFour{p}{d}{He3}{A} and light-ion collisions, nPDF effects are expected to be smaller in absolute terms but can represent a larger share of the overall nuclear modification, since final-state effects are also reduced. 
The measurement of the prompt photon $R_{AB}$ in \coll{p}{A} collisions \cite{ATLAS:2019ery} is consistent with unity to within $20\%$, leaving potential scope for initial-state effects to be as large as $20\%$. For example, theoretical calculations that include the effects of nPDFs predict $\sim 10\%$ deviations from one in light-ion  \cite{Gebhard:2024flv, Huss:2020whe} and \collFour{p}{d}{He3}{A} \cite{Helenius:2012wd} collisions. Other initial-state effects include the Cronin effect, which leads to enhancement for $p_T \lesssim 10 ~\mathrm{GeV}$, and initial-state energy loss \cite{Vitev:2007ve}, which predicts $\sim 10\%$ suppression for $p_T \gtrsim 10 ~\mathrm{GeV}$ \cite{Ke:2022gkq}. In summary, initial-state effects may individually lead to $\mathcal{O}(10\text{--}20\%)$ deviations from unity for $R_{AB}$ in both small and large systems. 
 However, because the sign and magnitude of these initial-state effects are similar in small and large systems, their effect on predictions (e.g.\ in small systems such as \coll{p}{A} and \coll{O}{O}) will be reduced since the $\alpha_s^{\text{eff.}}$ extracted by comparison to large system data will similarly be affected.
We therefore expect that the net impact of initial state effects on our $R_{AB}$ calculations in both \coll{A}{A} and \collFour{p}{d}{He3}{A} is likely much smaller than the naive $\sim 10\%$ that one might expect.

In addition to neglecting initial-state effects, we further assume that the modification of the \coll{A}{B} production spectrum $d N^{q}_{AB \text{, i}}$ is due only to energy loss,
\begin{equation}
	d N^q_{AB; \text{ final}}\left(p_T\right)=\int  d \epsilon \; d p_i \; d N^q_{AB, i}\left(p_i\right) P_{\text{tot}}\left(\epsilon \mid p_i\right) \delta \big ( p_T - (1 - \epsilon) p_i\big ),
\end{equation}
where $P_{\text{tot}}(\epsilon | p_i)$ is the  probability of losing a fraction of transverse momentum $\epsilon$ given an initial transverse momentum $p_i$. Given these assumptions, the expression for the partonic $R^q_{AB}$ may be written as
\begin{equation}
    R^q_{AB} = \frac{1}{f^q\left(p_T\right)} \int \frac{d \epsilon}{1-\epsilon} f^q\left(\frac{p_T}{1-\epsilon}\right) P_{\text {tot}}\left(\epsilon \left| \frac{p_T}{1-\epsilon}\right.\right),
    \label{eqn:full_raa_spectrum_ratio}
\end{equation}
where the $1 / 1 - \epsilon$ factor is due to changing integration variables, and we have introduced the notation $f^q(p_T) \equiv dN^q / d p_T (p_T)$. In the literature \cite{Wicks:2005gt, Horowitz:2010dm} %
and in our previous work \cite{Faraday:2023mmx}, an assumption was made that the production spectra followed a slowly-varying power law $f^q(p_T) \simeq A p_T^{-n(p_T)}$, which resulted in further simplifications. In this work, we calculate all presented $R_{AB}$ results according to \cref{eqn:full_raa_spectrum_ratio}. The validity of the slowly varying power law approximation is discussed in detail in \cite{Faraday:2024gzx}.

	We calculate heavy quark production spectra $f^q(p_T)$ using FONLL \cite{Cacciari:2001td}, and we compute gluons and light quark production spectra at leading order with \cite{wang_private_communication}, as in \cite{Vitev:2002pf,Wicks:2005gt, Horowitz:2011gd}. 

The spectrum $d N^h / d p_h$ for a hadron $h$ is related to the spectrum $d N^q / d p_q$ for a parton $q$ via \cite{Horowitz:2010dm}
\begin{equation}
	\frac{d N^h}{d p_h}\left(p_h\right) =\int \frac{dz}{z}\frac{d N^q}{d p_q}\left(\frac{p_h}{z}\right) D^h_q(z, Q),
    \label{eqn:parton_to_hadron_spectrum}
\end{equation}
where $z \equiv p_h / p_q \in (0,1]$, $p_h$ is the observed hadron momentum, $D^h_q(z,Q)$ is the fragmentation function for the process $q \mapsto h$, and $Q$ is the hard scale of the problem taken to be $Q = p_q =  p_h / z$ \cite{Horowitz:2010dm}. 
Fragmentation functions are taken from \cite{deFlorian:2007aj} for pions, and from \cite{Cacciari:2005uk} for $D$ and $B$ mesons. For simplicity, we compute theoretical results only for $\pi^0$ mesons (assumed to have the same fragmentation as $(\pi^+ + \pi^-)/2$ \cite{deFlorian:2007aj}), $D$ mesons (with $D \equiv 0.7 D^0 + 0.3 D^+$ \cite{Cacciari:2005uk}), and $B$ mesons (treating all $B$ mesons identically \cite{Cacciari:2005uk}).
The hadronic $R^h_{AB}$ is then found in terms of the partonic $R^q_{AB}$ (\cref{eqn:full_raa_spectrum_ratio}) as \cite{Horowitz:2010dm}
\begin{align}
    R_{AB}^h\left(p_T\right)&=\frac{\sum_q \int d z \frac{1}{z} D_q^h(z) f^q\left(\frac{p_T}{z}\right) R_{AB}^q\left(\frac{p_T}{z}\right)}{\sum_q \int d z \frac{1}{z} D_q^h(z) f^q\left(\frac{p_T}{z}\right)}.
    \label{eqn:parton_to_hadron_raa}
\end{align}
For details of the derivation needed for \cref{eqn:parton_to_hadron_raa}, refer to Appendix B of \cite{Horowitz:2010dm}.

\begin{figure}[!t]
	\centering
	\includegraphics[width=0.65\linewidth]{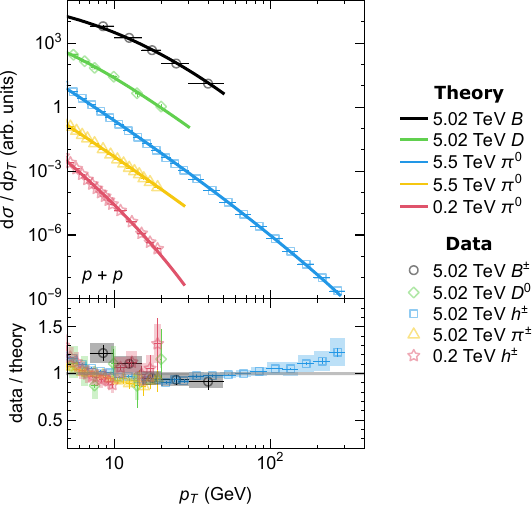}
	\caption{The top panel shows the $p_T$-differential production cross section $d \sigma / d p_T$ in \coll{p}{p} collisions for final state hadrons as a function of final hadron $p_T$. Experimental data are shown for 
charged hadrons at $\sqrt{s_{NN}} = 200 ~\mathrm{GeV}$ \cite{PHENIX:2007kqm}, charged hadrons \cite{ATLAS:2022kqu} and charged pions \cite{ALICE:2019hno} at $\sqrt{s_{NN}} = 5.02 ~\mathrm{TeV}$, $D^0$ mesons at $\sqrt{s_{NN}} = 5.02 ~\mathrm{TeV}$ \cite{ALICE:2016yta}, and $B^{\pm}$ mesons at $\sqrt{s_{NN}} = 5.02 ~\mathrm{TeV}$ \cite{CMS:2017uoy}.
		Theoretical calculations are shown for pions at $\sqrt{s_{NN}} = 200 ~\mathrm{GeV}$ and $\sqrt{s_{NN}} = 5.5 ~\mathrm{TeV}$, as well as $D$ and $B$ mesons at $\sqrt{s_{NN}} = 5.02 ~\mathrm{TeV}$. 
		Both experimental data and theoretical curves are shifted vertically for visual clarity. The bottom panel shows the ratio of the same experimental data to the theoretical predictions as a function of $p_T$.
Statistical uncertainties are represented by bars while systematic uncertainties are represented by shaded boxes.}
	\label{fig:pp_spectrum}
\end{figure}

\Cref{fig:pp_spectrum} plots the hadron $h$ production spectrum $d \sigma^{p + p \to h + X} / d p_T$ as a function of the final hadron momentum $p_T$ for the process $p + p \to h + X$. The spectra produced by our model are shown for pions at $\sqrt{s_{NN}} = 200 ~\mathrm{GeV}$ and $\sqrt{s_{NN}} = 5.5 ~\mathrm{TeV}$, as well as $D$ and $B$ mesons at $\sqrt{s_{NN}} = 5.02 ~\mathrm{TeV}$. We also show the $p + p \to h + X$ data for charged hadrons at $\sqrt{s_{NN}} = 200 ~\mathrm{GeV}$ \cite{PHENIX:2007kqm}, charged hadrons \cite{ATLAS:2022kqu} and charged pions \cite{ALICE:2019hno} at $\sqrt{s_{NN}} = 5.02 ~\mathrm{TeV}$, $D^0$ mesons at $\sqrt{s_{NN}} = 5.02 ~\mathrm{TeV}$ \cite{ALICE:2016yta}, and $B^{\pm}$ mesons at $\sqrt{s_{NN}} = 5.02 ~\mathrm{TeV}$ \cite{CMS:2017uoy}.
	The theoretical results are fit to the \coll{p}{p} data by multiplying by a constant $K$-factor for each data set, and both the theoretical curves and experimental data points are shifted vertically for visual clarity. Since $R_{AB}$ is insensitive to multiplicative differences in the spectra, these changes have no impact on our predictions.

	We observe that all of the measured \coll{p}{p} spectra are well-described in our model. 
Notably, the approximation that charged hadron spectra share similar $p_T$ dependence with $\pi^0$ meson spectra, and likewise for $D^0$ and $D$ mesons and $B^\pm$ and $B$ mesons, is justified by the agreement with \coll{p}{p} data for $p_T \gtrsim 5 ~\mathrm{GeV}$. The small difference between $\sqrt{s_{NN}} = 5.02~\mathrm{TeV}$ for the measured charged hadron and pion data and $\sqrt{s_{NN}} = 5.5~\mathrm{TeV}$ for the theoretical neutral pion cross section has a negligible effect, as seen by the agreement between the experimental data and theoretical calculation.

\subsection{Geometry} 
\label{sec:geometry}

The computation of an $R_{AA}$ that can be compared to experiment requires a model for the geometry and thermodynamic properties of the QGP. In \cref{sec:brick_model}, we discuss the canonical static ``brick" of QGP \cite{Armesto:2011ht} wherein the analytic pQCD-based energy loss calculations described in \cref{sec:radiative_energy_loss,sec:collisional_energy_loss} are performed. In \cref{sec:realistic_geometry}, we discuss how the fluctuating initial conditions that more accurately describe realistic heavy-ion collisions can be mapped to the simplified brick geometry. Finally, in \cref{sec:boundary_conditions}, we describe how we model energy loss before thermalization and after freeze-out.

\subsubsection{Brick model}
\label{sec:brick_model}

The energy loss formulae in \cref{sec:radiative_energy_loss,sec:collisional_energy_loss} are all derived under the assumption of a static ``brick" of QGP \cite{Armesto:2011ht}. This idealized ``brick" is specified by three parameters: the length $L$, the temperature $T$, and the distribution of scattering centers $\bar{\rho}(\Delta z)$. The brick is static, in the sense that there is no time evolution of the medium, and infinite in the plane transverse to the direction of propagation. The distribution of scattering centers $\bar{\rho}$ controls changes in the longitudinal density of scattering centers. In this paper, we follow previous work and assume an exponential distribution of scattering centers \cite{Djordjevic:2003zk,Armesto:2011ht}; the distribution of scattering centers was discussed in more detail in \cref{sec:dglv_radiative_energy_loss}.

The energy loss formulae depend on various thermal quantities---the gluon mean free path $\lambda_g$, the Debye mass $\mu_D$, and the thermal light quark $m_l$ and gluon $m_g$ masses---that may be calculated from the temperature $T$. From HTL perturbation theory \cite{Blaizot:2001nr}, one may derive the leading order expression for the Debye mass $\mu_D$,
\begin{equation}
	\mu_D  = g T \sqrt{\frac{2 N_c  + n_f}{6}},
	\label{eqn:debye_mass}
\end{equation}
where $g \equiv \sqrt{4 \pi \alpha_s}$. We proceed by modeling the QGP as an ultrarelativistic mixture of a Fermi and Bose gas with zero chemical potential, following \cite{Wicks:2005gt, Horowitz:2010dm, Faraday:2023mmx, Wicks:2008zz}. The relevant thermodynamic quantities are then given by \cite{Horowitz:2010dm}
\begin{subequations}
    \label{eqn:thermodynamic_quantities}
		\begin{alignat}{2}
			\sigma_{gg}  =& \frac{C_A^2 \pi \alpha_s^2}{2 \mu^2} \quad \text{and} \quad \sigma_{qg}  = \frac{C_F}{C_A} \sigma_{gg},&\\
			\rho_g =& 2 (N_c^2 - 1) \frac{\zeta(3)}{\pi^2} T^3 \quad \text{and} \quad \rho_q = 3 N_c n_f \frac{\zeta(3)}{\pi^2} T^3 \label{eqn:rho_thermal_g},&\\
			\rho =& \rho_g + \frac{\sigma_{q g}}{\sigma_{gg}} \rho_{q} =\frac{\zeta(3) (N_c^2 - 1)}{\pi^2} T^3 \left( 2 + \frac{3n_f}{2N_c}\right)\label{eqn:density},&\\
			\lambda_g^{-1}  =& \rho_g \sigma_{g g}+\rho_q \sigma_{q g} = \sigma_{gg} \rho,& \label{eqn:mean_free_path}
    \end{alignat}
\end{subequations}
where $\zeta$ is the Riemann zeta function, $T$ is the temperature, $\rho_q$ ($\rho_g$) is the density of quarks (gluons), $\sigma_{qg}$ ($\sigma_{gg}$) is the gluon-gluon (quark-gluon) collisional cross section, $n_f$ is the number of active quark flavors (taken to be $n_f = 2$ throughout), and $N_c=3$ is the number of colors. We denote the cross section weighted density as $\rho$, which we will subsequently refer to as the density for simplicity. We use the asymptotic light quark and gluon thermal masses \cite{Braaten:1989mz, Djordjevic:2003be} of $m_l = \mu_D / 2$ and $m_g = \mu_D / \sqrt{2}$, respectively.

\subsubsection{Realistic collision geometries}
\label{sec:realistic_geometry}

Real heavy-ion collisions produce highly non-uniform, fluctuating temperature profiles that evolve in time. In order to capture these important properties of heavy-ion collisions, while still being able to use the simplified energy loss formulae from \cref{sec:radiative_energy_loss,sec:collisional_energy_loss}, we require a mapping from the realistic collision profiles to the simplified brick medium.

To model the initial temperature distribution $T(\mathbf{x}, \tau_0)$ we use fluctuating IP-Glasma initial conditions \cite{Schenke:2012hg, Schenke:2012wb}. The IP-Glasma profiles are taken from \cite{Schenke:2020mbo}\footnote{All collision profiles, except for \coll{O}{O} and \coll{Ne}{Ne}, were obtained from \cite{shen_private_communication}. The \coll{O}{O} and \coll{Ne}{Ne} profiles were generated by us using the same parameters as in \cite{Schenke:2020mbo}, but with an updated $\sqrt{s_{NN}} = 5.36~\mathrm{TeV}$ to match the LHC experimental data. We validated our implementation by reproducing the \coll{Pb}{Pb} results from \cite{Schenke:2020mbo}.}, where the original authors demonstrated that, after full evolution with \textsc{MUSIC} viscous relativistic hydrodynamics \cite{Schenke:2010rr,Schenke:2011bn,Schenke:2010nt} and UrQMD microscopic hadron transport \cite{Bass:1998ca,Bleicher:1999xi}, a variety of measured bulk observables were accurately reproduced.

We extract the initial temperature profile $T(\mathbf{x}, \tau_0)$ at the thermalization time $\tau_0 = 0.4~\mathrm{fm}$. The temperature profiles may be mapped to the density using \cref{eqn:density}. Rather than using the full hydrodynamic evolution, we model the subsequent medium expansion using longitudinal Bjorken expansion \cite{Bjorken:1982qr} for consistency with previous work \cite{Djordjevic:2004nq,Djordjevic:2005db,Wicks:2005gt,Faraday:2023mmx,Faraday:2024gzx} and for numerical simplicity. We will discuss the details of the implementation of Bjorken expansion shortly. Work in progress \cite{Bert:2024} will incorporate the full viscous relativistic hydrodynamics evolution.

We follow WHDG \cite{Wicks:2005gt} and our previous work \cite{Faraday:2023mmx,Faraday:2024gzx}, and define the effective path length as
\begin{equation}
	L_{\text{eff}} (\mathbf{x}_i, \boldsymbol{\hat{\phi}}) = \frac{1}{\rho_{\text{eff}}} \int_{0}^\infty \mathrm{d}z \; \rho(\mathbf{x}_i + z \boldsymbol{\hat{\phi}}, \tau_0),
    \label{eqn:effective_length}
\end{equation}
and the effective density as
\begin{equation}
  \rho_{\text{eff}}(\tau_0) \equiv \frac{\int \mathrm{d}^2 \mathbf{x} \; \rho^2(\mathbf{x}, \tau_0)}{\int \mathrm{d}^2 \mathbf{x} \; \rho(\mathbf{x}, \tau_0)} \iff T_{\text{eff}}^{3}(\tau_0) \equiv \frac{\int \mathrm{d}^2 \mathbf{x} \; T^6(\mathbf{x}, \tau_0)}{\int \mathrm{d}^2 \mathbf{x} \; T^3(\mathbf{x}, \tau_0)}.
  \label{eqn:effective_density}
\end{equation}
In \cref{eqn:effective_length,eqn:effective_density}, the effective path length $L_{\text{eff}}$ includes all $(\mathbf{x}_i, \phi)$ dependence, and $\rho_{\text{eff}}$ is a constant for all paths that a parton takes through the plasma for a fixed centrality class. A weighting of the effective brick properties by the density is natural as the energy loss is proportional to the density in both the radiative and collisional case [see \cref{eqn:DGLV_dndx,eqn:elastic_energy_loss_low,eqn:elastic_energy_loss_high}]. However, there is no unique mapping from realistic collision geometries to simple brick geometries and more options are explored in \cite{Wicks:2008zz} and our previous work \cite{Faraday:2023mmx}. 

Longitudinal Bjorken expansion \cite{Bjorken:1982qr} is taken into account by approximating
\begin{equation}
T_{\text{eff}}(\tau) \approx T_{\text{eff}}(\tau_0) \left( \frac{\tau_0}{\tau} \right)^{1/3} \approx T_{\text{eff}}(\tau_0) \left( \frac{2 \tau_0}{L_{\text{eff}}} \right)^{1/3} \equiv T_{\text{eff}}\left(L_{\text{eff}}\right)
  \label{eqn:bjorken_expansion}
\end{equation}
where in the last step we have evaluated $T( \mathbf{x}, \tau)$ at the average time $\tau_{\text{avg}}=L_{\text{eff}} / 2$, following what was done in \cite{Wicks:2005gt, Djordjevic:2005db, Djordjevic:2004nq}. Transverse expansion is expected to negligibly impact the energy loss results and so is ignored \cite{Gyulassy:2001kr}. In \cite{Djordjevic:2004nq,Wicks:2005gt}, this average time approximation was found to be a good approximation to the full integration through the Bjorken expanding medium.

The starting locations of partons are assumed to be proportional to the binary collision density $n_{\text{coll}}(\mathbf{x}_i)$, which is calculated on an event-by-event basis from the initial collision profiles. Assuming further that hard partons are produced isotropically in the transverse plane, we compute the distribution of effective path lengths 
	\begin{equation}
		P^j_L\left(L_{\text{eff}}\right) \equiv \frac{\int d^2 \mathbf{x}_i d \phi \; n^j(\mathbf{x}_i) L^j_{\text{eff}}(\mathbf{x}_i)}{\int d^2 \mathbf{x}_i d \phi \; n^j(\mathbf{x_i})},
		\label{eqn:path_length_distribution}
	\end{equation}
where $j$ indexes a particular event. 
The geometry averaged $R_{AB}$ can then be computed in terms of the $R_{AB}$ at fixed length and temperature [calculated in \cref{eqn:full_raa_spectrum_ratio}]
	\begin{equation}
		R_{AB} = \frac{\sum_j \int dL_{\text{eff}} \; P^j_L\left(L_{\text{eff}}\right) \; R_{AB}\left[L = L_{\text{eff}}, T = T_{\text{eff}}\left(L_{\text{eff}}\right) \right]}{\sum_j\int dL_{\text{eff}} \; P^j_L\left(L_{\text{eff}}\right)},
		\label{eqn:geometry_averaged_rab}
	\end{equation}
	where the sum over $j$ is over all events in a specific centrality class, $P^i_L\left(L_{\text{eff}}\right)$ is calculated in \cref{eqn:path_length_distribution}, $L_{\text{eff}}$ is calculated in \cref{eqn:effective_length} and $T_{\text{eff}}(L_{\text{eff}})$ is calculated in \cref{eqn:bjorken_expansion}. 
	Centrality is classified according to the total charged particle multiplicity at mid-rapidity $| \eta | < 0.5$. 

	We use all available profiles that were provided to use from simulation \cite{shen_private_communication}, which amounted to between 2000 and 4000 events per collision system. There were more events in the smaller collision systems, as event-by-event fluctuations are more pronounced \cite{Schenke:2020mbo}. Parton starting points are distributed on a $20 \times 20$ grid, and each parton propagates isotropically along 20 equally spaced angles from $0$ to $2 \pi$.

\subsubsection{Energy loss before thermalization and after freeze-out}
\label{sec:boundary_conditions}

As an incident high-$p_T$ parton propagates through the QGP formed during a collision, it first traverses through a non-thermalized medium, then a thermalized QGP, and finally a hadronic medium. The first and last of these three phases are not rigorously modeled by pQCD energy loss, even though they likely contribute to the energy loss, which means we must make a phenomenological choice on how to perform energy loss during this period of the collision. 

We follow our previous work \cite{Faraday:2023mmx,Faraday:2024gzx,Faraday:2024qtl} and assume that the temperature distribution takes the form $T(\mathbf{x}, \tau) \equiv T(\mathbf{x}) \left(\tau_0/\tau\right)^{1 / 3}$ for a transverse position $\mathbf{x}$, proper time $\tau$, and thermalization time $\tau_0$. This assumption extrapolates back the post-thermalization time behavior of Bjorken expansion to the pre-thermalization time. Future work might treat pre-thermalization time energy loss as a theoretical uncertainty \cite{Xu:2014ica,Ilic:2021ezl} or perform first principles calculations of the pre-thermalization time energy loss \cite{Ipp:2020mjc,Carrington:2021dvw,Hauksson:2021okc,Avramescu:2023qvv,Boguslavski:2023waw,Barata:2024xwy}.

There is also theoretical uncertainty in the calculation of the energy loss after freeze-out. While the hydrodynamics temperature profiles have a turn-off at $T = 0.155$ GeV \cite{Schenke:2020mbo}, no such turn-off exists for the Bjorken expansion formula for the temperature. In our calculations we do not artificially add such a cutoff to the Bjorken expansion formula and, therefore, to a good approximation we extrapolate the energy loss into the hadronic phase.

\section{Statistical analysis}
\label{sec:statistical_analysis}

In this section we describe the method used to constrain our model to high-$p_T$, central and semi-central heavy-ion collision data from RHIC and LHC. \Cref{sec:chi2_minimization_procedure} describes the details of the $\chi^2$ minimization procedure that is used to constrain our model in this analysis. In \cref{sec:discussion_modeling_of_correlations_in_systematic_uncertainties}, we describe the assumptions that we have made in our modeling of the correlations and discuss how this might impact our results.
	In \cref{sec:experimental_data}, we describe the experimental selection criteria for the global extraction of $\alpha_s^{\text{eff.}}$ as well as the full set of experimental data that is used for other analyses in \crefrange{sec:applicability_of_pqcd_energy_loss_from_data}{sec:running_coupling}. \Cref{sec:theoretical_unc} describes how we treat various theoretical uncertainties in this work. 
	In \cref{sec:coverage} we discuss the sensitivity in our statistical analysis to the modeling of the covariance matrix, the ability for our chosen range of $\alpha_s^{\text{eff.}}$ to sufficiently cover the experimental data, and the range of lengths and temperatures that are probed by the experimental data considered in our analysis.

\subsection{Minimization procedure}
\label{sec:chi2_minimization_procedure}

We extract the effective strong coupling $\alpha_s^{\text{eff.}}$ in our model by minimizing the $\chi^2 (\alpha_s^{\text{eff.}})$. Generally, one may calculate the $\chi^2$ in terms of the covariance matrix $C$, theoretical model vector $\vec{\mu}(\alpha_s^{\text{eff.}})$, effective strong coupling $\alpha_s^{\text{eff.}}$, and the vector of data points $\vec{y}$ as \cite{ParticleDataGroup:2022pth}
\begin{equation}
	\chi^2(\alpha_s^{\text{eff.}}) \equiv \sum_{i,j}^N\left[y_i - \mu_i (\alpha_s^{\text{eff.}})\right] (C^{-1})_{ij} \left[y_j - \mu_j(\alpha_s^{\text{eff.}})\right],
	\label{eqn:chi2_definition}
\end{equation}
where $N$ is the number of experimental data points (i.e.\ the length of the $\vec{\mu}$ and $\vec{y}$ vectors), and the theoretical model values $\vec{\mu}$ are evaluated at the same momenta as the experimental values $\vec{y}$. The method to calculate the theoretical model vector $\mu_i(\alpha_s^{\text{eff.}})$ is described in \cref{sec:model}, and experimental data points $\vec{y}$ are reported by experiments; however, the covariance matrix $C$ is not reported by experiments and so some degree of modeling is required.

Contributions to the covariance matrix stem from measurement uncertainties and their correlations. Experimental measurements of the $R_{AB}$ generally report three types of uncertainties \cite{PHENIX:2008ove,PHENIX:2008saf,ATLAS:2022kqu}:

\begin{itemize}
    \item \emph{Type A}: Statistical and systematic uncertainties that are uncorrelated in $p_T$.
    \item \emph{Type B}: Systematic uncertainties that are correlated in $p_T$, though the form of the correlation is unknown.
    \item \emph{Type C}: Normalization uncertainties, primarily those from the estimation of $\left\langle N_{\text{coll}} \right\rangle$ in the Glauber model \cite{Miller:2007ri} and from luminosity, which are fully correlated in $p_T$.
\end{itemize}

For a dataset from a single experiment with a single centrality class, which we index by $k$, the contributions of Type A and Type C uncertainties to the covariance matrix are rigorously understood. We will assume a length-correlated form for the contribution of Type B uncertainties to the covariance matrix \cite{JETSCAPE:2021ehl,JETSCAPE:2024cqe}. The validity of the assumptions about the form of the covariance matrix is discussed in \cref{sec:discussion_modeling_of_correlations_in_systematic_uncertainties}.

If the total standard deviation associated with Type A, Type B, and Type C uncertainties for the data point $y_{k,i}$ are $\sigma^{\text{A}}_{k,i}$, $\sigma^{\text{B}}_{k,i}$, and $\sigma^{\text{C}}_{k,i}$, respectively, then the covariance matrix $C_{k}$ for a given experiment and centrality class indexed by $k$ may be written as
\begin{align}
	C_{k,ij} =& C^{\text{\emph{A}}}_{k,ij} + C^{\text{\emph{B}}}_{k,ij} + C^{\text{\emph{C}}}_{k,ij} \label{eqn:correlation_matrix}\\
	C_{k,ij}^{\text{\emph{A}}} =& \sigma_{k,i}^{\text{\emph{A}}} \sigma_{k, j}^{\text{\emph{A}}} \delta_{ij} \label{eqn:correlation_matrix_A}\\
	C_{k, ij}^{\text{\emph{B}}} =& \exp\left[- \left| \frac{p_{k,i} - p_{k,j}}{\ell_k(p_k^{\text{max}} - p_k^{\text{min}})}\right|^\alpha\right] \sigma_{k,i} \sigma_{k,j}\label{eqn:correlation_length},\\
	C_{k,ij}^{\text{\emph{C}}} =& \sigma_{k,i}^{\text{\emph{C}}} \sigma_{k, j}^{\text{\emph{C}}},
\end{align}
where in \cref{eqn:correlation_matrix_A,eqn:correlation_length} there is no implicit sum over $i$ and $j$, $\ell_k$ is the correlation length, $\alpha$ is the correlation exponent, $p_{k,i}$ is the momentum of the $i^{th}$ data point in the $k^{th}$ dataset, $p_k^{\text{min}}$ ($p_k^{\text{max}}$) is the minimum (maximum) momentum in the $k^{th}$ dataset that is used in the analysis. Typically $\alpha \sim 2$ is assumed [82, 83]; we take $\alpha= 1.9$ following JETSCAPE \cite{JETSCAPE:2021ehl,JETSCAPE:2024cqe}. Consistent with \cite{JETSCAPE:2024cqe,JETSCAPE:2021ehl}, we take $\ell_k = 0.2$; the sensitivity of our results to this choice is computed in \cref{sec:coverage}.

For a dataset that includes contributions from multiple experiments and centrality classes, we must make further approximations about the form of the covariance matrix. In particular, we will assume that contributions from different centrality classes and experiments are completely uncorrelated. If we index a particular experimental measurement for a particular centrality class by $k$, then the total covariance matrix is block diagonal
\begin{equation}
	C_{ij} = \sum_{k=1}^{N_{\text{exp.}}} C_{k, ij},
	\label{eqn:covariance_matrix_in_terms_of_experiments}
\end{equation}
where $N_{\text{exp.}}$ is the number of experimental datasets in the analysis. The block diagonal form of the covariance matrix allows the $\chi^2$ in \cref{eqn:chi2_definition} to simplify to
\begin{equation}
	\chi^2 = \sum_k^{N_{\text{exp.}}} \chi^2_k \quad \text{with} \quad \chi^2_k \equiv \sum_{i,j}^{N_k} \left( y_{k,i} - \mu_{k,i} \right) \left( C^{-1}\right)_{k,ij}  \left( y_{k,j} - \mu_{k,j} \right),
	\label{eqn:chi2}
\end{equation}
where $N_k$ is the number of data points in the dataset indexed by $k$ and $N = \sum_k^{N_{\text{exp.}}} N_k$.

	In the analyses in this work, we will find it useful to have both an extracted value $\alpha_s^{\text{eff.}}$ as well as a corresponding one standard deviation confidence interval. The one standard deviation confidence interval corresponds to the values of $\alpha_s^{\text{eff.}}$ for which $\chi^2$ increases by one relative to its minimum value \cite{ParticleDataGroup:2022pth}. Additionally, we will present $p$-values while interpreting the goodness-of-fit of our model to experimental data. The $p$-value is calculated as \cite{ParticleDataGroup:2022pth}

\begin{equation}
	p = \int_{\chi^2}^{\infty} f(z, N - 1) dz,
	\label{eqn:p_value}
\end{equation}
where $f(z, N-1)$ is the $\chi^2$ probability density function with $N-1$ degrees of freedom, $N$ is the number of data points, and $\chi^2$ is the minimum value of $\chi^2(\alpha_s^{\text{eff.}})$.

\subsection{Experimental data}
\label{sec:experimental_data}

In this work, we consider different sets of  selection criteria for the experimental data used in extracting $\alpha_s^{\text{eff.}}$. Our primary analysis considers data from collision systems where the formation of a QGP is well established and in a $p_T$ range where we expect our model is valid. This data is used to extract the global value of the effective strong coupling $\alpha_s^{\text{eff.}}$ in our model, separately for $\sqrt{s_{NN}} = 0.2 ~\mathrm{TeV}$ \coll{Au}{Au} collisions at RHIC and $\sqrt{s_{NN}} = 5.02 ~\mathrm{TeV}$ \coll{Pb}{Pb} collisions at LHC.
Additionally, we consider various subsets of the data used in the global extraction of $\alpha_s^{\text{eff.}}$ and data from various other collision systems and kinematic regions where either the formation of QGP or the presumed applicability of pQCD-based energy loss is less certain. 
These analyses aim to 1) assess the consistency of our model by performing separate extractions as a function of centrality and flavor, 2) make predictions for small collision systems and compare the $\alpha_s^{\text{eff.}}$ extracted to that from large systems, 3) qualitatively estimate medium modifications to hadronization by varying the minimum $p_T$ of the considered data, and 4) qualitatively estimate the scales at which the couplings run, and its implications for the collisional energy loss mechanism, by performing extractions at different hadron momentum ranges and $\sqrt{s_{NN}}$.

Our primary global extraction of $\alpha_s^{\text{eff.}}$ uses $R_{AB}$ data for inclusive charged hadrons, $\pi^\pm$, $\pi^0$, $D$, and $B$ mesons produced in $0\text{--}50\%$ centrality \coll{Au}{Au} collisions at $\sqrt{s_{NN}} = 0.2 ~\mathrm{TeV}$ and \coll{Pb}{Pb} collisions at $\sqrt{s_{NN}} = 5.02~\mathrm{TeV}$, with a final hadron momentum range of between $8~\mathrm{GeV}$ and $50~\mathrm{GeV}$. We use all data reported before March 2025 that meet these criteria for our global extraction of $\alpha_s^{\text{eff.}}$; a full list is given in \cref{tab:raa_experiments}.
This choice of datasets results in a total of 197 data points from LHC and 98 data points from RHIC being used as input in this global analysis. %
We refer to this analysis as our \emph{global extraction} of $\alpha_s^{\text{eff.}}$ and the results using this extraction as \emph{large system constrained} model results. We exclude lower-$p_T$ data (below $8~\mathrm{GeV}$) due to the inapplicability of pQCD-based energy loss and vacuum hadronization models in this regime. The exact minimum $p_T$ which should be used is not easy to determine from first principles, and we discuss the applicability of our model as a function of the minimum $p_T$ in \cref{sec:applicability_of_pqcd_energy_loss_from_data}.
Additionally, we restrict our analysis to $0\text{--}50\%$ heavy-ion collisions where a QGP is agreed to be produced \cite{ALICE:2022wpn,CMS:2024krd,PHENIX:2004vcz,STAR:2005gfr} and there are no significant selection biases or model dependencies when interpreting the suppression result \cite{ALICE:2018ekf}.

Further analyses using subsets of the global dataset and additional collision systems are discussed in \crefrange{sec:applicability_of_pqcd_energy_loss_from_data}{sec:running_coupling}. These analyses include data covering extended $p_T$ and centrality ranges and measurements from small collision systems. A complete list of the data used in these analyses is provided in \cref{tab:raa_experiments_all_data}. We also present model predictions from our large system constrained model for all data sets listed in \cref{tab:raa_experiments_all_data} in \cref{fig:all_data_vs_theory} , which contains 826 data points with $p_T \geq 5 ~\mathrm{GeV}$. %

\setlength{\tabcolsep}{9pt}
\begin{table*}[!t]
    \centering
		\caption{Leading hadron $R_{AB}$ data used in the global extraction of the effective strong coupling $\alpha_s^{\text{eff.}}$ from our pQCD-based energy loss model. The dataset includes $R_{AB}$ measurements for light- and heavy-flavor mesons in $0\text{–}50\%$ centrality \coll{Pb}{Pb} collisions at $\sqrt{s_{NN}} = 5.02 ~\mathrm{TeV}$ and \coll{Au}{Au} collisions at $\sqrt{s_{NN}} = 0.2 ~\mathrm{TeV}$.}
    \label{tab:raa_experiments}
    \begin{tabular}{l|| c c c c c c}
        \toprule
				\makecell{Collaboration\\\textnormal{[reference]}} & \makecell{$\sqrt{s_{\mathrm{NN}}}$\\(TeV)} & \makecell{Collision\\system} & \makecell{Hadron\\species} & \makecell{Centrality\\(\%)} & \makecell{$p_{T}$-range\\($\mathrm{GeV}/c$)} \\
        \midrule
        \midrule
				  STAR \cite{STAR:2003fka}  & 0.2 & \coll{Au}{Au}& $h^{\pm}$ & 0--40 & 8--9.5 \\
        \midrule
				  STAR \cite{STAR:2009fqa}  & 0.2 & \coll{Au}{Au}& $\pi^0$& 0--40 & 8--10.5 \\
        \midrule
				  PHENIX \cite{PHENIX:2008saf}  & 0.2 & \coll{Au}{Au}& $\pi^0$ & 0--50 & 8--19 \\
        \midrule
          PHENIX \cite{PHENIX:2012jha}  & 0.2 & \coll{Au}{Au}& $\pi^0$ & 0--50 & 8--19 \\
        \midrule
          STAR \cite{STAR:2018zdy}  & 0.2 & \coll{Au}{Au}& $D^0$ & 0--40 & 8--9 \\
        \midrule
				  ALICE \cite{ALICE:2018vuu}  & 5.02 & \coll{Pb}{Pb}& $h^{\pm}$ & 0--50 & 8--40 \\
        \midrule
				  ATLAS \cite{ATLAS:2022kqu}  & 5.02 & \coll{Pb}{Pb}& $h^{\pm}$ & 0--50 & 8--50 \\
        \midrule
				  CMS \cite{CMS:2016xef}  & 5.02 & \coll{Pb}{Pb}& $h^{\pm}$ & 0--50 & 8--50 \\
        \midrule
				  ALICE \cite{ALICE:2019hno}  & 5.02 & \coll{Pb}{Pb}& $\pi^{\pm}$ & 0--50 & 8--50 \\
        \midrule
				  ALICE \cite{ALICE:2018lyv}  & 5.02 & \coll{Pb}{Pb}& $D^0$ & 0--50 & 8--43 \\
        \midrule
          CMS \cite{CMS:2017qjw}  & 5.02 & \coll{Pb}{Pb}& $D^0$ & 0--50 & 8--50 \\
        \bottomrule
    \end{tabular}
\end{table*}

\setlength{\tabcolsep}{9pt}
\begin{table*}[!t]
    \centering
		\caption{Leading hadron $R_{AB}$ measurements used in various extractions of $\alpha_s^{\text{eff.}}$ in this work, based on different data subsets. These include data from systems where QGP formation is not well established and kinematic regions where the model’s applicability may be limited.}
    \label{tab:raa_experiments_all_data}
		\begin{tabular}{@{} l||c c c c c c @{}}
        \toprule
				\makecell{Collaboration\\\textnormal{[reference]}} & \makecell{$\sqrt{s_{\mathrm{NN}}}$\\(TeV)} & \makecell{Collision\\system} & \makecell{Hadron\\species} & \makecell{Centrality\\(\%)} & \makecell{$p_{T}$-range\\($\mathrm{GeV}/c$)} & \makecell{Analysis\\(Sec.)}\\
        \midrule
        \midrule
			 STAR \cite{STAR:2003fka}  & 0.2 & \coll{Au}{Au}& $h^{\pm}$ & 0--80 & 5--9.5 & \labelcref{sec:global_extraction}, \labelcref{sec:centrality}\\
        \midrule
			 STAR \cite{STAR:2009fqa}  & 0.2 & \coll{Au}{Au}& $\pi^0$ & 0--80 & 5--10.5 & \labelcref{sec:global_extraction}, \labelcref{sec:centrality}\\
        \midrule
			 PHENIX \cite{PHENIX:2008saf}  & 0.2 & \coll{Au}{Au}& $\pi^0$ & 0--92 & 5--19 & \labelcref{sec:global_extraction}, \labelcref{sec:centrality}, \labelcref{sec:applicability_of_pqcd_energy_loss_from_data}\\
        \midrule
		 PHENIX \cite{PHENIX:2012jha}  & 0.2 & \coll{Au}{Au}& $\pi^0$ & 0--93 & 5--19 & \labelcref{sec:global_extraction}, \labelcref{sec:centrality}, \labelcref{sec:applicability_of_pqcd_energy_loss_from_data}\\
        \midrule
		 STAR \cite{STAR:2018zdy}  & 0.2 & \coll{Au}{Au}& $D^0$ & 0--80 & 5--9 & \labelcref{sec:global_extraction}, \labelcref{sec:centrality}\\
        \midrule
				 PHENIX \cite{PHENIX:2023dxl}  & 0.2 & \coll{d}{Au}& $\pi^0$ & 0--5 & 7.75--17 & \labelcref{sec:small_systems}\\
        \midrule
				ALICE \cite{ALICE:2018vuu}  & 5.02 & \coll{Pb}{Pb}& $h^{\pm}$ & 0--80 & 5--40 & \makecell{\labelcref{sec:global_extraction}, \labelcref{sec:flavor},\\\labelcref{sec:centrality}, \labelcref{sec:applicability_of_pqcd_energy_loss_from_data}} \\
        \midrule
				ATLAS \cite{ATLAS:2022kqu}  & 5.02 & \coll{Pb}{Pb}& $h^{\pm}$ & 0--80 & 5--270 & \makecell{\labelcref{sec:global_extraction}, \labelcref{sec:flavor}, \labelcref{sec:centrality},\\\labelcref{sec:applicability_of_pqcd_energy_loss_from_data}, \labelcref{sec:running_coupling}} \\
        \midrule
				CMS \cite{CMS:2016xef}  & 5.02 & \coll{Pb}{Pb}& $h^{\pm}$ & 0--90 & 5--325 & \makecell{\labelcref{sec:global_extraction}, \labelcref{sec:flavor},\\\labelcref{sec:centrality}, \labelcref{sec:applicability_of_pqcd_energy_loss_from_data}} \\
        \midrule
				ALICE \cite{ALICE:2018ekf}  & 5.02 & \coll{Pb}{Pb}& $h^{\pm}$ & 0--100 & 8--20 & \makecell{\labelcref{sec:small_systems}} \\
        \midrule
				ALICE \cite{ALICE:2019hno}  & 5.02 & \coll{Pb}{Pb}& $\pi^{\pm}$ & 0--80 & 5--19 & \makecell{\labelcref{sec:global_extraction}, \labelcref{sec:flavor},\\\labelcref{sec:centrality}, \labelcref{sec:applicability_of_pqcd_energy_loss_from_data}} \\
        \midrule
				ALICE \cite{ALICE:2018lyv}  & 5.02 & \coll{Pb}{Pb}& $D^0$ & 0--80 & 5--43 & \makecell{\labelcref{sec:global_extraction}, \labelcref{sec:flavor},\\\labelcref{sec:centrality},\labelcref{sec:applicability_of_pqcd_energy_loss_from_data}}\\
        \midrule
				CMS \cite{CMS:2017qjw}  & 5.02 & \coll{Pb}{Pb}& $D^0$ & 0--100 & 5--80 & \makecell{\labelcref{sec:global_extraction}, \labelcref{sec:flavor},\\\labelcref{sec:centrality}, \labelcref{sec:applicability_of_pqcd_energy_loss_from_data}} \\
        \midrule
				 CMS \cite{CMS:2017uoy}  & 5.02 & \coll{Pb}{Pb}& $B^\pm$ & 0--100 & 8--40 & \labelcref{sec:flavor}\\
        \midrule
				 CMS \cite{CMS:2024vip}  & 5.02 & \coll{Pb}{Pb}& $B^+$ & 0--90 & 8--40 & \labelcref{sec:flavor}\\
        \midrule
				 CMS \cite{CMS:2024vip}  & 5.02 & \coll{Pb}{Pb}& $B^0_s$ & 0--90 & 10--40 & \labelcref{sec:flavor}\\
        \midrule
				 ALICE \cite{ALICE:2014xsp}  & 5.02 & \coll{p}{Pb}& $h^{\pm}$ & 0--5\% & 5--30 & \labelcref{sec:small_systems}\\
        \midrule
				 ATLAS \cite{ATLAS:2022kqu}  & 5.02 & \coll{p}{Pb}& $h^{\pm}$ & 0--5 & 5--170 & \labelcref{sec:small_systems}\\
        \midrule
				 ALICE \cite{ALICE:2019fhe}  & 5.02 & \coll{p}{Pb}& $D^0$ & 0--10 & 5--30 & \labelcref{sec:small_systems}\\
        \bottomrule
    \end{tabular}
\end{table*}

\subsection{Theoretical uncertainties}
\label{sec:theoretical_unc}

There are a number of theoretical uncertainties present in the modeling of energy loss of high-$p_T$ partons. There are two sources of model uncertainty which we include in this work:
\begin{enumerate}
	\item \emph{Cutoff on transverse radiated gluon momentum.} We set the upper bound on the magnitude of the transverse radiated gluon momentum $|\mathbf{k}|$ such that neither the collinear nor large formation time approximations are violated, which leads to 
	\begin{equation}
		|\mathbf{k}|_{\text{max}} \sim \operatorname{Min}\left[2 x E(1-x), \sqrt{2 x E \sqrt{\mu^2+\mathbf{q}_1^2}}\right].
	\end{equation}
	We will vary the $|\mathbf{k}|_{\text{max}}$ by a factor, which we will call the ``$|\mathbf{k}|_{\text{max}}$ multiplier", in the range $[0.5, 2]$ to explore the sensitivity of our results to the exact value chosen for the upper bound. The motivation and implications of this cutoff are discussed in \cref{sec:large_formation_time_approximation} and \cite{Faraday:2023uay,Faraday:2023mmx}.
	\item \emph{HTL vs Vacuum propagators.} There is an uncertainty in the crossover between HTL propagators and vacuum propagators as a function of exchanged momentum \cite{Romatschke:2004au,Gossiaux:2008jv,Wicks:2008zz, Faraday:2024gzx} (see \cref{sec:collisional_energy_loss}). In our model, this propagator uncertainty is included for the collisional energy loss only, but in principle is also a contributing factor to the uncertainty in the radiative energy loss. 
		In our model of the radiative energy loss, which utilizes the Gyulassy-Wang potential, it is not possible to estimate this uncertainty.
		We estimate this propagator uncertainty in the collisional energy loss by calculating results for two choices that reasonably explore the possibilities: HTL-only \cite{Wicks:2008zz} collisional energy loss, which uses only the HTL propagators, and BT \cite{Braaten:1991jj,Braaten:1991we} collisional energy loss which changes between the HTL and vacuum propagators. Consult \cref{sec:collisional_energy_loss} for details.
\end{enumerate}

While many theoretical uncertainties could be considered, we focus here only on the two described above, for several reasons. First, these particular uncertainties are relatively straightforward to estimate theoretically, allowing us to quantitatively estimate their effect. Second, although these uncertainties represent only a subset of the full theoretical uncertainty, they provide a qualitative sense of how sensitive our predictions are to theoretical uncertainties more broadly. Finally, as we will show, these uncertainties are significant, varying the extracted $\alpha_s^{\text{eff.}}$ within the range $0.25\text{–}0.55$. The size of these uncertainties indicate that they likely form a large contribution to the full theoretical uncertainty. Interestingly, we will find that although the uncertainties on $\alpha_s^{\text{eff.}}$ are large, the uncertainty on the model predictions for $R_{AB}$ are relatively insensitive to the theoretical uncertainties, similar to the findings of \cite{Horowitz:2009eb}. 

One way to incorporate theoretical uncertainties is by adding a contribution to the covariance matrix in \cref{eqn:correlation_matrix} based on the uncertainties discussed above \cite{Jaiswal:2025hyp}. However, this approach is challenging since many of these uncertainties lack a well-defined statistical distribution and, therefore, a notion of a one-sigma variation. An alternative approach is to introduce model parameters that account for these theoretical uncertainties and fit the model using an expanded parameter space. Instead, because of its flexibility compared to extending the parameter space, we choose a third strategy: we treat each theoretical curve generated under different assumptions independently and we apply the full statistical analysis to each curve separately. This strategy allows us to quantify theoretical uncertainty by treating different model implementations as a fundamental model uncertainty. 
Additionally, this strategy allows us to make physical conclusions about relevant aspects of the model beyond just a theoretical uncertainty, for example the relative importance of collisional vs.\ radiative energy loss.

\subsection{Model coverage and sensitivity}
\label{sec:coverage}

In our global extraction of $\alpha_s^{\text{eff.}}$, we evaluate our energy loss model described in \cref{sec:model} at set values $\alpha_s^{\text{eff.}} \in \{0.2, 0.25, 0.3, 0.35, 0.4, 0.45, 0.5, 0.55 \}$ for a variety of experimental collision systems and final states discussed in \cref{sec:experimental_data}. 
The energy loss model is additionally varied in order to estimate the theoretical uncertainties as described in \cref{sec:theoretical_unc}. Before presenting the results of the global extraction of $\alpha_s^{\text{eff.}}$, we first demonstrate that our chosen range of $\alpha_s^{\text{eff.}}$ produces model results which adequately span the experimental measurements.

\begin{figure}[!b]
	\centering
	\includegraphics[width=\linewidth]{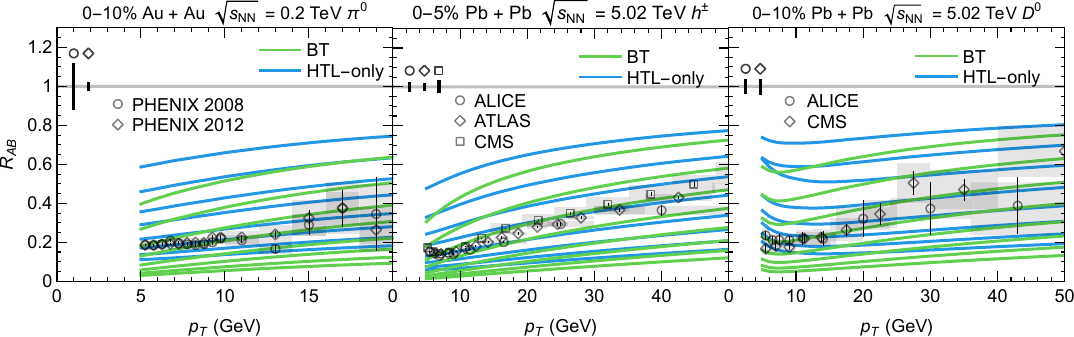}
	\caption{$R_{AB}$ as a function of $p_T$ for (left to right)  $\pi^0$ mesons produced in $0\text{--}10\%$ centrality \coll{Au}{Au} collisions, charged hadrons produced in $0\text{--}5\%$ centrality \coll{Pb}{Pb} collisions at $\sqrt{s_{NN}} = 5.02 ~\mathrm{TeV}$, and $D^0$ mesons produced in $0\text{--}10\%$ centrality \coll{Pb}{Pb} collisions at $\sqrt{s_{NN}} = 5.02 ~\mathrm{TeV}$. Theoretical model curves are shown for BT (green) and HTL-only (blue) collisional energy loss implementations. The top-most $R_{AB}$ curves of each color correspond to $\alpha_s = 0.2$. Each subsequent curve increases $\alpha_s$ by $0.05$ until the bottom-most curve, which corresponds to $\alpha_s = 0.55$. Experimental data from PHENIX \cite{PHENIX:2008saf,PHENIX:2012jha}; ALICE, ATLAS, and CMS \cite{ALICE:2018vuu,ATLAS:2022kqu,CMS:2016xef} (left); and ALICE and CMS \cite{ALICE:2018lyv,CMS:2017qjw} are also shown.}
	\label{fig:raa_with_changing_alphas}
\end{figure}

\Cref{fig:raa_with_changing_alphas} shows model calculations of $R_{AB}$ as a function of $p_T$ across our $\alpha_s^{\text{eff.}}$ parameter space compared with a representative sample o the experimental data. Model results are shown for BT collisional energy loss and HTL-only collisional energy loss, with the $k_{\text{max}}$ multiplier fixed at one for purposes of illustration; see \cref{sec:theoretical_unc} for a description of these theoretical uncertainties. Theoretical curves are generated with $\alpha_s^{\text{eff.}}$ from $0.2$ to $0.55$ in steps of $0.05$, where larger $\alpha_s^{\text{eff.}}$ values produce stronger energy loss and thus smaller $R_{AB}$ values. Results are shown for $\pi^0$ mesons produced in $0\text{--}10\%$ centrality $\sqrt{s_{NN}} = 0.2 ~\mathrm{TeV}$ \coll{Au}{Au} collisions (left), charged hadrons produced in $0\text{--}5\% $ centrality $\sqrt{s_{NN}} = 5.02 ~\mathrm{TeV}$ \coll{Pb}{Pb} collisions (middle), and $D^0$ mesons produced in $0\text{--}10\%$ centrality $\sqrt{s_{NN}} = 5.02 ~\mathrm{TeV}$ \coll{Pb}{Pb} collisions (right). These systems represent a subset of the experimental measurements used in our global extraction, as detailed in \cref{tab:raa_experiments}. As evident from the figure, the range $\alpha_s^{\text{eff.}} \in [0.2, 0.55]$ sufficiently covers the experimental data points, and the model varies smoothly as a function of $\alpha_s^{\text{eff.}}$. 
We verified that the model’s smooth dependence on $\alpha_s^{\text{eff.}}$ and its coverage of the experimental data persist across all datasets listed in \cref{tab:raa_experiments_all_data}, including those not shown in \cref{fig:raa_with_changing_alphas}.
Thus our selected parameter grid provides appropriate coverage for global extraction of $\alpha_s^{\text{eff.}}$ in our model from the data that we are considering.

	In both the JETSCAPE and PHENIX implementations for extracting $\alpha_s^{\text{eff.}}$ there are specific choices made in the modeling of the covariance matrix. 
In the JETSCAPE formulation, the correlation length $\ell_{\text{corr.}}$ governs the primary uncertainty in modeling the covariance matrix, while in the PHENIX implementation, the fully uncorrelated and same-sign correlated limits emerge as special cases of a length-correlated model.
	To investigate how these modeling choices affect our results, we extract various values of $\alpha_s^{\text{eff.}}$ and examine goodness-of-fit parameters as functions of the correlation length $\ell_{\text{corr.}}$. 
We use a reduced $p_T$ range in this section of $p_T \in [10, 20] ~\mathrm{GeV}$ compared to the range $8 ~\mathrm{GeV} \leq p_T \leq  50 ~\mathrm{GeV}$ used in our global extraction, in order to have $p$-values not too different from one. Larger $p_T$ ranges are not well-predicted in our model, likely because we neglect non-perturbative effects at lower-$p_T$ and the influence of running coupling at higher-$p_T$; we discuss this point further in \cref{sec:missing_physics}.

\begin{figure}[!t]
	\centering
	\includegraphics[width=0.5\linewidth]{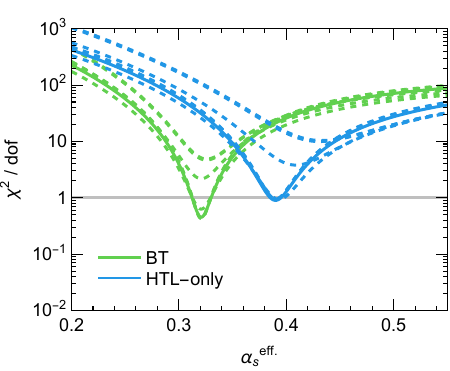}
	\caption{$\chi^2$ per degree of freedom ($\chi^2 / \text{dof}$) as a function of the effective strong coupling in our model. The dashed curves correspond to different correlation lengths $\ell_{\text{corr.}}$ between $10^{-2}$ and $10^3$; consult text for details. The solid curve corresponds to $\ell_{\text{corr.}} = 0.2$, which is used throughout the rest of this work. }
	\label{fig:chi_square_vs_alpha_s_different_correlation_lengths}
\end{figure}

\Cref{fig:chi_square_vs_alpha_s_different_correlation_lengths} plots the $\chi^2 / \text{dof}$ for all experimental data from LHC that are used in the global analysis as a function of $\alpha_s^{\text{eff.}}$. Results are shown for BT (green) and HTL-only (blue) collisional energy loss implementations and for varying correlation lengths $\ell_{\text{corr.}} \in [0.01, 0.1, 0.2, 1, 10, 1000]$. 
Recall that $\ell_{\text{corr.}} \to 0$ corresponds to treating the Type B uncertainties as fully uncorrelated and $\ell_{\text{corr.}} \to \infinity$ corresponds to treating the Type B uncertainties as fully same-sign correlated.
The curve with $\ell_{\text{corr.}} = 0.2$, which is the value used for this analysis, is solid while the other curves are dashed. We note that a $|\mathbf{k}|_{\text{max}}$ multiplier of $1$ is used for all curves for simplicity.
We observe that the $\chi^2 / \text{dof}$ changes fairly significantly as $\ell_{\text{corr.}}$ changes. This strong sensitivity to $\ell_{\text{corr.}}$ seems to warrant concern in the choice of $\ell_{\text{corr.}}$, which is not determined \emph{a priori}. However, we will show next that the quantities which are important in this statistical analysis---the extracted value of $\alpha_s^{\text{eff}}$ and its confidence interval as well as the $p$-value---are significantly less sensitive to the correlation length used. 
Comparing the $\chi^2/\text{dof}$ values for the BT and HTL-only collisional energy loss implementations, we find that BT yields a smaller $\chi^2/\text{dof}$ and appears to better describe the experimental data. This better agreement for BT stems from the faster rise of the $R_{AA}$ predicted by the model which includes BT compared to that which includes HTL-only collisional energy loss. As we discussed previously, we do not interpret this as evidence that the BT implementation is the ``correct" implementation, since we neglect running coupling effects. We examine these effects in \cref{sec:running_coupling}.

\begin{figure}[!b]
	\centering
	\includegraphics[width=\linewidth]{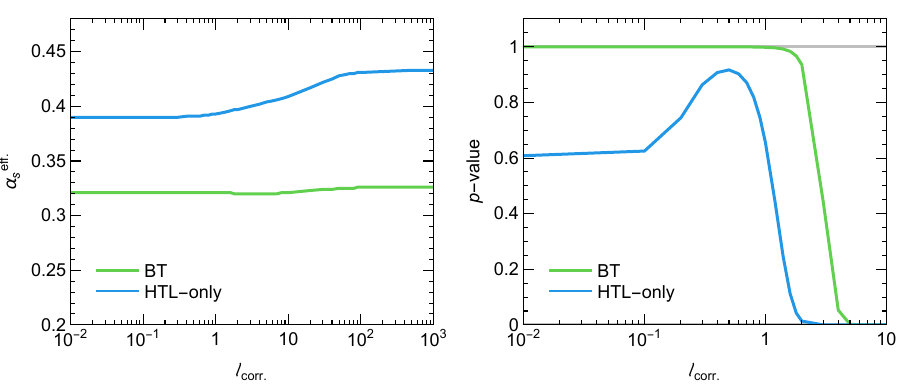}
	\caption{(left) Extracted $\alpha_s^{\text{eff.}}$ as a function of the correlation length $\ell_{\text{corr.}}$. (right) The $p$-value as a function of $\ell_{\text{corr.}}$.
		All theoretical curves are computed with $|\mathbf{k}|_{\text{max}}$ multiplier of one and for both BT (green) and HTL-only (blue) collisional model implementations. Data used for the model comparison is all LHC data from \cref{tab:raa_experiments} with a $p_T$ of between $10 ~\mathrm{GeV}$ and $20 ~\mathrm{GeV}$.}
	\label{fig:correlation_length}
\end{figure}

The left panel of \cref{fig:correlation_length} plots the extracted value of $\alpha_s^{\text{eff.}}$ as a function of the correlation length $\ell_{\text{corr.}}$ for BT (green) and HTL-only (blue) collisional energy loss implementations, both evaluated with a $|\mathbf{k}|_{\text{max}}$ multiplier of one. 
Uncertainties on the extracted values of $\alpha_s^{\text{eff.}}$ are not shown in the figure but are $\mathcal{O}(0.5\%)$ and largely independent of the choice of $\ell_{\text{corr.}}$.
From the left panel of \cref{fig:correlation_length}, we observe that for $\ell_{\text{corr.}} \lesssim 1$, the extracted $\alpha_s^{\text{eff.}}$ and its confidence interval are relatively independent of $\ell_{\text{corr.}}$. However, for larger values of $\ell_{\text{corr.}}$, the extracted $\alpha_s^{\text{eff.}}$ varies by $\mathcal{O}(15\%)$ for HTL-only energy loss, while the extracted $\alpha_s^{\text{eff.}}$ for BT energy loss is relatively constant as a function of $\ell_{\text{corr.}}$. We visually compared the model fits to the data for $\ell_{\text{corr.}} \gtrsim 10$ and found that the model clearly failed to describe the data, requiring normalization shifts by several standard deviations of the Type C uncertainties. We interpret this as a breakdown of the fully correlated approximation used for the Type B uncertainties, an issue made more evident by the small statistical errors in the dataset. We note that the visual over-suppression of our model compared to data when the Type B uncertainties are fully correlated ($\ell_{\text{corr.}} \to \infty$) is similar to what was found in \cite{Soltz:2024gkm}, where the authors concluded that the visual mismatch was due to a subtle difference in the shape of the model and the experimental data.
Considering the $\ell_{\text{corr.}} \to 0$ limit, the extracted $\alpha_s^{\text{eff.}}$ changes by only $\mathcal{O}(0.1\%)$ for $\ell_{\text{corr.}} \lesssim 1$, indicating that ignoring the correlations in the Type B uncertainty is likely to minimally impact the extracted parameters.

Finally, the right panel of \cref{fig:correlation_length} plots the $p$-value as a function of the correlation length $\ell_{\text{corr.}}$, calculated according to \cref{eqn:p_value}.
We observe from the figure that the $p$-value is more sensitive to changes in $\ell_{\text{corr.}}$ than the extracted value of $\alpha_s^{\text{eff.}}$. The sensitivity to the choice of $\ell_{\text{corr.}}$ is also strongly dependent on the collisional energy loss implementation. We observe that the BT collisional energy loss implementation has a large $p$-value, independent of $\ell_{\text{corr.}}$ except for $\ell_{\text{corr.}} \gtrsim 1$; however, the HTL-only implementation has a maximum at $\ell_{\text{corr.}} \simeq 0.5$ before decreasing. 
This dependence of the $p$-value on the collisional energy loss mechanism suggests that a quantitative understanding of the experimental covariance matrix could impose stronger constraints on the energy loss mechanism. Specifically, larger correlation lengths allow models that fail to accurately describe the $p_T$ dependence in the data to remain viable, whereas smaller correlation lengths rule out such models.
We conclude from \cref{fig:correlation_length} that for $\ell_{\text{corr.}} \lesssim 1$ the extracted value of $\alpha_s^{\text{eff.}}$, the uncertainty on this extracted value, and the $p$-value of the fit are relatively independent of $\ell_{\text{corr.}}$. We therefore expect that there is not a significant sensitivity of our results to the model of the correlation in the Type B uncertainties as a function of $p_T$.

\begin{figure}[!th]
	\centering
	\includegraphics[width=0.9\linewidth]{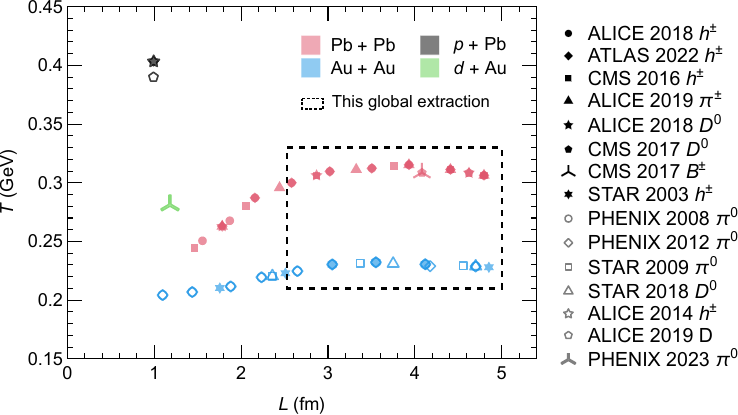}
	\caption{Average temperature $T$ vs average length $L$ calculated in our model for collisions probed by various high-$p_T$ suppression measurements; see \cref{tab:raa_experiments_all_data} for a full list. 
	Points within the dashed box are included in the global extraction of $\alpha_s^{\text{eff.}}$ (see \cref{tab:raa_experiments}). Different points for a given experiment correspond to different centrality classes; more central collisions correspond to larger $L$.}
	\label{fig:phase_space_experiments}
\end{figure}

\Cref{fig:phase_space_experiments} shows the length-temperature phase space probed by all experimental data in \cref{tab:raa_experiments_all_data}. Each point is placed at the average temperature $T$ and average length $L$ that a high-$p_T$ parton traverses through the medium formed in the relevant collision. Lengths and temperatures are computed as described in \cref{sec:geometry}.
Multiple data points are shown for each experimental dataset to represent different centrality classes, with more central collisions corresponding to larger path lengths.
The experimental data which is used in our global extraction of $\alpha_s^{\text{eff.}}$ (from \cref{tab:raa_experiments}) falls within the dashed rectangle in the figure. We observe that while changing the centrality in \coll{A}{A} collisions probes a large range of lengths $L \in [1, 5] ~\mathrm{fm}$, there is a significantly smaller, and bifurcated, set of temperatures that are probed by changing the centrality, $T \in 0.2\text{--}0.3 ~\mathrm{GeV}$. While more central collisions do have higher initial temperature, %
Bjorken expansion leads to the average temperature being relatively independent of centrality. The range of lengths probed that are input into our global extraction of $\alpha_s^{\text{eff.}}$, $L \in [2.5, 5] ~\mathrm{fm}$, is smaller than the full range of heavy-ion collision data since we exclude collisions with centrality $> 50 \%$ in the global extraction of $\alpha_s^{\text{eff.}}$.
In more peripheral collisions, which are not included in our global extraction but are discussed in \cref{sec:centrality}, there is a larger change in temperature as a function of centrality. 

\subsection{Discussion: modeling of the covariance matrix}
\label{sec:discussion_modeling_of_correlations_in_systematic_uncertainties}

	In this section, we contrast our modeling of contributions to the covariance matrix with that of our previous work \cite{Faraday:2024qtl} and related studies in the literature \cite{JETSCAPE:2021ehl,JETSCAPE:2024cqe,PHENIX:2008saf,PHENIX:2008ove}. 
To motivate the procedure outlined in \cref{sec:chi2_minimization_procedure} and to clarify the assumptions underlying our treatment of the covariance matrix, we describe how different uncertainties may be correlated in $p_T$, centrality, and across experimental datasets. Type A uncertainties are generally uncorrelated across these dimensions and will not be discussed further.

We begin with the modeling of the covariance matrix contributions from correlations in $p_T$ within the Type B uncertainties. In \cite{Faraday:2024qtl}, we followed a modified version of the procedure used by PHENIX \cite{PHENIX:2008ove,PHENIX:2008saf} to determine the correlation in $p_T$ between Type B uncertainties. In this work, we model the correlations in the Type B uncertainties in a more flexible way according to \cref{eqn:correlation_length}, following JETSCAPE \cite{JETSCAPE:2021ehl,JETSCAPE:2024dgu}.%

In the PHENIX procedure \cite{PHENIX:2008saf,PHENIX:2008ove}, it is assumed that there is a single underlying Gaussian-distributed, random variable $z_B$, which is responsible for the Type B variation of all data points for a given experiment and centrality class.
By definition, therefore, the Type B uncertainties are fully correlated in $p_T$. Under this assumption, the $\chi^2$ can be written as
\begin{equation}
	\tilde{\chi}^2(\epsilon_B; \alpha_s^{\text{eff.}})  = \chi^2_{\text{shift}}(\epsilon_B; \alpha_s^{\text{eff.}}) + \epsilon_B^2,
	\label{eqn:chi_2_epsilon_B_phenix}
\end{equation}
where
\begin{align}
	\chi^2_{\text{shift}}(\epsilon_B; \alpha_s^{\text{eff.}}) = \sum_{i=1}^n \frac{\left[y_i+f_i(\epsilon_B) \sigma_{B_i}-\mu_i(\alpha_s^{\text{eff.}})\right]^2}{\tilde{\sigma}_i^2} \quad \text{with} \quad \tilde{\sigma}_i=\sigma_i\left(\frac{y_i+ f_i(\epsilon_B) \sigma_{B_i}}{y_i}\right),
	\label{eqn:chi_squared_shifted}
\end{align}
with $f_i(\epsilon_B)$ modeling the form of the correlation. One may view this procedure as shifting the data point $y_{k,i}$ by a fraction $f_i(\epsilon_B)$ of the Type B uncertainty $\sigma^{\text{B}}_{k,i}$, and then evaluating the $\chi^2$ as usual at these shifted data points. The shift has an associated cost to the $\chi^2$ of $\epsilon_B^2$ due to the random variable $z_B$ being Gaussian distributed and centered at $z_B = 0$.
One then minimizes the $\chi^2(\epsilon_B; \alpha_s^{\text{eff.}})$ by varying both $\alpha_s^{\text{eff.}}$ and $\epsilon_B$ simultaneously. 
In \cite{Faraday:2024qtl}, we considered three different models for how the Type B uncertainties are correlated in $p_T$ to induce this shift.
\begin{enumerate}
	\item \emph{Same-sign correlated}: the Type B uncertainties are fully correlated with every other data point, $f_i(\epsilon_B) = \epsilon_B$.
	\item \emph{Tilted}: the Type B uncertainties are correlated such that the lower-$p_T$ and higher-$p_T$ points shift in opposite directions, with a linear interpolation between them, $f_i(\epsilon_B) = \frac{(y_n + \epsilon_B) - (y_1 - \epsilon_B)}{x_n - x_1} (x_i - x_1) - \epsilon_B$.
	\item \emph{Uncorrelated}: the Type B uncertainties are uncorrelated, $f_i(\epsilon_B) = 0$, and the Type B uncertainties are added in quadrature with the Type A uncertainties.
\end{enumerate}
The method used in \cite{PHENIX:2008ove} included only the same-signed correlated and tilted model above. The uncorrelated model was added by us \cite{Faraday:2024qtl} as it was clear that the significantly reduced statistical uncertainties at LHC lead to extremely large $\chi^2$ per degree of freedom ($\chi^2 / \text{dof}$) values of $\mathcal{O}(1000)$ 
for all reasonable $\alpha_s^{\text{eff.}}$ when only models 1 and 2 were used. 

To contrast our approach with previous modeling, we consider limiting cases of the correlation length $\ell_k$ that recover earlier prescriptions. For $\ell_k = 0$, the covariance reduces to $C_{k,ij}^{\text{B}} = \sigma_{k,i}\sigma_{k,j}\delta_{i,j}$, corresponding to uncorrelated Type B uncertainties. In the opposite limit, $\ell_k \to \infty$, we have $C_{k,ij}^{\text{B}} = \sigma^{\text{B}}{k,i}\sigma^{\text{B}}{k,j}$, where Type B uncertainties are fully same-sign correlated, as in the PHENIX procedure \cite{PHENIX:2008ove}. Importantly, the length-correlated framework does not reproduce the ``tilt" prescription of \cite{PHENIX:2008ove}. We discussed the effects of different correlation lengths on our results in \cref{sec:coverage}.

We will now discuss more generally the types of correlations that one might expect across centrality and between experiments. The correlations across centrality and between experimental results for Type B uncertainties is not well understood. Studies of low-$p_T$ observables have employed a length-correlated approach to model centrality correlations \cite{Bernhard:2019bmu}, similar to how we model correlations in $p_T$ in this work. This approach introduces sufficient correlation to avoid overfitting while remaining flexible enough for models to provide a reasonable description of the data. While this approach is convenient, direct input from experiments on the structure of Type B correlations across centrality, $p_T$, experiments, and observables would be extremely valuable. As shown in \cref{fig:correlation_length}, these correlations can substantially affect the goodness-of-fit. Similarly, comparisons of simple theoretical models to data indicate that the choice of correlation modeling can significantly affect parameter extraction, changing extracted values by up to 100\% \cite{Soltz:2024gkm}.

	Future experimental measurements that provide the full covariance matrix will significantly better constrain theoretical models, as has been done extensively in \coll{p}{p} collisions and more recently for jets in heavy-ion collisions \cite{CMS:2024zjn}. Additionally, providing signed uncertainties for each bin allows for more accurate modeling of the covariance matrix \cite{ALICE:2022vsz,ALICE:2024fip, ALICE:2024jtb}, and reporting results in the principal-component basis \cite{Bhalerao:2014mua,CMS:2017mzx}---which, by construction, produces a diagonal covariance matrix---would further improve constraints on theoretical models.

We now discuss the Type C correlations between centrality classes within one experiment. As noted in \cite{JETSCAPE:2024cqe}, Type C uncertainties from luminosity are clearly correlated in centrality, as they are identical between centrality classes. Since the contribution from luminosity is small compared to other contributions to the uncertainty, ignoring the correlation in luminosity between centrality classes is a small effect \cite{JETSCAPE:2024cqe}. Additionally, it seems reasonable to be concerned about correlations between centrality classes in the Type C Glauber uncertainties. For example, the number of binary collisions $N_{\text{coll}}$ is typically estimated with the Glauber model \cite{Miller:2007ri}, and the systematic uncertainties on $N_{\text{coll}}$ are estimated by varying the model parameters, including the inelastic nucleon-nucleon cross section and the parameters which describe the nuclear profile \cite{ALICE:2014xsp,ATLAS:2016tor,ATLAS:2015hkr}. Since the same model parameters are used for all centrality classes, one therefore expects significant correlation between the determination of $N_{\text{coll}}$ in different centrality classes. However, this correlation is not reported by experiments. In central and semi-central collisions the Type C uncertainties are typically significantly smaller than the Type B uncertainties, and so the correlation from the Glauber model is not likely to affect our results significantly. In more peripheral centrality classes, which are treated in \cref{sec:centrality}, the Type C Glauber uncertainties are much larger, and so any correlation might affect our results. 

Further, one might anticipate correlations between Type C uncertainties across experiments. Different experiments often use similar implementations of the Glauber model with similar parameters for the nuclear profiles and the inelastic nucleon-nucleon cross section at the same $\sqrt{s_{NN}}$, which may lead to significant correlation in  the Type C Glauber uncertainties between different experiments. 
We have followed previous work \cite{JETSCAPE:2024cqe,JETSCAPE:2021ehl,Karmakar:2024jak,JET:2013cls,Casalderrey-Solana:2018wrw} in ignoring the correlations between centrality classes and different experiments for all uncertainties. 
Future extractions might further model the correlations between experiments and centrality classes to understand the impact of this uncertainty on the results.

It is important to note that our approach of minimizing the sum of $\chi^2$ unevenly weights data from different physical regions of interest due to differences in the sizes of the associated uncertainties. In particular, lower-$p_T$, light-flavor, and central collision data exert a disproportionately strong influence on the extraction of $\alpha_s^{\text{eff.}}$ compared to higher-$p_T$, heavy-flavor, and semi-central collision data, due to the significantly smaller uncertainties of the former compared to the latter.
To account for the disproportionate weighting of different regions of the experimental phase space, we consider our model results and our extraction procedure on various subsets of the experimental data in \cref{sec:model_robustness}.

\section{Global extraction of the strong coupling}\label{sec:global_extraction}

We now present the results from our global analysis of central- and semi-central heavy-ion collision data at RHIC and LHC. This analysis extracts the value of the effective strong coupling $\alpha_s^{\text{eff.}}$ from our energy loss model described in \cref{sec:model}. In particular, we minimize the $\chi^2(\alpha_s^{\text{eff.}})$ by comparing our model predictions to both light- and heavy-flavor leading hadron $R_{AA}$ data from $0\text{--}50\%$ centrality $\sqrt{s_{NN}} = 5.02 ~\mathrm{TeV}$ \coll{Pb}{Pb} and $\sqrt{s_{NN}} = 0.2 ~\mathrm{TeV}$ \coll{Au}{Au} collisions with $8 ~\mathrm{GeV} \leq p_T  \leq 50 ~\mathrm{GeV}$. Since LHC collisions at $\sqrt{s_{NN}} = 5.02 ~\mathrm{TeV}$ are $\sim \! 30 \%$ hotter than RHIC collisions at $\sqrt{s_{NN}} = 0.2 ~\mathrm{TeV}$ (see \cref{fig:phase_space_experiments}), the effective strong coupling $\alpha_s^{\text{eff.}}$ may be significantly different at LHC compared to RHIC, and so we perform the extraction separately on LHC and RHIC data. 
This choice of datasets results in a total of 197 data points from LHC and 98 data points from RHIC being used as input in this analysis; consult \cref{tab:raa_experiments} for details. %
We discussed the motivation for the experimental inclusion criteria in detail in \cref{sec:experimental_data}.

As discussed in \cref{sec:theoretical_unc}, we consider contributions from two important theoretical uncertainties: 1) the exact value of the kinematic cutoff on the transverse radiated gluon momentum, which captures lingering sensitivity related to the large formation time approximation, and 2) the collisional energy loss implementation that is used, which captures an uncertainty in the transition between HTL and vacuum propagators. To estimate these theoretical uncertainties, we evaluate our model for seven values of the upper bound on the transverse radiated gluon momentum, $|\mathbf{k}|_{\text{max}} \in \{2^{-1}, 2^{-2 / 3}, 2^{-1 / 3}, 2^0, 2^{1 /3}, 2^{2 / 3}, 2^1\}$, and two collisional energy loss implementations, HTL-only and BT, resulting in 14 different model implementations for which our analysis is conducted.

\begin{figure}[!b]
	\centering
	\includegraphics[width=\linewidth]{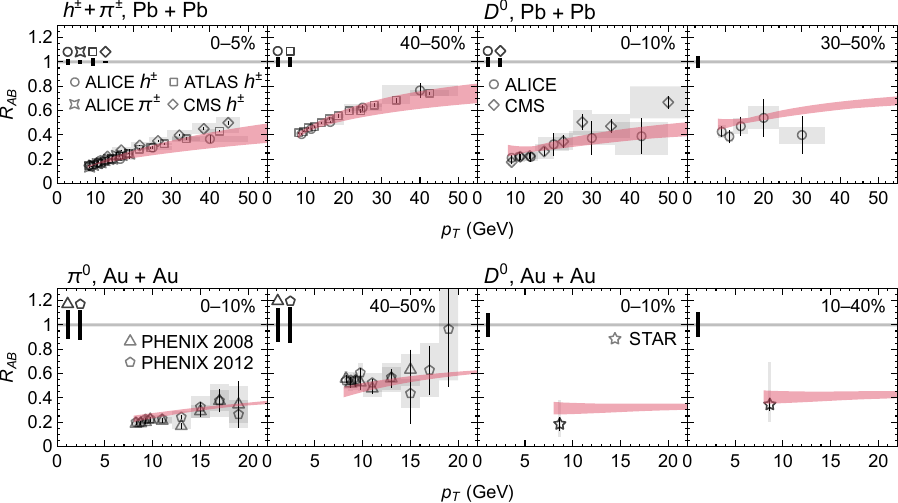}
	\caption{$R_{AB}$ as a function of $p_T$ for a representative sample of the experimental data that is used in our global extraction of $\alpha_s^{\text{eff.}}$. Experimental data is shown with open markers and our constrained model results are shown as a red band. From left-to-right and top-to-bottom results are shown for $h^{\pm}$ and $\pi^{\pm}$ mesons produced in $0\text{--}5\%$ and $40\text{--}50\%$ centrality $\sqrt{s_{NN}} = 5.02 ~\mathrm{TeV}$ \coll{Pb}{Pb} collisions \cite{ALICE:2018vuu, ALICE:2019hno, ATLAS:2022kqu, CMS:2016xef}, $D^0$ mesons produced in $0\text{--}10\%$ and $30\text{--}50\%$ centrality $\sqrt{s_{NN}} = 5.02 ~\mathrm{TeV}$ \coll{Pb}{Pb} collisions \cite{ALICE:2018lyv, CMS:2017qjw}, $\pi^0$ mesons produced in $0\text{--}10\%$ and $40\text{--}50\%$ centrality $\sqrt{s_{NN}}= 0.2 ~\mathrm{TeV}$ \coll{Au}{Au} collisions \cite{STAR:2009fqa, PHENIX:2008saf, PHENIX:2012jha}, and $D^0$ mesons produced in $0\text{--}10\%$ and $10\text{--}40\%$ centrality $\sqrt{s_{NN}} = 0.2 ~\mathrm{TeV}$ \coll{Au}{Au} collisions \cite{STAR:2018zdy}.
	Statistical experimental uncertainties are represented by error bars, systematic uncertainties by shaded boxes, and global normalization uncertainties by bars at unity. The width of the model bands corresponds to the theoretical model uncertainty and uncertainty from the extraction, added in quadrature. }
	\label{fig:representative_raa_global}
\end{figure}

\Cref{fig:representative_raa_global} shows the measured nuclear modification factor $R_{AB}$ as a function of $p_T$ for a representative sample of the experimental data that was used in our global extraction of $\alpha_s^{\text{eff.}}$ (open markers). The corresponding model results, computed using the globally extracted value of $\alpha_s^{\text{eff.}}$, are shown as red bands. The band width represents the theoretical uncertainties and the uncertainties on the extraction, added in quadrature.
We observe in \cref{fig:representative_raa_global} that, within the theoretical and experimental uncertainties, there is good visual agreement between the data and our theoretical model.  A full comparison of the globally constrained model with all data used in the extraction is provided in \cref{fig:all_data_vs_theory_global} in \cref{sec:app_global}  (see \cref{tab:raa_experiments}). An extended comparison, including data from kinematic regions and systems that are not included in our global extraction of $\alpha_s^{\text{eff.}}$, is shown in \cref{fig:all_data_vs_theory} in \cref{sec:app_all} (see \cref{tab:raa_experiments_all_data}).

\begin{figure}[!t]
	\centering
	\includegraphics[width=\linewidth]{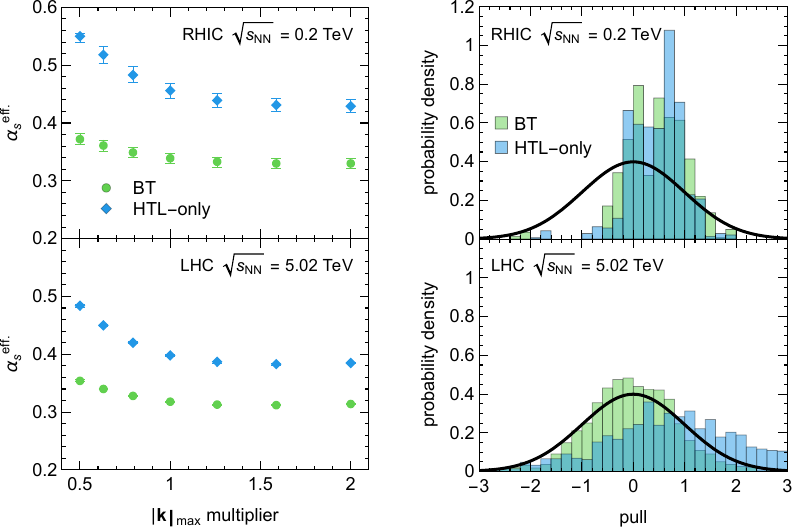}\hfill
	\caption{(left) Extracted $\alpha_s^{\text{eff.}}$ from comparison of our theoretical model to heavy-ion collision data (consult text for details) as a function of the $|\mathbf{k}|_{\text{max}}$ multiplier used in the theoretical energy loss model, and for both BT (green circles) and HTL-only (blue diamonds) elastic energy loss. The top panel shows the results for $\sqrt{s_{NN}} = 0.2 ~\mathrm{TeV}$ \coll{Au}{Au} collisions at RHIC and the bottom panel shows the results for $\sqrt{s_{NN}} = 5.02 ~\mathrm{TeV}$ \coll{Pb}{Pb} collisions at LHC. Error bars on the data points indicate the uncertainty associated with the extraction of $\alpha_s^{\text{eff.}}$. For LHC data, the uncertainties are smaller than the marker size. (right) Distribution of pulls for BT (green) and HTL-only (blue) collisional energy loss implementations at RHIC (top) and LHC (bottom). A normal distribution is overlaid in black.}
	\label{fig:extracted_alpha_s_global_many_theory}
\end{figure}

The left panel of \cref{fig:extracted_alpha_s_global_many_theory} plots the extracted effective strong coupling $\alpha_s^{\text{eff.}}$ at both RHIC (top) and LHC (bottom) as a function of the $|\mathbf{k}|_{\text{max}}$ multiplier that is used, which reflects the theoretical uncertainty related to the large formation time approximation. Additionally, the theoretical uncertainty associated with the transition between HTL and vacuum propagators is captured by two limiting implementations of the collisional energy loss: BT (green circles) and HTL-only (blue diamonds). Error bars indicate the statistical uncertainty associated with the extraction procedure; consult \cref{sec:statistical_analysis} for details. The right panel of \cref{fig:extracted_alpha_s_global_many_theory} shows the distribution of \textit{pulls} \cite{Demortier:2008} for all data used in the global extraction of $\alpha_s^{\text{eff.}}$ at RHIC (top) and at LHC (bottom), shown separately for BT (green) and HTL-only (blue) collisional energy loss implementations. 
Pulls are computed as $(y_i - \mu_i)/\sigma_i$, where $y_i$ is the experimental value, $\mu_i$ the constrained model prediction for the experimental value $\mu_i$, and $\sigma_i$ the total experimental uncertainty. 
Pulls provide a useful qualitative visual comparison between theoretical models and experimental data, as they are normally distributed if the theoretical model correctly describes the experimental data.
While our global extraction of $\alpha_s^{\text{eff.}}$ treats correlations in systematic and normalization uncertainties in a sophisticated way (see \cref{sec:statistical_analysis}), the pull distributions are computed with all uncertainties added in quadrature for visual purposes.

We see in the left panel of \cref{fig:extracted_alpha_s_global_many_theory} that HTL-only collisional energy loss leads to an $\mathcal{O}(30\text{--}50)\%$ increase in the extracted $\alpha_s^{\text{eff.}}$ compared to BT collisional energy loss. The increased extracted $\alpha_s^{\text{eff.}}$ from HTL-only compared to BT is due to the smaller energy loss of HTL-only \cite{Faraday:2024gzx}.
We additionally observe that varying the $|\mathbf{k}|_{\text{max}}$ multiplier from $0.5$ to $2$ leads to an $\mathcal{O}(15\text{--}30)\%$ decrease in the extracted value of $\alpha_s^{\text{eff.}}$. The decrease in the extracted $\alpha_s^{\text{eff.}}$ as a function of the $|\mathbf{k}|_{\text{max}}$ multiplier is due to the relaxing of the restriction on the radiated phase space. As $|\mathbf{k}|_{\text{max}}$ is increased, there is more phase space for the radiated gluon, which leads to the energy loss increasing, and therefore to a smaller $\alpha_s^{\text{eff.}}$ \footnote{We note that in \cref{sec:large_formation_time_approximation}, we saw that an increase in $|\mathbf{k}|_{\text{max}}$ multiplier can result in a decrease in energy loss, due to negative contributions from the short pathlength correction for specific values of the length and temperature. The results for the globally extracted $\alpha_s^{\text{eff.}}$ appears to show that in the full, realistic geometry case these negative contributions are negligible. }.
In the figure, the BT collisional energy loss implementation $\alpha_s^{\text{eff.}}$ is significantly less sensitive to the exact value of $|\mathbf{k}|_{\text{max}}$ used in comparison to the HTL-only results. This difference in sensitivity is because the BT results have a larger contribution of collisional vs.\ radiative energy loss, in comparison to the HTL-only results. Since the collisional energy loss contribution does not depend on $|\mathbf{k}|_{\text{max}}$, the larger contribution of collisional energy loss reduces the sensitivity of the final results to the exact value of $|\mathbf{k}|_{\text{max}}$ used.  %

The right panel of \cref{fig:extracted_alpha_s_global_many_theory} shows good overall agreement between the globally-constrain\-ed model and the experimental data used in the global extraction at both RHIC and LHC.
At RHIC, the pull distribution is slightly biased toward positive values, which we attribute to a weaker centrality dependence of our model results compared to that of experimental data.
In particular, the model tends to overpredict suppression in semi-central collisions, resulting in positively-biased pulls.
We discuss the centrality dependence of $R_{AB}$ and the extracted $\alpha_s^{\text{eff.}}$ in detail in \cref{sec:centrality,sec:peripheral_large_system_suppression_predictions}.
At LHC, the HTL-only results are systematically oversuppressed compared to experimental data, leading to positively-biased pulls. The systematic oversuppression of the HTL-only results compared to data occurs at higher-$p_T$ and is due to the weaker $p_T$-dependence of the HTL-only results compared to the BT results. We attribute this discrepancy to the absence of running coupling effects, which we address in detail in \cref{sec:running_coupling}.

While it is not clear how one should associate a probabilistic interpretation to the theoretical uncertainties treated in this work, one can naively estimate the $68\%$ confidence interval associated with the extracted value of $\alpha_s^{\text{eff.}}$. 
We assume that the extracted values of $\alpha_s^{\text{eff.}}$ across the 14 different model implementations are a one-sigma deviation in the theoretical uncertainty, leading to $\alpha_s^{\text{eff.}} = \num{0.41(14:10)}$ at RHIC and $\alpha_s^{\text{eff.}} = \num{0.37(11:8)}$ at LHC. %
We once again emphasize that one should take care in interpreting the one sigma confidence interval since, for instance, there is no obvious reason that varying the $|\mathbf{k}|_{\text{max}}$ multiplier by factors of two should lead to a one sigma variation. We believe that quoting these extracted confidence intervals is still instructive as a simple estimate of the sensitivity of our extractions to these theoretical uncertainties. We note that the extracted values of $\alpha_s^{\text{eff.}}$ at RHIC and LHC agree within the theoretical uncertainties; however, the RHIC $\alpha_s^{\text{eff.}}$ is $\sim \! 8\%$ larger, qualitatively consistent with the coupling running, at least partially, at the temperature scale of the plasma. It is difficult to make any conclusions at this fully averaged level, and we discuss the compatibility of the extracted $\alpha_s^{\text{eff.}}$ at RHIC and LHC with running coupling estimates in more depth in \cref{sec:running_coupling}.

\section{Model robustness from data subsets}
\label{sec:model_robustness}

As noted in \cref{sec:chi2_minimization_procedure}, different regions of the experimental parameter space contribute unevenly to the $\chi^2$ minimization procedure. Central collisions produce more hard particles than semi-central collisions, typically leading to smaller uncertainties and a stronger influence on $\chi^2$. Similarly, charged hadrons---mostly light-flavor pions---are significantly more abundant than heavy-flavor hadrons, resulting in smaller uncertainties and a stronger influence on $\chi^2$. To explore some of the regions of the experimental parameter space that are less constraining on our extracted value of $\alpha_s^{\text{eff.}}$, we now extract $\alpha_s^{\text{eff.}}$ separately in different centrality regions and for heavy-flavor hadrons. We expect that our model should have similar extracted $\alpha_s^{\text{eff.}}$ in these different regions of the experimental phase space because 1) changing centrality changes mostly length and not temperature (see \cref{fig:phase_space_experiments}) and the coupling is not expected to run with length, and 2) 
the typical values of $\mathbf{k}$ and $\mathbf{q}$ for light and heavy flavors are similar---on the order of the Debye mass $\mu_D$---and so any running coupling effects are likely to be relatively independent of flavor.

\subsection{Centrality dependence}
\label{sec:centrality}

One topic that has been significantly discussed in the literature, both through first principles calculations and phenomenological extractions, is the path length dependence of energy loss \cite{Djordjevic:2018ita,Wu:2023azi,Beattie:2022ojg,Arleo:2022shs,Horowitz:2011gd,Bass:2008rv, Betz:2014cza, Chesler:2008uy, Dominguez:2008vd, PHENIX:2010nlr, Shuryak:2001me, Noronha-Hostler:2016eow}.
As shown in \cref{fig:phase_space_experiments}, centrality primarily changes the average path length in the collision as opposed to the average temperature. Since the strong coupling is not expected to run with path length, we anticipate that the extracted strong coupling $\alpha_s^{\text{eff.}}$ should be approximately the same in different centrality classes. 
Therefore, examining the centrality dependence of the extracted $\alpha_s^{\text{eff.}}$ serves as a self-consistency test of how well the model captures the path length dependence of energy loss.
Other similar tests of the path length dependence might include accurately describing high-$p_T$ azimuthal anisotropy and dihadron correlations, which we will treat in future work \cite{Bert:2024}. 

In this section, we consider final state light- and heavy-flavor hadrons with $8 ~\mathrm{GeV} \leq p_T \leq 50 ~\mathrm{GeV}$ produced in both $\sqrt{s_{NN}} = 5.02 ~\mathrm{TeV} $ \coll{Pb}{Pb} and $\sqrt{s_{NN}} = 0.2 ~\mathrm{TeV}$ \coll{Au}{Au} collisions. We separately extract $\alpha_s^{\text{eff.}}$ in $0\text{--}10\%$, $10\text{--}30\%$, and $30\text{--}50\%$ centrality collisions. 
	We consider only systems with centrality $\leq 50\%$ currently, as we are evaluating the consistency of our model description across the experimental data that was used in the global extraction of $\alpha_s^{\text{eff.}}$ in \cref{sec:global_extraction}.
We will discuss more peripheral collisions in \cref{sec:small_systems}.

\begin{figure}[!th]
	\centering
	\includegraphics[width=\linewidth]{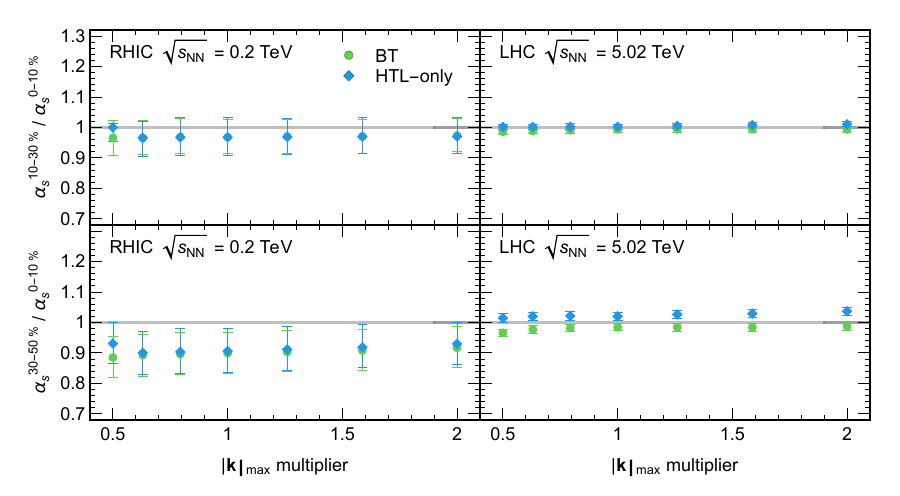}
	\caption{Extracted $\alpha_s^{\text{eff.}}$ from $10\text{--}30\%$ centrality (top row) and $30\text{--}50\%$ centrality (bottom row) collision systems divided by the extracted $\alpha_s^{\text{eff.}}$ from $0\text{--}10\%$ centrality collision systems as a function of the $|\mathbf{k}|_{\text{max}}$ multiplier. The left column shows results using data from $\sqrt{s_{NN}} = 0.2 ~\mathrm{TeV}$ \coll{Au}{Au} collisions at RHIC and the right column shows the results using data from $\sqrt{s_{NN}} = 5.02 ~\mathrm{TeV}$ \coll{Pb}{Pb} collisions at LHC. Results are shown separately for BT (green circles) and HTL-only (blue diamonds) collisional energy loss implementations. Error bars correspond to the uncertainty on the extracted value of $\alpha_s^{\text{eff.}}$.}
	\label{fig:centrality}
\end{figure}

\Cref{fig:centrality} plots the ratio of the $\alpha_s^{\text{eff.}}$ extracted in $10\text{--}30\%$ centrality collisions (top) and $\alpha_s^{\text{eff.}}$ extracted in $30\text{--}50\%$ centrality collisions (bottom) to that extracted in $0\text{--}10\%$ centrality collisions as a function of the $|\mathbf{k}|_{\text{max}}$ multiplier. The left column plots the results for hadrons produced in $\sqrt{s_{NN}} = 0.2 ~\mathrm{TeV}$ \coll{Au}{Au} collisions at RHIC, and the right column for hadrons produced in $\sqrt{s_{NN}} = 5.02 ~\mathrm{TeV}$ \coll{Pb}{Pb} collisions at LHC. Results are shown for both BT (green circles) and HTL-only (blue diamonds) collisional energy loss implementations and as a function of the $|\mathbf{k}|_{\text{max}}$ multiplier, both of which capture theoretical model uncertainties (see \cref{sec:theoretical_unc}). We observe from the top-right panel that the $10\text{--}30\%$ centrality extracted $\alpha_s^{\text{eff.}}$ is in percent-level agreement with the $0\text{--}10\%$ centrality extracted $\alpha_s^{\text{eff.}}$ at LHC, largely independent of the collisional energy loss implementation and $|\mathbf{k}|_{\text{max}}$ multiplier that are used. At RHIC, the extracted $\alpha_s^{\text{eff.}}$ from $10\text{--}30\%$ and $0\text{--}10\%$ centrality systems (top-left) agree to within one standard deviation.
At LHC, the extracted $\alpha_s^{\text{eff.}}$ from $30\text{--}50\%$ centrality collisions is in good agreement with that extracted from $0\text{--}10\%$ centrality collisions (bottom-right), with the central values from the two extractions differing by $\lesssim 2\%$. The bottom left plot shows that the extracted value from $30\text{--}50\%$ centrality collisions at RHIC is $\sim \! 10\%$ smaller than that in $0\text{--}10\%$ centrality collisions, and the values agree within $1\text{--}2$ standard deviations.

We conclude that our model self-consistently describes the centrality dependence of $R_{AB}$ in $0\text{--}50\%$ centrality heavy-ion collisions at RHIC and LHC. The extracted $\alpha_s^{\text{eff.}}$ shows excellent, percent-level agreement at LHC and is within $10\%$ at RHIC. The difference in the ability of our model to consistently describe the centrality dependence at RHIC compared to LHC may suggest that the strong couplings run with centrality at RHIC more prominently than at LHC, missing physics in our model, or mismodeling of the correlations in the uncertainties between centrality classes.  Future extractions might consider data from systems of varying size, including \coll{O}{O}, \coll{Ne}{Ne}, \coll{Ar}{Ar}, and \coll{Xe}{Xe}, to further test how well our model can describe the system size dependence of energy loss.

\subsection{Mass and color representation dependence}
\label{sec:flavor}

In this section we apply our extraction procedure of $\alpha_s^{\text{eff.}}$ separately to the heavy- and light-flavor data shown in \cref{tab:raa_experiments}. The hadron flavor dependence of suppression provides insight into the mass and color representation dependence of partonic energy loss; for a review see \cite{Andronic:2015wma}. 
While our global extraction of $\alpha_s^{\text{eff.}}$ from heavy-ion data (\cref{sec:global_extraction}) includes heavy-flavor data, the smaller uncertainties of light-flavor data---due to its relative abundance compared to heavy-flavor data---make light-flavor data significantly more constraining than heavy-flavor data on the extracted $\alpha_s^{\text{eff.}}$. Since pions, which dominate the charged hadron spectrum, fragment from light quarks and gluons, $D$ mesons fragment from charm quarks, and $B$ mesons fragment from bottom quarks, we will consider these data separately in order to probe the mass and color representation dependence of our energy loss model.

We perform our statistical analysis procedure, described in \cref{sec:statistical_analysis}, on $R_{AB}$ data from $8 ~\mathrm{GeV} \leq p_T \leq 50 ~\mathrm{GeV}$ hadrons produced in $0\text{--}50\%$ centrality \coll{Pb}{Pb} collisions; consult \cref{tab:raa_experiments_all_data} for details. 
Since our current model implementation can make predictions for only a limited set of heavy-flavor data from RHIC, this analysis will focus solely on LHC data. 
Additionally, we note that the lowest $p_T$ point from CMS 2017 the $B^{\pm}$ meson data \cite{CMS:2017uoy}, $p_T = 8.5 ~\mathrm{GeV}$, is not particularly large in comparison to the bottom quark mass of $m_B \sim \! 4.75 ~\mathrm{GeV}$. Therefore, one might expect that various assumptions in the energy loss models that we have used \cite{Djordjevic:2003zk,Kolbe:2015suq,Kolbe:2015rvk,Braaten:1991jj,Braaten:1991we,Wicks:2008zz} may not be applicable, for instance $E \simeq p_T$.
For this reason, all local fits of the $B^{\pm}$ meson data are performed without this data point. Excluding this data point had little impact on the extracted value of $\alpha_s^{\text{eff.}}$ for the local best fit but resulted in a higher $p$-value and larger uncertainties on $\alpha_s^{\text{eff.}}$.

\begin{figure}[!b]
	\centering
	\includegraphics[width=0.5\linewidth]{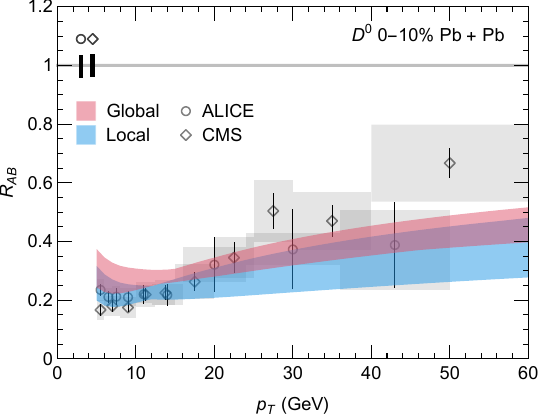}
	\caption{Nuclear modification factor $R_{AB}$ for $D^0$ mesons produced in $0\text{--}10\%$ centrality $\sqrt{s_{NN}} = 5.02 ~\mathrm{TeV}$ \coll{Pb}{Pb} collisions as a function of transverse momentum $p_T$. Experimental data for $D^0$ mesons from ALICE \cite{ALICE:2018lyv} (open circles) and CMS \cite{CMS:2017qjw} (open diamonds) are shown. Error bars indicate statistical uncertainties, gray shaded squares indicate systematic uncertainties, and solid bars at $R_{AA} = 1$ indicate global normalization uncertainties. Model results using our global extraction of $\alpha_s^{\text{eff.}}$ (red) as well as a local extraction of $\alpha_s^{\text{eff.}}$ (blue) are shown for both final states. The band width represents both theoretical uncertainties and the uncertainty stemming from the extraction procedure, added in quadrature. }
	\label{fig:d_meson_raa_local_vs_global}
\end{figure}

\begin{figure}[!t]
	\centering
	\includegraphics[width=\linewidth]{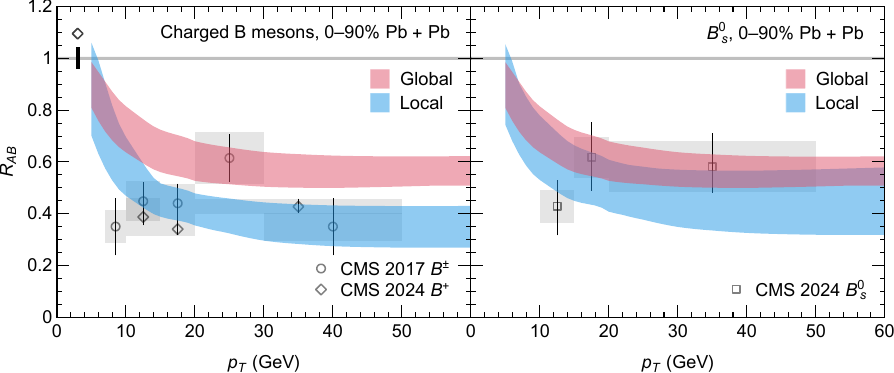}
	\caption{Theoretical model results for the nuclear modification factor $R_{AB}$ for charged $B$ mesons (left) and $B^0_s$ mesons (right) produced in $0\text{--}90\%$ centrality $\sqrt{s_{NN}} = 5.02 ~\mathrm{TeV}$ \coll{Pb}{Pb} collisions as a function of transverse momentum $p_T$. Model results using our global extraction of $\alpha_s^{\text{eff.}}$ (red) as well as a local extraction of $\alpha_s^{\text{eff.}}$ (blue) are shown. 
Experimental data for $B^{\pm}$ \cite{CMS:2017qjw} (open diamonds) and $B^+$ mesons \cite{CMS:2024vip} (open circles) as well as $B^0_s$ mesons (open squares) are shown. Note that the $B^{\pm}$ meson data is from $0\text{--}100\%$ centrality collisions, while the $B^{0}_s$ and $B^+$ meson data is from $0\text{--}90\%$ centrality collisions.
	Error bars indicate statistical uncertainties, gray shaded squares indicate systematic uncertainties, and solid bars at $R_{AA} = 1$ indicate global normalization uncertainties. The band width represents both theoretical uncertainties and the uncertainty stemming from the extraction procedure, added in quadrature. }
	\label{fig:b_meson_raa_local_vs_global}
\end{figure}

\Cref{fig:d_meson_raa_local_vs_global} plots the experimentally measured $R_{AB}$ as a function of $p_T$ for $D^0$ mesons produced in $0\text{--}10\%$ centrality $\sqrt{s_{NN}} = 5.02 ~\mathrm{TeV}$ \coll{Pb}{Pb} collisions \cite{ALICE:2018lyv,CMS:2017qjw}. Theoretical model results for $D^0$ mesons are also shown, computed with both the global extraction of $\alpha_s^{\text{eff.}}$ (red) and the local extraction of $\alpha_s^{\text{eff.}}$ (blue) from all $D^0$ meson $R_{AB}$ data satisfying the previously mentioned criteria. The width of the model bands represents the theoretical uncertainty and the uncertainty on the extracted value of $\alpha_s^{\text{eff.}}$, added in quadrature. We observe from \cref{fig:b_meson_raa_local_vs_global} that the global extraction of $\alpha_s^{\text{eff.}}$ produces $D^0$ meson $R_{AB}$ results which are moderately under suppressed compared to experimental data. The local best fit describes the experimental data well within the experimental and theoretical uncertainties. We note that the faster rise of the experimentally measured $R_{AB}$ with $p_T$ is likely due to running coupling effects which are not included in our model; these effects are investigated in \cref{sec:running_coupling}.

\Cref{fig:b_meson_raa_local_vs_global} plots the experimentally measured $R_{AB}$ as a function of $p_T$ for $B^{+}$ and $B^{\pm}$ mesons (left), and $B^0_s$ mesons (right) produced in $0\text{--}90\%$ centrality $\sqrt{s_{NN}} = 5.02 ~\mathrm{TeV}$ \coll{Pb}{Pb} collisions \cite{CMS:2017uoy,CMS:2024vip}. Theoretical model results for $B$ mesons are also shown, computed with both the global extraction of $\alpha_s^{\text{eff.}}$ (red) and the local extraction of $\alpha_s^{\text{eff.}}$ (blue) from all $B$ meson $R_{AB}$ data satisfying the previously mentioned criteria. We see from the left panel of \cref{fig:b_meson_raa_local_vs_global} that the $B^{\pm}$ meson results produced with the globally extracted $\alpha_s^{\text{eff.}}$ are significantly under suppressed compared to experimental data. The local best fit describes all but the lowest (excluded) $p_T$ data point, $p_T = 8.5 ~\mathrm{GeV}$, to within the experimental and theoretical uncertainties. As noted previously, the Eikonal assumption in our model likely breaks down for this lowest $p_T$ data point since the mass $m \simeq 4.75 ~\mathrm{GeV}$ is not significantly smaller than the $p_T = 8.5 ~\mathrm{GeV}$. The right panel of \cref{fig:b_meson_raa_local_vs_global} is the same as the left panel, but for $B_s^0$ mesons. In our theoretical framework, all $B$ hadrons hadronize in the same way, and so the theoretical model predictions for the left and right panels of \cref{fig:b_meson_raa_local_vs_global} are identical. We see in the right panel of \cref{fig:b_meson_raa_local_vs_global} that the experimentally measured $B^0_s$ $R_{AB}$ is significantly less suppressed than the $B^{+}$ $R_{AB}$ and is consistent with our model results within the theoretical and experimental uncertainties. The fairly large difference between the experimentally measured $B^0_s$ and $B^{+}$ $R_{AB}$ may be indicative of an unrealistic $B$ hadron hadronization procedure in our model. To investigate the consistency of our model results across flavor more quantitatively, we now examine the extracted values of $\alpha_s^{\text{eff.}}$ as a function of flavor.

\begin{figure}[!t]
	\centering
	\includegraphics[width=\linewidth]{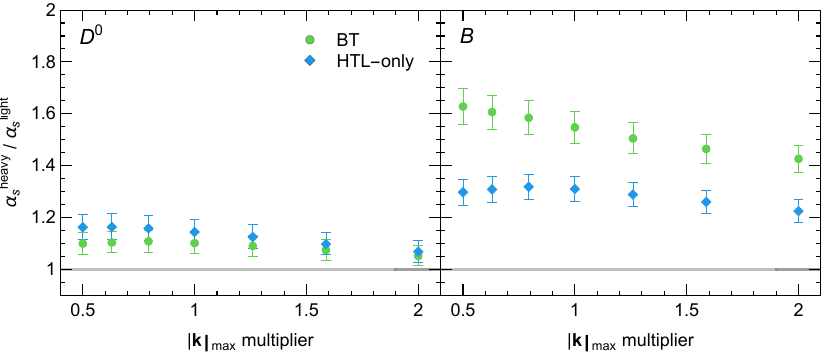}
	\caption{Extracted $\alpha_s^{\text{eff.}}$ from $D^0$ mesons (left) and $B^{\pm}$ mesons (right) divided by the extracted $\alpha_s^{\text{eff.}}$ from charged hadrons and charged pions as a function of the $|\mathbf{k}|_{\text{max}}$ multiplier. All data is from $0\text{--}50\%$ centrality $\sqrt{s_{NN}} = 5.02 ~\mathrm{TeV}$ \coll{Pb}{Pb} collisions at LHC; see \cref{tab:raa_experiments_all_data} for the full list of data included.
	Results are shown separately for BT (green circles) and HTL-only (blue diamonds) collisional energy loss implementations. Error bars correspond to the uncertainty on the extracted value of $\alpha_s^{\text{eff.}}$.}
	\label{fig:heavy_over_light_extracted_alpha}
\end{figure}

The left panel of \cref{fig:heavy_over_light_extracted_alpha} plots the ratio of the extracted $\alpha_s^{\text{eff.}}$ from $D^0$ mesons to that extracted from charged hadrons, as a function of the $|\mathbf{k}_{\text{max}}|$ multiplier and for both BT (green circles) and HTL-only (blue diamonds) collisional energy loss implementations. The right panel presents the same ratio but for $B$ mesons (including $B^{\pm}$, $B^+$, and $B^0_s$ mesons) relative to charged hadrons. The error bars indicate the uncertainty on the extracted value of $\alpha_s^{\text{eff.}}$.

Considering the left panel of \cref{fig:heavy_over_light_extracted_alpha}, we observe that the $D^0$ meson suppression data yields an $\alpha_s^{\text{eff.}}$ that is $\mathcal{O}(5\text{–}15)\%$ larger and $\mathcal{O}(1\text{–}3)$ standard deviations higher than that extracted from charged hadron $R_{AB}$. %
The best agreement between theory and data is obtained with a $|\mathbf{k}|_{\text{max}}$ multiplier of $2$ and with the HTL-only collisional energy loss implementation, yielding a $p$-value of $0.91$. The typical $p$-value for the various model implementations is $p = 0.01\text{--}0.5$, indicating that the agreement of light- and heavy-flavor suppression is extremely sensitive to the theoretical uncertainties that we have considered.

Considering the right panel of \cref{fig:heavy_over_light_extracted_alpha}, we observe that the $\alpha_s^{\text{eff.}}$ extracted from $B^{\pm}$ meson data is $\mathcal{O}(15\text{--}60)\%$ larger and $\mathcal{O}(5\text{--}15)$ standard deviations higher than that extracted from charged hadron data. The many-standard-deviations disagreement of the $\alpha_s^{\text{eff.}}$ extracted from $B$ meson and charged hadron data, respectively, prompts us to conclude that there is important physics missing in our understanding of the $B$ meson $R_{AB}$.

We conclude that HTL-only and large $|\mathbf{k}|_{\text{max}}$ multiplier lead to the best description of the flavor dependence of energy loss; however, the agreement is not particularly good for any of the model implementations in this work. 
The difference between model implementations of the ratio of extracted $\alpha_s^{\text{eff.}}$ from heavy-flavor mesons to charged hadrons shows qualitatively how heavy-flavor $R_{AB}$ data may provide valuable insight into the amount of collisional vs.\ radiative energy loss. However, our conclusions are fairly sensitive to the uncertainty arising from the large formation time kinematic cut on the transverse radiated gluon momentum.
The disagreement between the $\alpha_s^{\text{eff.}}$ values extracted from heavy- and light-flavor hadrons---particularly the multiple-standard-deviation difference for $B$ mesons---suggests shortcomings in our modeling of the flavor dependence of suppression. The disagreement may indicate that our model misrepresents the relative contributions of radiative and collisional energy loss compared to reality, that different quark mass values are more appropriate, or that additional mechanisms---such as heavy-meson dissociation \cite{Adil:2006ra}---contribute to open heavy-flavor suppression.

\section{Extrapolation to central small and peripheral large systems}
\label{sec:small_systems}

We now consider the extrapolation of our model, which is tuned on central- and semi-central heavy-ion collision data, to peripheral large \coll{A}{A} and central small \collFour{p}{d}{He3}{A} collisions at RHIC and LHC. As discussed in \cref{sec:model}, our energy loss model receives short path length corrections to both the radiative and collisional energy loss, which allows for a justifiable comparison to central small and peripheral large systems.
In addition, we will contrast our global model predictions and extracted $\alpha_s^{\text{eff.}}$ to the locally extracted $\alpha_s^{\text{eff.}}$ from central small and peripheral large systems.

\subsection{Centrality bias}
\label{sec:centrality_bias}

Small system and peripheral large system suppression data is difficult to interpret due to a collection of event selection biases that potentially non-trivially correlate the hard and soft modes in the collision \cite{PHENIX:2013jxf,ALICE:2014xsp,ALICE:2018ekf,PHENIX:2023dxl,ATLAS:2023zfx}. In addition to these selection biases, there is also a significant dependence on the model \cite{ATLAS:2016xpn} that is used to map from the determined centrality to the number of binary collisions $N_{\text{coll}}$ used to normalize the $R_{AB}$. We will refer to this collection of selection biases and model dependencies as \emph{centrality bias}, which we will consider carefully when interpreting the agreement between our results and experimental data in central small and peripheral large systems.
Because this work is our first quantitative comparison with small system suppression data, we do not attempt to include any of these biases of unknown size in our theoretical predictions. 
While these additional effects may be important, we have focused in this work on other key advances to our theoretical model, in particular: the restriction of the phase space for the radiated gluon emission, the quantitative estimate of various theoretical uncertainties, and the full statistical analysis and comparison to experimental data. Even though we have not included the effects of centrality bias, we still feel that it is fruitful to compare our predictions to experimental data, and leave it to future work to further quantitatively include the potential effects from centrality bias.

To understand the origin and potential mitigation of centrality bias, it is important to examine how centrality and \( N_{\text{coll}} \) are determined in practice.
Centrality and \( N_{\text{coll}} \) estimates based on total charged particle multiplicity are particularly susceptible to centrality bias \cite{ALICE:2014xsp}.
Such determinations of centrality are typically performed with the total multiplicity, multiplicity at mid-rapidity, or the multiplicity in the forward, $\mathrm{Pb}$-going direction \cite{ATLAS:2014cpa,ALICE:2014xsp}. An alternative approach is to use the total energy deposited by slow nucleons down the beam-axis as a measure of the wounded nucleons \cite{Miller:2007ri}, which may then be used to estimate the centrality and $N_{\text{coll}}$ \cite{ALICE:2014xsp}. This wounded-nucleon based approach is not expected to suffer significantly from selection bias; however, further modeling is required to relate the deposited energy by nucleons down the beam axis to geometrical quantities.
This mapping is required to determine $N_{\text{coll.}}$, which is crucial for the normalization of $R_{p / d A}$, and thus in determining whether there is suppression or enhancement in these small system collisions.
This mapping in small systems is performed using a Glauber model or some modification to a Glauber model, where the applicability of the Glauber model is less well tested or understood than in large system collisions.
ATLAS measurements \cite{ATLAS:2022kqu,ATLAS:2014cpa} have so far used the total transverse energy deposited in the forward, $\mathrm{Pb}$-going direction to determine centrality and geometrical quantities; however, there has been recent work in determining the cause and magnitude of centrality bias in \coll{p}{Pb} collisions by ATLAS \cite{ATLAS:2024qsm,ATLAS:2023zfx}, which may lead to less biased estimates for $N_{\text{coll.}}$ in the future. ALICE measurements \cite{ALICE:2014xsp, ALICE:2019fhe} have used the total energy deposited by nucleons down the beam axis to determine centrality and geometrical quantities.

One way to reduce the effect of centrality bias in a model-agnostic way is to use weakly interacting probes to normalize the $R_{AB}$. The PHENIX collaboration performed such a measurement \cite{PHENIX:2023dxl}, where they calculated the $R_{AB}$ as
\begin{equation}
	R_{A B, \gamma}^h\left(p_T\right)\equiv\frac{Y_{A B}^h\left(p_T\right)}{N_{\text {coll}}^{\text{EXP}} Y^h_{p p}\left(p_T\right)} \quad \text{with} \quad N_{\mathrm{coll}}^{\mathrm{EXP}}\left(p_T\right)\equiv\frac{Y_{AB}^{\gamma^{\mathrm{dir}}}\left(p_T\right)}{Y_{p p}^{\gamma^{\mathrm{dir}}}\left(p_T\right)},
\end{equation}
where $Y^h_{AB}$ ($Y^h_{pp}$) is the yield of the hadron $h$ in the collision \coll{A}{B} (\coll{p}{p}) and $Y^{\gamma^{\text{dir}}}_{AB}$ ($Y^{\gamma^{\text{dir}}}_{pp}$) is the yield of direct photons in the collision \coll{A}{B} (\coll{p}{p}). Importantly, the Glauber model is not used in the determination of the number of binary collisions. It is expected, and verified in heavy-ion collisions \cite{CMS:2012oiv,ALICE:2024yvg,ATLAS:2015rlt}, that the yield of direct photons is not modified by the presence of a medium in a heavy-ion collision, even in centrality-cut measurements. Therefore, one expects that $N_{\mathrm{coll}}^{\mathrm{EXP}}\left(p_T\right)$ is an experimental, model-independent measure of the number of binary collisions. There are, however,  concerns that such a measurement is still sensitive to centrality bias. For instance, the QCD color fluctuation model \cite{Alvioli:2014eda,Alvioli:2017wou} predicts $R_{AB,\text{EXP}}^{\pi^0} < 1$ \cite{Perepelitsa:2024eik}, qualitatively consistent with the PHENIX measured $R_{AB,\text{EXP}}^h$ \cite{PHENIX:2023dxl}. In the color fluctuation model, $R_{AB,\text{EXP}}^h < 1$ stems from the different Bjorken $x$ values that are probed by the prompt photon and the $\pi^0$ measurements due to fragmentation effects. 
		Alternatively, selection biases related to global energy conservation may also lead to $R_{pA} < 1$ in central collisions \cite{Kordell:2016njg}.
		A repetition of the PHENIX analysis for \coll{p}{Au} and \coll{He3}{Au} could potentially disentangle initial-state centrality bias from final state energy loss as sources of high-$p_T$ suppression, since energy loss predicts $R_{{}^3\mathrm{He}\mathrm{Au}} < R_{d\mathrm{Au}} < R_{p\mathrm{Au}}$ while initial-state effects predict the opposite scaling \cite{Perepelitsa:2024eik}.

\subsection{Peripheral large system suppression predictions}
\label{sec:peripheral_large_system_suppression_predictions}

We now consider suppression data from peripheral large system collisions at RHIC and LHC. 
We showed previously \cite{Faraday:2024qtl} that peripheral $60\text{--}80\%$ centrality \coll{A}{A} collisions form a plasma which is larger but cooler than that which may be formed in $0\text{--}5\%$ centrality \collThree{p}{d}{A} collisions, leading to similar expected suppression in both systems. Additionally, since the peripheral \coll{A}{A} collisions involve $\sim 5$ times more binary collisions, they should also be significantly less sensitive to centrality bias.
In this section, we consider all experimental $R_{AB}$ data from $60\text{--}80\%$ centrality heavy-ion collisions, separately for RHIC and LHC; consult \cref{tab:raa_experiments_all_data} for details. 

\begin{figure}[!t]
	\includegraphics[width=\linewidth]{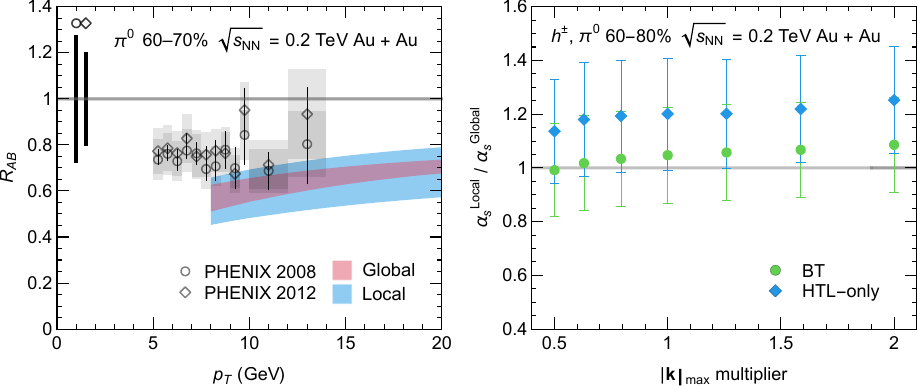}
	\caption{(left) Measured $R_{AB}$ as a function of $p_T$ for charged hadrons produced in $60\text{--}70\%$ centrality $\sqrt{s_{NN}} = 0.2 ~\mathrm{TeV}$ \coll{Au}{Au} collisions by PHENIX in \cite{PHENIX:2012jha,PHENIX:2008saf}. Model results are shown as bands for global best fit $\alpha_s^{\text{eff.}}$ (red) and local best fit $\alpha_s^{\text{eff.}}$ (blue) constrained by $R_{AB}$ data from $60\text{--}80\%$ centrality \coll{Au}{Au} collisions; consult text for details. The thickness of the theoretical uncertainty bands correspond to the theoretical uncertainties and statistical extraction uncertainties on $\alpha_s^{\text{eff.}}$, added in quadrature. Error bars on data points represent the statistical uncertainty, shaded boxes represent the systematic uncertainty, and the solid bar at $R_{AB} = 1$ represents the relative normalization uncertainty.
		(right) Ratio of the $\alpha_s^{\text{eff.}}$ extracted in $60\text{--}80\%$ centrality $\sqrt{s_{NN}} = 0.2 ~\mathrm{TeV}$ \coll{Au}{Au} collisions ($\alpha_s^{\text{Local}}$) to the $\alpha_s^{\text{eff.}}$ extracted from $0\text{--}50\%$ centrality heavy-ion collisions ($\alpha_s^{\text{Global}}$); consult text for details. This ratio is plotted as a function of the $|\mathbf{k}|_{\text{max}}$ multiplier and for both BT (green circles) and HTL-only (blue diamonds) collisional energy loss implementations.}
	\label{fig:raa_alpha_ratio_combined_peripheral_RHIC}
\end{figure}

\Cref{fig:raa_alpha_ratio_combined_peripheral_RHIC} (left) shows the $R_{AB}$ as a function of final hadron $p_T$ for $\pi^0$ mesons produced in $60\text{--}70\%$ centrality $\sqrt{s_{NN}} = 0.2 ~\mathrm{TeV}$ \coll{Au}{Au} collisions measured by PHENIX \cite{PHENIX:2008saf,PHENIX:2012jha}. Model predictions from our globally constrained model on central and semi-central heavy-ion collision data (red) as well as from a local best fit to $60\text{--}80\%$ centrality heavy-ion data from RHIC (blue) are shown as bands. 
The $\chi^2$ minimization procedure includes all data within the $60\text{--}80\%$ centrality interval in order to extract a single $\alpha_s^{\text{eff.}}$ for each model implementation; the left panel shows the comparison of our results to $R_{AB}$ for the $60\text{--}70\%$ centrality class as a representative example.
The width of the model bands corresponds to the theoretical uncertainty and uncertainty on the extracted $\alpha_s^{\text{eff.}}$, added in quadrature. \Cref{fig:raa_alpha_ratio_combined_peripheral_RHIC} (right) shows the extracted $\alpha_s^{\text{eff.}}$ in $60\text{--}80\%$ centrality $\sqrt{s_{NN}} = 0.2 ~\mathrm{TeV}$ \coll{Au}{Au} collisions at RHIC ($\alpha_s^{\text{local}}$) divided by the globally extracted $\alpha_s^{\text{eff.}}$ from $0\text{--}50\%$ centrality \coll{Au}{Au} collisions described in \cref{sec:global_extraction} ($\alpha_s^{\text{global}}$). This ratio is plotted as a function of the $|\mathbf{k}|_{\text{max}}$ multiplier and for both BT (green circles) and HTL-only (blue diamonds) collisional energy loss implementations.

We observe from the left panel of  \cref{fig:raa_alpha_ratio_combined_peripheral_RHIC} that the global and local model results are in good qualitative agreement with each other and that both models visually appear to slightly overestimate the observed suppression. 
It is particularly striking that the local best fit appears visibly over-suppressed relative to the experimental data. This apparent over-suppression is a consequence of the sizable $\sim 15\text{–}30\%$ normalization uncertainties on the data, which allow a vertical shift of the model relative to the data without significantly affecting the $\chi^2$. The local best fit apparently more accurately captures the $p_T$ dependence of the data at this level of suppression, despite the overall change in the normalization.
We additionally compared our model results for $\pi^0$ mesons produced in $70\text{--}80\%$ \coll{Au}{Au} collisions \cite{PHENIX:2008saf,PHENIX:2012jha} and $h^{\pm}$ hadrons produced in $60\text{--}80\%$ \coll{Au}{Au} collisions \cite{STAR:2003fka} and found good qualitative agreement between the global model predictions and the experimental data. We do not show this figure here as it is qualitatively similar to that already shown; these plots are shown in \cref{fig:all_data_vs_theory} in \cref{sec:app_all}.
Considering the right panel of \cref{fig:raa_alpha_ratio_combined_peripheral_RHIC}, we see that the locally and globally extracted $\alpha_s^{\text{eff.}}$ agree to within one standard deviation for both BT and HTL-only collisional energy loss implementations, largely independent of the $|\mathbf{k}|_{\text{max}}$ multiplier.

\begin{figure}[!t]
	\includegraphics[width=\linewidth]{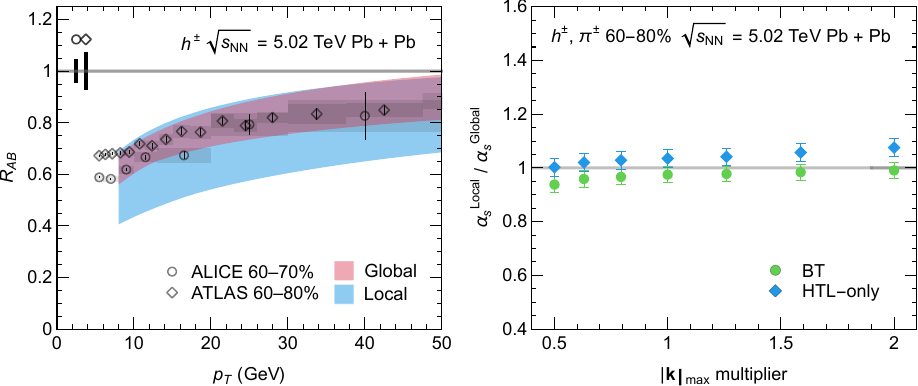}
	\caption{(left) Measured $R_{AB}$ as a function of $p_T$ for charged hadrons produced in $60\text{--}70\%$ and $60\text{--}80\%$ centrality $\sqrt{s_{NN}} = 5.02 ~\mathrm{TeV}$ \coll{Pb}{Pb} collisions by ALICE \cite{ALICE:2018vuu} and ATLAS \cite{ATLAS:2022kqu}, respectively, and for $\pi^{\pm}$ mesons produced in $60\text{--}80\%$ centrality \coll{Pb}{Pb} collisions. Model results for the $60\text{--}80\%$ centrality class are shown as bands for global best fit $\alpha_s^{\text{eff.}}$ (red) and local best fit $\alpha_s^{\text{eff.}}$ (blue), which is constrained by $R_{AB}$ data from $60\text{--}80\%$ centrality \coll{Pb}{Pb} collisions; consult text for details. The thickness of the theoretical uncertainty bands correspond to the theoretical uncertainties and statistical extraction uncertainties on $\alpha_s^{\text{eff.}}$ added in quadrature. Error bars on data points represent the statistical uncertainty, shaded boxes represent the systematic uncertainty, and the solid bars at $R_{AB} = 1$ represent the relative normalization uncertainties.
		(right) Ratio of the $\alpha_s^{\text{eff.}}$ extracted in $60\text{--}80\%$ centrality $\sqrt{s_{NN}} = 5.02 ~\mathrm{TeV}$ \coll{Pb}{Pb} collisions ($\alpha_s^{\text{local}}$) to the $\alpha_s^{\text{eff.}}$ extracted from $0\text{--}50\%$ centrality heavy-ion collisions ($\alpha_s^{\text{global}}$); consult text for details. This ratio is plotted as a function of the $|\mathbf{k}|_{\text{max}}$ multiplier and for both BT (green circles) and HTL-only (blue diamonds) collisional energy loss implementations.}
	\label{fig:raa_alpha_ratio_combined_peripheral_LHC}
\end{figure}

	\Cref{fig:raa_alpha_ratio_combined_peripheral_LHC} (left) shows the $R_{AB}$ as a function of final hadron $p_T$ for charged hadrons produced in $60\text{--}70\%$ and $60\text{--}80\%$ centrality $\sqrt{s_{NN}} = 5.02 ~\mathrm{TeV}$ \coll{Pb}{Pb} collisions measured by ALICE \cite{ALICE:2018vuu} and ATLAS \cite{ATLAS:2022kqu}, respectively, and for $\pi^{\pm}$ mesons produced in $60\text{--}80\%$ centrality \coll{Pb}{Pb} collisions measured by ALICE \cite{ALICE:2019hno}. Model predictions from our globally constrained model on central and semi-central heavy-ion collision data (red) as well as from a local best fit to peripheral heavy-ion data from LHC (blue) are shown as bands. The width of the model bands corresponds to the theoretical uncertainty and uncertainty on the extracted $\alpha_s^{\text{eff.}}$, added in quadrature. \Cref{fig:raa_alpha_ratio_combined_peripheral_LHC} (right) shows the extracted $\alpha_s^{\text{eff.}}$ in $60\text{--}80\%$ centrality $\sqrt{s_{NN}} = 5.02 ~\mathrm{TeV}$ \coll{Pb}{Pb} collisions at LHC ($\alpha_s^{\text{local}}$) divided by the globally extracted $\alpha_s^{\text{eff.}}$ from semi-central and central \coll{Pb}{Pb} collisions described in \cref{sec:global_extraction} ($\alpha_s^{\text{global}}$). This ratio is plotted as a function of the $|\mathbf{k}|_{\text{max}}$ multiplier and for both BT (green circles) and HTL-only (blue diamonds) collisional energy loss implementations.

In \cref{fig:raa_alpha_ratio_combined_peripheral_LHC}, we see that the predictions from our model using the global extracted $\alpha_s^{\text{eff.}}$ for $60\text{--}80\%$ centrality \coll{Pb}{Pb} collisions are in good agreement with experimental data. From the right panel of \cref{fig:raa_alpha_ratio_combined_peripheral_LHC}, the local and global extracted $\alpha_s^{\text{eff.}}$ differ by $\sim 10 \%$ and agree within two standard deviations for BT, while for HTL-only the local and global extracted $\alpha_s^{\text{eff.}}$ differ by $\sim 1 \%$ and agree within one standard deviation. 
	The dependence of the ratio $\alpha_s^{\text{Local}} / \alpha_s^{\text{Global}}$ is largely independent of the $|\mathbf{k}|_{\text{max}}$ multiplier used.

\begin{figure}[!b]
	\centering
	\includegraphics[width=0.6\linewidth]{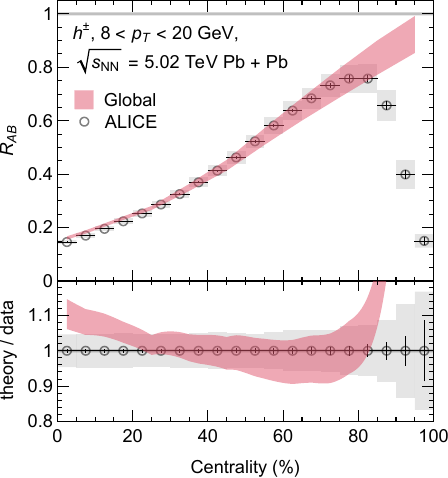}
	\caption{(top) $R_{AB}$ as a function of centrality for $8 < p_T < 20 ~\mathrm{GeV}$ charged hadrons produced in \coll{Pb}{Pb} collisions as measured by ALICE \cite{ALICE:2018ekf}. Also shown are predictions from our global extraction of $\alpha_s^{\text{eff.}}$ (red band). (bottom) Ratio of the model results for the $R_{AB}$ to the experimental data. The width of the band represents the theoretical uncertainties and the uncertainties from the extraction, added in quadrature. Statistical experimental uncertainties are represented by error bars and systematic uncertainties by shaded squares.}
	\label{fig:raa_pbpb_vs_centrality}
\end{figure}

Having examined the $p_T$-dependence of $R_{AB}$ at RHIC and LHC, we show in \cref{fig:raa_pbpb_vs_centrality} the $R_{AB}$ as a function of centrality to examine the onset of centrality bias.
The top panel of \cref{fig:raa_pbpb_vs_centrality} shows the $R_{AB}$ as a function of centrality for $8 < p_T < 20 ~\mathrm{GeV}$ charged hadrons produced in \coll{Pb}{Pb} collisions as measured by ALICE \cite{ALICE:2018ekf} (open black circles) and model predictions from our global extraction of $\alpha_s^{\text{eff.}}$ (red bands). 
The width of the band represents the theoretical uncertainties and the uncertainties from the extraction, added in quadrature. Statistical experimental uncertainties are represented by error bars and systematic uncertainties by shaded squares.
The bottom panel shows the ratio of the theoretical predictions to the experimental data. We find good agreement between the global model predictions and experimental data for $0\text{–}80\%$ centrality collisions, within the $\mathcal{O}(10\%)$ theoretical and experimental uncertainties. The dramatic disagreement between the model predictions and experimental data for $80\text{--}100\%$ centrality collisions is likely due to centrality bias \cite{ALICE:2018ekf}. 

We conclude that our global model predictions for peripheral \coll{A}{A} collisions are in good agreement with the experimental data. This agreement is largely independent of the $|\mathbf{k}|_{\text{max}}$ multiplier, indicating that our results are insensitive to the large formation time approximation.

\subsection{Central small system suppression predictions}
\label{sec:central_small_system_suppression_predictions}

In the preceding sections we presented our global extraction of $\alpha_s^{\text{eff.}}$, showed that our model passed a variety of consistency checks based on the flavor and centrality dependence of heavy-ion experimental data, and finally showed that our model results for peripheral \coll{A}{A} collisions were in good agreement with experimental data; now we go to the central question of the manuscript: whether we predict measurable suppression from energy loss in central small systems and how our predictions compare to experimental measurements.

\begin{figure}[!t]
	\centering
	\includegraphics[width=\linewidth]{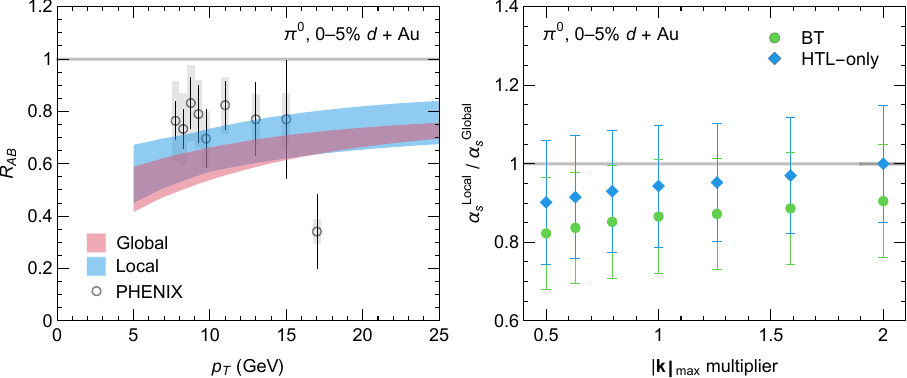}
	\caption{(left) Measured $R_{AB}$ as a function of $p_T$ for $\pi^0$ mesons produced in $0\text{--}5\%$ centrality $\sqrt{s_{NN}} = 0.2 ~\mathrm{TeV}$ \coll{d}{Au} collisions by PHENIX \cite{PHENIX:2023dxl}. Model results are shown as bands for global best fit $\alpha_s^{\text{eff.}}$ (red) and local best fit $\alpha_s^{\text{eff.}}$ (blue) constrained by $R_{AB}$ data from $0\text{--}5\%$ centrality \coll{d}{Au} collisions; consult text for details. The thickness of the theoretical uncertainty bands correspond to the theoretical uncertainties and statistical extraction uncertainties on $\alpha_s^{\text{eff.}}$, added in quadrature. Error bars on data points represent the statistical uncertainty, shaded boxes represent the systematic uncertainty, and the solid bar at $R_{AB} = 1$ represents the relative normalization uncertainty.
		(right) Ratio of the $\alpha_s^{\text{eff.}}$ extracted in $0\text{--}5\%$ centrality $\sqrt{s_{NN}} = 0.2 ~\mathrm{TeV}$ \coll{d}{Au} collisions ($\alpha_s^{\text{Local}}$) to the $\alpha_s^{\text{eff.}}$ extracted from $0\text{--}50\%$ centrality \coll{Au}{Au} collisions at $\sqrt{s_{NN}} = 0.2 ~\mathrm{TeV}$ ($\alpha_s^{\text{Global}}$); consult text for details. This ratio is plotted as a function of the $|\mathbf{k}|_{\text{max}}$ multiplier and for both BT (green circles) and HTL-only (blue diamonds) collisional energy loss implementations.}
	\label{fig:raa_dau_0005_global_vs_local_fit_no_legend}
\end{figure}

\Cref{fig:raa_dau_0005_global_vs_local_fit_no_legend} (left) plots the $R_{AB,\text{EXP}}$ as a function of $p_T$ for $\pi^0$ mesons produced in $0\text{--}5\%$ centrality \coll{d}{Au} collisions as measured by PHENIX \cite{PHENIX:2023dxl}. Model predictions using our global extraction of $\alpha_s^{\text{eff.}}$ are shown as a red band. Additionally, a blue band is produced with the local best fit of $\alpha_s^{\text{eff.}}$ to the PHENIX \coll{d}{Au} data \cite{PHENIX:2023dxl}. The width of the bands correspond to the theoretical uncertainty and the uncertainty on the extracted value of $\alpha_s^{\text{eff.}}$, added in quadrature. The right panel of \cref{fig:raa_dau_0005_global_vs_local_fit_no_legend} plots the ratio of the locally extracted $\alpha_s^{\text{eff.}}$ from the $0\text{--}5\%$ centrality \coll{d}{Au} $R_{AB,\text{EXP}}$ data  to the globally extracted $\alpha_s^{\text{eff.}}$ from central and semi-central \coll{Au}{Au} data; see \cref{sec:global_extraction} for details. The error bars indicate the uncertainty on the extracted value of $\alpha_s^{\text{eff.}}$. We note that in the local best fit, all data points are used, including the single data point at $7 ~\mathrm{GeV}$.

In the left panel of \cref{fig:raa_dau_0005_global_vs_local_fit_no_legend}, we observe that the large-system constrained model prediction (red) is in good agreement with the data. We emphasize that no small or peripheral system high-$p_T$ suppression data was used in constraining the global model result. %
In \cref{fig:raa_dau_0005_global_vs_local_fit_no_legend} the majority of the uncertainty in the local best fit to \coll{d}{Au} data is due to the uncertainty on the extracted value of $\alpha_s^{\text{eff.}}$ and not due to the theoretical uncertainties. In contrast, the uncertainties on the global model are primarily theoretical uncertainties.
We see from the right panel of \cref{fig:raa_dau_0005_global_vs_local_fit_no_legend} that the global and local extracted values of $\alpha_s^{\text{eff.}}$ agree within one standard deviation for all model implementations.
The $p$-values of the global model predictions compared to the experimental data are $p \simeq 0.3$ for all 14 model implementations.
We note that a significant fraction of the deviation of this $p$-value from one is due to the highest $p_T$ data point. Excluding the highest $p_T$ data point yields a $p$-value of $p \simeq 0.6\text{--}0.95$ for our global model predictions compared to the \coll{d}{Au} data.
We additionally observe that the ratio of the globally extracted effective strong coupling to the locally extracted effective strong coupling is largely insensitive, within the large extraction uncertainties, to the collisional energy loss implementation and $|\mathbf{k}|_{\text{max}}$ multiplier that are used.

\begin{figure}[!t]
	\centering
	\includegraphics[width=0.8\linewidth]{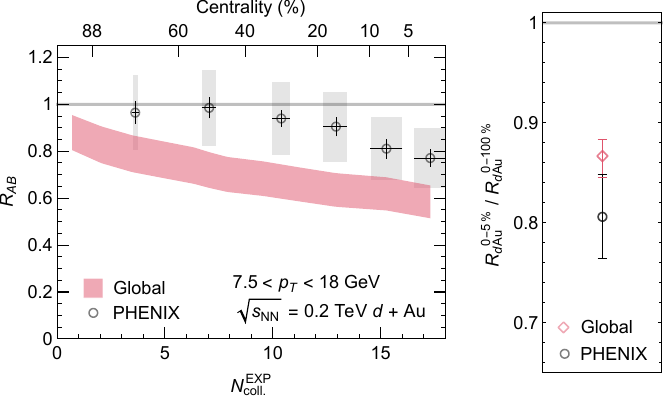}
	\caption{(left) $R_{AB}$ as a function of the experimentally measured number of binary collisions $N_{\text{coll.}}^{\text{EXP}}$ for $7.5 ~\mathrm{GeV} \leq p_T \leq 18 ~\mathrm{GeV}$ $\pi^0$ mesons produced in \coll{d}{Au} collisions as measured by PHENIX \cite{PHENIX:2023dxl} (open markers). Statistical uncertainties are shown as error bars and a global systematic uncertainty of $16.5\%$, which is common to all points, is represented as shaded boxes. The predictions from our global extraction of $\alpha_s^{\text{eff.}}$ are shown as a red band. The band width indicates the theoretical uncertainties and the uncertainties from the extraction, added in quadrature. Theoretical model results are plotted as a function of centrality (top axis) and compared to experimental data at corresponding centrality values.
		(right) Double ratio $R_{d \mathrm{Au}}^{0\text{--}5\%} / R_{d \mathrm{Au}}^{0\text{--}100\%}$ for $7.5 < p_T < 18 ~\mathrm{GeV}$ $\pi^0$ mesons produced in $\sqrt{s_{NN}} = 0.2 ~\mathrm{TeV}$ \coll{d}{Au} collisions as measured by PHENIX \cite{PHENIX:2023dxl} (open black circle) and as predicted by our globally constrained model (open red diamond). The black error bars corresponds to the statistical uncertainty and the red error bars correspond to the theoretical uncertainty and uncertainty from the extraction, added in quadrature.
	}
	\label{fig:dau_rab_vs_ncoll}
\end{figure}

The left panel of \cref{fig:dau_rab_vs_ncoll} plots the $R_{AB,\text{EXP}}$ as a function of the number of the measured binary collisions $N_{\text{coll.}}^{\text{EXP}}$ for $\pi^0$ mesons produced in \coll{d}{Au} collisions with $7.5 ~\mathrm{GeV} \leq p_T \leq 18 ~\mathrm{GeV}$ as measured by PHENIX \cite{PHENIX:2023dxl}. Model predictions using our global extraction of $\alpha_s^{\text{eff.}}$ are shown as a red band as a function of centrality (top axis). 
The quantity $N_{\text{coll.}}^{\text{EXP}}$ cannot be directly calculated in our model, so we compare theory and data at the same centrality to minimize potential centrality biases in the theoretical prediction.
The right panel of \cref{fig:dau_rab_vs_ncoll} plots the double ratio $R_{d \mathrm{Au}}^{0\text{--}5\%} / R_{d \mathrm{Au}}^{0\text{--}100\%}$ for $7.5 < p_T < 18 ~\mathrm{GeV}$ $\pi^0$ mesons produced in $\sqrt{s_{NN}} = 0.2 ~\mathrm{TeV} $\coll{d}{Au} collisions as measured by PHENIX (black circle) and as predicted by our global model (red diamond).

We observe in the left panel of \cref{fig:dau_rab_vs_ncoll} that our global predictions are oversuppressed compared to the experimental data; however, they agree within $\sim 1$ standard deviation. Additionally, the $N_{\text{coll.}}$ dependence of the model predictions and the experimental data are similar.
Note especially that the systematic uncertainties shown as shaded squares are fully correlated between points, which means that an overall normalization difference between data and theory is not unexpected. 
The oversuppression in our results compared to data may stem from assumptions about pre-thermalization time energy loss; we assume a pre-thermalization temperature scaling as $T \propto (\tau_0 / \tau)^{1/3}$, an extrapolation of the later-time Bjorken expansion, which may overestimate energy loss, especially in small systems. If a QGP forms in \coll{d}{Au} collisions then fragmentation photons may also be suppressed, contributing to an underestimate in the experimental suppression \cite{PHENIX:2023dxl}. 
In the right panel of \cref{fig:dau_rab_vs_ncoll}, the double ratio has reduced experimental and theoretical uncertainties compared to $R_{AB}$, and the double ratio exhibits a statistically significant deviation from unity.
Our result agrees with data within the experimental and theoretical uncertainties; the difference between our result and the data can be understood from experimental $R_{dA}^{0\text{--}100\%} \sim 1$  while the presence of QGP formation in all centrality classes in our model leads to $R_{dA}^{0\text{--}100\%} < 1$.

\begin{figure}[!t]
	\centering
	\includegraphics[width=\linewidth]{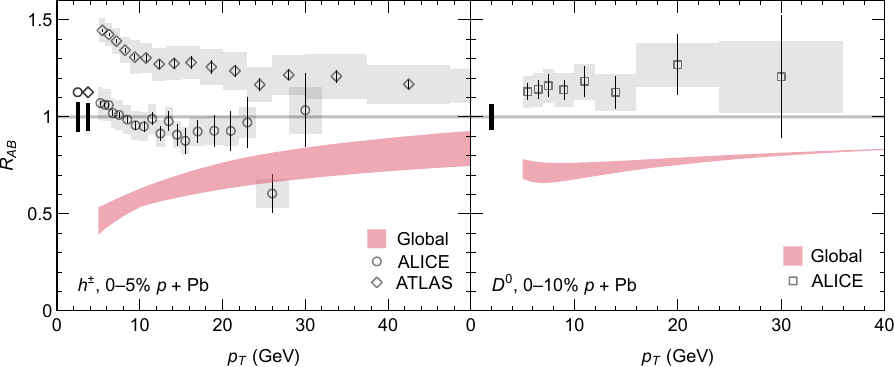}
	\caption{Plot of the experimentally measured $R_{AB}$ as a function of $p_T$ for charged hadrons produced in $0\text{--}5\%$ centrality $\sqrt{s_{NN}} = 5.02 ~\mathrm{TeV}$ \coll{p}{Pb} collisions (left) measured by ALICE \cite{ALICE:2014xsp} (open circles) and ATLAS \cite{ATLAS:2022kqu} (open diamonds) and for $D^0$ mesons produced in $0\text{--}10\%$ centrality \coll{p}{Pb} collisions (right) measured by ALICE \cite{ALICE:2019fhe} (open squares). 
		The ALICE data quoted were obtained using the hybrid method for centrality determination and the charged-particle multiplicity in the $\mathrm{Pb}$-going direction \cite{ALICE:2014xsp}.
		Also shown are the model predictions using the globally extracted $\alpha_s^{\text{eff.}}$ (red) for both final states. The width of the model band indicates the theoretical uncertainty and the uncertainty from the extracted value of $\alpha_s^{\text{eff.}}$. The bars on the open circles represent the statistical uncertainties, the gray boxes represent the systematic uncertainties, and the bars at $R_{AB} = 1$ represent the normalization uncertainty. }
	\label{fig:rab_pPb_combined}
\end{figure}

We now turn our attention to measurements at the LHC of the leading hadron nuclear modification factor in central \coll{p}{Pb} collisions. Currently, no similar photon-normalized $R_{AB,\text{EXP}}$ measurement has been performed at the LHC and all measurements are therefore likely sensitive to (potentially significant) centrality bias. \Cref{fig:rab_pPb_combined} plots the $R_{AB}$ as a function of the hadron transverse  momentum $p_T$ for charged hadrons produced in $0\text{--}5\%$ centrality $\sqrt{s_{NN}} = 5.02 ~\mathrm{TeV}$ \coll{p}{Pb} collisions (left) and for $D^0$ mesons produced in $0\text{--}10\%$ centrality $\sqrt{s_{NN}} = 5.02 ~\mathrm{TeV}$ \coll{p}{Pb} collisions (right). Measured experimental charged hadron data from ALICE \cite{ALICE:2014xsp} (open circles) and ATLAS \cite{ATLAS:2022kqu} (open diamonds) and $D^0$ meson data from ALICE \cite{ALICE:2019fhe} (open squares) are shown. As discussed in \cref{sec:centrality_bias}, the ALICE data utilizes the energy deposited by slow nucleons down the beam line to determine centrality and geometric properties while the ATLAS results use the transverse energy deposited in the forward, $\mathrm{Pb}$-going direction. The different methods for centrality categorization lead to the expectation that the ATLAS results suffer more from centrality bias than the ALICE results \cite{ALICE:2014xsp, ATLAS:2023zfx,ATLAS:2024qsm}.
Predictions from our model based on our global analysis of central and semi-central high-$p_T$ heavy-ion data is shown in red, where the width of the band represents the theoretical uncertainty and the uncertainty on the extraction, added in quadrature. 
We do not perform a local best fit, shown in previous plots as a blue band, since such a fit does not converge to an $\alpha_s^{\text{eff.}}$ within the range for which the model was run $\alpha_s^{\text{eff.}} \in [0.2, 0.55]$.
We observe from the figure that, for both $D^0$ mesons and charged hadrons, the experimental data is in stark disagreement with the model predictions. 
Quantitatively, the $p$-value for the ALICE charged hadron data is $p \simeq 0.1$, the ATLAS charged hadron data is $p \simeq 10^{-7}$ and the $p$-value for the ALICE $D^0$ meson data is $p \simeq 0.01$. 
We note that an exploration of very generic energy loss models, $\Delta E \sim L^a T^b f(E)$, tuned to produce $60\text{--}80\%$ centrality \coll{A}{A} suppression data, also could not reproduce the lack of suppression observed in \coll{p}{Pb} collisions.

	\subsection{Predictions for \texorpdfstring{\coll{O}{O}}{O + O} and \texorpdfstring{\coll{Ne}{Ne}}{Ne + Ne} collisions at the LHC}
\label{sec:predictions_for_oo_nene}

\begin{figure}[!htbp]
	\centering
	\includegraphics[width=\linewidth]{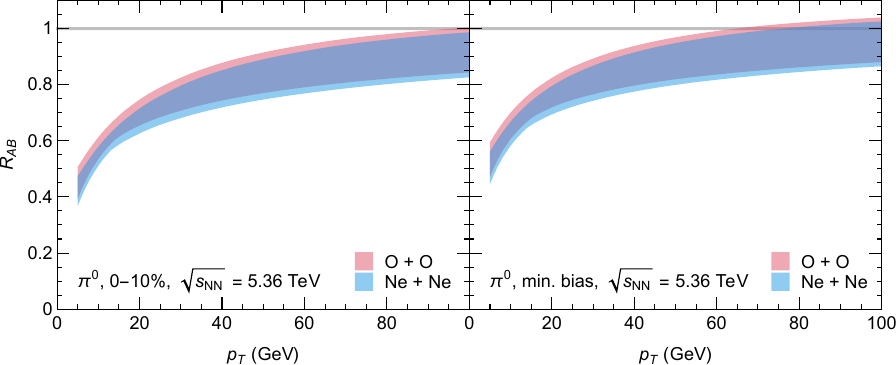}
	\caption{Plot of model predictions for $R_{AB}$ as a function of $p_T$ for charged hadrons produced in $0\text{--}10\%$ (left) and $0\text{--}100\%$ (right) centrality $\sqrt{s_{NN}} = 5.36 ~\mathrm{TeV}$ \coll{O}{O} (red) and \coll{Ne}{Ne} collisions (blue). The width of the model band indicates the theoretical uncertainty and the uncertainty from the extracted value of $\alpha_s^{\text{eff.}}$, added in quadrature.}
	\label{fig:OONeNeRAA}
\end{figure}

In \cref{fig:OONeNeRAA}, we present predictions for $\pi^0$ mesons produced in \coll{O}{O} (red) and \coll{Ne}{Ne} (blue) collisions at $\sqrt{s_{NN}} = 5.36 ~\mathrm{TeV}$ that were recently conducted at LHC. The left panel presents results for $0\text{--}10\%$ centrality collisions and the right panel for minimum bias collisions ($0\text{--}100\%$ centrality). The band width indicates the theoretical model uncertainty and the uncertainty from the extracted value of $\alpha_s^{\text{eff.}}$, added in quadrature. We observe from the figure that there is fairly significant suppression $R_{AB} \sim 0.4\text{--}0.6$ for $5\text{--}20 ~\mathrm{GeV}$ charged hadrons produced in central \coll{O}{O} and \coll{Ne}{Ne} collisions. The $R_{AB}$ for \coll{Ne}{Ne} collisions is systematically smaller than that of \coll{O}{O} collisions; however, the difference is only $\mathcal{O}(1\text{--}5)\%$. The $R_{AB}$ in minimum bias collisions is a particularly nice observable from both a theoretical and experimental standpoint. Theoretically, modifications to the $R_{AB}$ from nuclear PDFs are known to an accuracy of $\sim 10\%$ for \coll{O}{O} collisions \cite{Huss:2020whe, Huss:2020dwe, Gebhard:2024flv}. Experimentally, the lack of centrality selection means that the Glauber model is not needed to make the measurement, eliminating the model dependence of the result \cite{Brewer:2021kiv,Huss:2020dwe,Huss:2020whe}\footnote{In practice, removing the dependence of the $R_{AB}$ on the Glauber model requires that the luminosity is well-known and so experiments will likely still use the Glauber-model-averaged minimum bias $R_{AB}$.}. We observe the same ordering of \coll{O}{O} and \coll{Ne}{Ne} $R_{AB}$ in minimum bias collisions (right panel of \cref{fig:OONeNeRAA}) as was observed in central collisions. 
The amount of suppression for $p_T \lesssim 40 ~\mathrm{GeV}$  particles is significantly more than the $\sim 10\%$ uncertainties in the baseline, meaning that experiment should be able to distinguish between the quenching and no-quenching scenarios \cite{Huss:2020whe, Huss:2020dwe, Gebhard:2024flv}.

\subsection{Predictions to disentangle final- from initial-state effects}
\label{sec:disentangling_final_from_initialstate_effects_predictions_for_the_rhic_system_size_scan}

Motivated by the qualitatively different system size dependence that is expected for final state energy loss compared to initial-state effects \cite{PHENIX:2023dxl}, we present predictions for the double ratio $R \equiv R_{{}^3 \text{He} A} / R_{p A}$. In an energy loss scenario, one anticipates that the larger plasma formed in \coll{He3}{Au} collisions compared to \coll{p}{Au} collisions leads to larger suppression in the former and consequently $R < 1$. On the other hand, initial-state effects, for instance the color fluctuation model \cite{Perepelitsa:2024eik}, and remaining selection biases are expected to have the opposite system size dependence \cite{PHENIX:2023dxl,Perepelitsa:2024eik}, leading to $R>1$.

\begin{figure}[!hb]
	\centering
	\includegraphics[width=\linewidth]{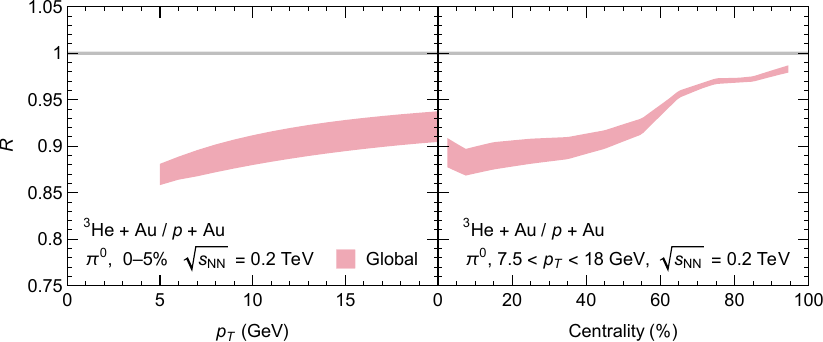}
	\caption{(left) Double ratio $R \equiv R_{{}^3 \text{He} A} / R_{p A}$ as a function of $p_T$ for $\pi^0$ mesons produced in $0\text{--}5\%$ centrality \coll{He3}{Au} and \coll{p}{Au} collisions at $\sqrt{s_{NN}} = 0.2 ~\mathrm{TeV}$, as predicted by our global model tuned on central and semi-central heavy-ion collision data. (right) Global model predictions for $R$ as a function of centrality for $7.5 < p_T < 18 ~\mathrm{GeV}$ $\pi^0$ mesons produced in \coll{He3}{Au} and \coll{p}{Au} collisions at $\sqrt{s_{NN}} = 0.2 ~\mathrm{TeV}$.
	The width of the band corresponds to theoretical uncertainties and uncertainties from the extraction, added in quadrature}
	\label{fig:he3au_over_pau}
\end{figure}

The left panel of \cref{fig:he3au_over_pau} shows the global model prediction (red band) for the double ratio $R_{{}^3 \text{He} A} / R_{p A}$ as a function of $p_T$ for $\pi^0$ mesons produced in $0\text{--}5\%$ centrality \coll{He3}{Au} and \coll{p}{Au} collisions at $\sqrt{s_{NN}} = 0.2 ~\mathrm{TeV}$. The right panel shows $R$ as a function of centrality for $7.5 < p_T < 18 ~\mathrm{GeV}$ $\pi^0$ mesons produced in \coll{He3}{Au} and \coll{p}{Au} collisions at $\sqrt{s_{NN}} = 0.2 ~\mathrm{TeV}$. The model predicts $R \simeq 0.9$ with $\mathcal{O}(1\text{--}5\%)$ theoretical uncertainties for $5\text{--}20 ~\mathrm{GeV}$ $\pi^0$ mesons produced in $\sim 0\text{--}50\%$ centrality collisions. 
The interplay of the different temperature and length distributions in \coll{p}{Au} and \coll{He3}{Au} collisions leads to a non-monotonic centrality dependence of $R$. 
Contrasting predictions of the double ratio $R$ from various initial-state models \cite{Kordell:2016njg,Perepelitsa:2024eik} with our energy-loss-based predictions would help assess whether measuring $R$ would disentangle final- from initial-state suppression within experimental uncertainties.

\section{Identifying and quantifying potential missing physics}
\label{sec:missing_physics}

In the preceding sections we compared our model to experimental data from kinematic regions where we expect that our model is applicable in order to extract a global value of $\alpha_s^{\text{eff.}}$, to check the consistency of our energy loss model as a function of centrality and flavor, and to make predictions for central small and peripheral large systems. 
In this section we consider extending the experimental data that we consider to include kinematic regions where we anticipate that our model will receive corrections due to currently missing physics from the model. In particular, in \cref{sec:applicability_of_pqcd_energy_loss_from_data} we will consider varying the minimum momentum of the data which we include in our statistical extraction in order to extract the scale at which nonperturbative effects, such as medium modification to hadronization, become important. In \cref{sec:running_coupling}, we consider our statistical extraction in different $p_T$ ranges and as a function of $\sqrt{s_{NN}}$ so as to constrain the scales at which the strong couplings in our model might run.%

\subsection{Applicability of pQCD energy loss from data}
\label{sec:applicability_of_pqcd_energy_loss_from_data}

We have thus far considered data with a minimum $p_T$ of $8 ~\mathrm{GeV}$, motivated by the potential onset of non-perturbative effects at lower momenta that are not considered in our model. In particular, our model assumes that partons hadronize according to vacuum fragmentation functions, all energy loss is computed from perturbative QCD, and we neglect contributions from, e.g., temperature gradients and medium flow \cite{Lekaveckas:2013lha,Casalderrey-Solana:2016jvj,Reiten:2019fta,He:2020iow,Antiporda:2021hpk,Bahder:2024jpa}. 
Experimental measurements of the enhanced baryon-to-meson ratio indicate that medium-induced modifications to hadronization is likely the dominant, low-$p_T$ effect and is important for $p_T \lesssim 5\text{--}10 ~\mathrm{GeV}$ \cite{ALICE:2021bib,ALICE:2022exq,CMS:2023frs,LHCb:2018weo,ALICE:2017thy,PHENIX:2003tvk}. 
There are proposals for modifications to hadronization due to the presence of a dense medium of strongly interacting matter, such as coalescence \cite{Sato:1981ez,Biro:1994mp,Biro:1998dm}. 
However, the flavor and $\sqrt{s_{NN}}$ dependence of these effects, as well as their implications for the nuclear modification factor, remain unclear.

In this section we attempt to understand the minimum $p_T$ for which our pQCD based energy loss model with vacuum fragmentation is applicable by comparing our model predictions to experimental data. In particular, we will consider a minimum $p_T$, $p_T^{\text{min}} \in [5, 12] ~\mathrm{GeV}$, and perform the statistical analysis described in \cref{sec:statistical_analysis} on hadrons produced in $0\text{--}50\%$ centrality heavy-ion collisions with $p_T^{\text{min}} \leq p_T \leq 20 ~\mathrm{GeV}$. We perform this analysis separately on light-flavor data from RHIC, light-flavor data from LHC, and heavy-flavor data from LHC; data matching these criteria are taken from \cref{tab:raa_experiments_all_data}. We exclude the STAR charged hadron \cite{STAR:2003fka} and $\pi^0$ \cite{STAR:2009fqa} data from this analysis, as they do not extend beyond $10~\mathrm{GeV}$---within the $p_T$ range ($5\text{--}12 ~\mathrm{GeV}$) over which we vary the lower $p_T$ cutoff---potentially biasing the results.
In this section, we restrict our maximum $p_T$ to $20 ~\mathrm{GeV}$, which results in a weak $p_T$ dependence of the $R_{AB}$ data. This restriction is enforced to limit contributions from running coupling effects, which are not included in our energy loss model. We may thus interpret deviations of the $p$-value from $\sim 1$ as missing low-$p_T$ physics in our model.%

\begin{figure}[!t]
	\includegraphics[width=\linewidth]{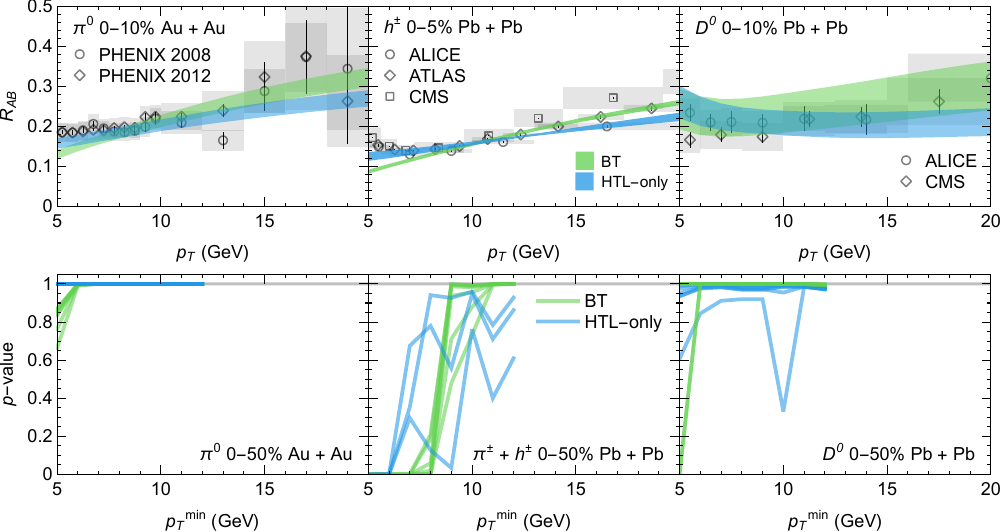}
	\caption{Top: Nuclear modification factor $R_{AB}$ as a function of hadron $p_T$ for $\pi^0$ mesons in $0\text{--}10\%$ centrality $\sqrt{s_{NN}} = 0.2$ TeV \coll{Au}{Au} collisions (left), charged hadrons in $0\text{--}5\%$ centrality $\sqrt{s_{NN}} = 5.02$ TeV \coll{Pb}{Pb} collisions (middle), and $D^0$ mesons in $0\text{--}10\%$ centrality $\sqrt{s_{NN}} = 5.02$ TeV \coll{Pb}{Pb} collisions (right). Theoretical model results with HTL-only collisional energy loss (blue) and BT collisional energy loss (green) are shown, with bands indicating theoretical and statistical uncertainties, added in quadrature. 
		Model bands are constrained from data with maximum $p_T$ of $20 ~\mathrm{GeV}$ and minimum $p_T$ of $5, 9,$ and $5 ~\mathrm{GeV}$ from left to right. These minimum $p_T$ values correspond to the smallest $p_T$ for which the fit produces a $p$-value $> 0.75$; see text for details.
		Experimental data for the respective final states (left to right) are also shown, as measured by PHENIX \cite{PHENIX:2008saf, PHENIX:2012jha}; ALICE \cite{ALICE:2018vuu}, ATLAS \cite{ATLAS:2022kqu}, and CMS \cite{CMS:2016xef}; and ALICE \cite{ALICE:2018lyv} and CMS \cite{CMS:2017qjw}.
		Statistical uncertainties are shown as error bars and systematic uncertainties are shown as gray shaded rectangles; global normalization uncertainties are not shown in the figure and are $\mathcal{O}(3\%)$ for all data except for the PHENIX 2008 data \cite{PHENIX:2008saf} where the global normalization uncertainty is $\mathcal{O}(10\%)$.
Bottom: $p$-value as a function of the minimum $p_T$ used in the analysis for HTL-only (blue) and BT (green) collisional energy loss. Different lines correspond to different $|\mathbf{k}|_{\text{max}}$ multipliers.
The colliding systems and final states considered in the bottom row are the same as the top row, but with an extended centrality range of $0\text{--}50\%$.
Model results that do not yield a $p$-value $> 0.75$ are omitted in the bottom panel for visual clarity.}
	\label{fig:min_pt_from_data}
\end{figure}

\Cref{fig:min_pt_from_data} (top) plots the $R_{AB}$ for $\pi^0$ mesons produced in $0\text{--}10\%$ centrality $\sqrt{s_{NN}} = 0.2 ~\mathrm{TeV}$ \coll{Au}{Au} collisions (left), charged hadrons produced in $0\text{--}5\%$ centrality  $\sqrt{s_{NN}} = 5.02 ~\mathrm{TeV}$ \coll{Pb}{Pb} collisions (middle), and $D^0$ mesons produced in $0\text{--}10\%$ centrality $\sqrt{s_{NN}} = 5.02 ~\mathrm{TeV}$ \coll{Pb}{Pb} collisions (right). 
Note that the top panel is a representative sample of the centrality ranges on which this analysis was conducted, all included data are listed in \cref{tab:raa_experiments_all_data}. 
Also shown are the theoretical model results with HTL-only (blue) and BT (green) collisional energy loss implementations. The band width corresponds to the uncertainty on the extracted value and the theoretical uncertainties, added in quadrature. 
Model results are obtained from a local best fit to the data shown in each respective panel, but using a broader centrality range of $0\text{–}50\%$.
The effective strong coupling $\alpha_s^{\text{eff.}}$ used to construct the model results is obtained for data with a maximum $p_T$ of $20 ~\mathrm{GeV}$ and a minimum $p_T$ of $5$, $9$, and $5 ~\mathrm{GeV}$ for each of the panels from left to right. 
The local effective coupling $\alpha_s^{\text{eff.}}$ is extracted from data with a maximum $p_T$ of $20~\mathrm{GeV}$ and minimum $p_T$ values of $5$, $9$, and $5~\mathrm{GeV}$ for the left, center, and right panels, respectively.
These minimum values correspond to the lowest $p_T$ at which the model provides an adequate description of the data, defined by a $p$-value greater than $0.75$.
The bottom panel shows the $p$-value as a function of the minimum $p_T$ used in the analysis, where both HTL-only (blue) and BT (green) collisional energy loss results are shown and different lines correspond to different values of the $|\mathbf{k}|_{\text{max}}$ multiplier\footnote{Note that model results which do not ever result in a $p$-value $> 0.75$ are not included in the bottom panel of the figure for visual clarity}.

The colliding systems and final states used to construct the plots in the bottom row are the same as that in the top row, but with an extended centrality range of $0\text{--}50\%$. Additionally $R_{AB}$ data measured by ALICE \cite{ALICE:2019hno} for $\pi^{\pm}$ meson produced in $0\text{--}50\%$ centrality \coll{Pb}{Pb} collisions are also included in the light-flavor LHC results (middle panel) even though they are not shown in the corresponding representative $R_{AB}$ plot.
When interpreting the goodness-of-fit of our energy loss mode to experimental data, we take $p > 0.75$ as an adequate description of the data by the model; we found that our results are not sensitive to the exact $p$-value chosen.
We note that the $p$-value curves are not particularly smooth as a function of $p_T^{\text{min}}$, which is because shifting the cutoff changes the included data points discretely. If the excluded data points form a large portion of the total dataset then there can be discontinuous changes in the $p$-values, particularly noticeable for the point at $p_T^{\text{min}} = 10 ~\mathrm{GeV}$ in the bottom-right panel of \cref{fig:min_pt_from_data}. We emphasize that these we checked that these discontinuous changes are not false convergences.

Considering the left panel of \cref{fig:min_pt_from_data}, we observe that our model accurately describes the $\pi^0$ meson data at RHIC down to $p_T^{\text{min}} = 5 ~\mathrm{GeV}$, largely independent of different theoretical model uncertainties.
We do not extend our model lower than $p_T = 5 ~\mathrm{GeV}$, due to concerns about the applicability of pQCD energy loss. We conclude that, within the $p_T$ range that we consider, there is no evidence in the $\pi^0$ $R_{AB}$ at RHIC for medium modifications to hadronization.
We observe from the middle panel of \cref{fig:min_pt_from_data}, that the minimum $p_T$ required for our model to adequately describe the $p_T$ dependence of charged hadron data at LHC is $p_T^{\text{min}} \simeq 9 ~\mathrm{GeV}$, largely independent of the different theoretical model uncertainties.
We interpret this minimum $p_T$ below which $p$-values drop approximately to zero as indirect evidence that there are medium modifications to pion hadronization below $p_T \simeq 9 ~\mathrm{GeV}$ at LHC.
We note that if we repeat this analysis on only the $\pi^{\pm}$ meson data from LHC \cite{ALICE:2019hno}, this minimum $p_T$ is less constrained: we find $p_T^{\text{min}}\simeq 6\text{--}9 ~\mathrm{GeV}$ depending on the $|\mathbf{k}|_{\text{max}}$ multiplier and collisional energy loss implementation that are used. We believe that the smaller minimum $p_T$ is because the identified hadron data is less constraining, due to the larger uncertainties, than the inclusive charged hadron data. %

We see that our model describes the RHIC data accurately down to a smaller minimum $p_T$ than at the LHC, consistent with the $\sim 30\%$ lower temperatures in RHIC collisions compared to LHC collisions leading to decreased effects of medium modifications to hadronization at RHIC.
Considering the right panel of \cref{fig:min_pt_from_data}, we see that our model accurately describes the $D^0$ meson data at LHC down to $p_T^{\text{min}} \simeq 5 ~\mathrm{GeV}$, largely independent of the different theoretical model uncertainties. The ability of our model to describe the $D^0$ meson data to lower $p_T$ than the charged hadron data may simply be a reflection of the larger uncertainties on the measured $D^0$ mesons compared to charged hadrons. Alternatively, this may indicate that medium modified hadronization does not extend to as high $p_T$ for heavy-flavor mesons compared to light-flavor mesons, presumably due to the mass of the charm quark. 
Either scenario is compatible with the current experimentally measured heavy-flavor meson to heavy-flavor baryon ratio \cite{ALICE:2022exq,CMS:2023frs,LHCb:2018weo}. 

We conclude that to extend our results to a lower $p_T$ range for pions at the LHC, medium modifications to hadronization, for instance coalescence, should be included in the energy loss model. It appears that heavy-flavor mesons at LHC and light-flavor mesons at RHIC are less sensitive to these effects, at least within the current experimental uncertainties.

\subsection{Running coupling}
\label{sec:running_coupling}

Up until now, we have considered data over a relatively small $p_T$ range in order to limit the effects of running coupling, which are not currently incorporated into our model.  In this section, we will constrain our model on data from different $p_T$ ranges in an attempt to estimate what scale(s) the strong couplings are expected to run from high-$p_T$ suppression data. Since the data for identified hadrons does not extend past $p_T \gtrsim 100 ~\mathrm{GeV}$ and has significantly larger uncertainties than the charged hadron data, we consider only charged hadron data from $\sqrt{s_{NN}} = 5.02 ~\mathrm{TeV}$ \coll{Pb}{Pb} collisions as measured by CMS \cite{CMS:2016xef} and ATLAS \cite{ATLAS:2022kqu}, which extend to $p_T \simeq 300 ~\mathrm{GeV}$. We will additionally restrict our analysis to $0\text{--}50\%$ centrality collisions as there is significant evidence that a QGP is formed in this centrality range \cite{ALICE:2022wpn,CMS:2024krd,PHENIX:2004vcz,STAR:2005gfr} and experimental measurements in this centrality range are insensitive to centrality bias \cite{ALICE:2018ekf}. For $p_T \gtrsim 50 ~\mathrm{GeV}$, there are non-negligible differences in the reported $R^{h^{\pm}}_{AB}$ by ATLAS \cite{ATLAS:2022kqu} and CMS \cite{CMS:2016xef}. To understand the different implications of the ATLAS and CMS data, we conduct our subsequent analyses on the ATLAS and CMS datasets separately.

We extract a separate value of $\alpha_s^{\text{eff.}}$ from data in each of the following disjoint $p_T$ ranges (in $\mathrm{GeV}$): $\{ [10, 17], (17, 30], (30, 55], (55, 132], (132, 330]\}$. These five $p_T$ ranges were chosen to contain a similar number of data points and to be relatively uniform on a logarithmic scale. %
The similar number of data points allows a similarly constrained fit for each $p_T$ range, albeit the higher $p_T$ ranges have larger experimental uncertainties. The approximate uniformity of the $p_T$ ranges on a logarithmic scale means that strong couplings that may run as $\alpha_s (p_T) \sim 1 / \ln p_T$ change by approximately the same relative amount between the bounds of each $p_T$ range.

\begin{figure}[!t]
	\includegraphics[width=\linewidth]{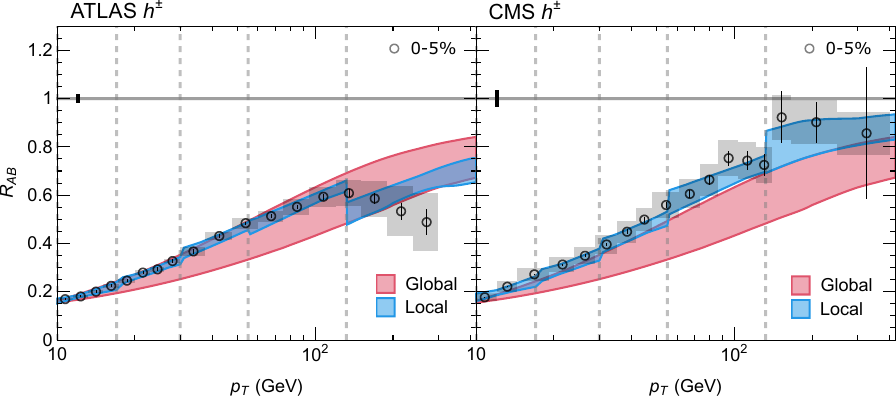}
	\caption{$R_{AB}$ as a function of $p_T$ for charged hadrons produced in $0\text{--}5\%$ centrality \coll{Pb}{Pb} collisions measured by ATLAS \cite{ATLAS:2022kqu} (left) and CMS \cite{CMS:2016xef} (right). 
		Model results from our global extraction of $\alpha_s^{\text{eff.}}$ (described in \cref{sec:global_extraction}) are shown in red, where the band width indicates the theoretical and extraction uncertainty, added in quadrature.
	The dashed vertical gray lines delineate the different $p_T$ ranges for which a separate local value of the effective strong coupling is extracted. Blue bands indicate the model results for each of these separate local best fits, where the band width indicates the theoretical and extraction uncertainty, added in quadrature..
}
	\label{fig:raa_CMS_ATLAS}
\end{figure}

\Cref{fig:raa_CMS_ATLAS} plots the measured $R_{AB}$ of charged hadrons in $0\text{--}5\%$ centrality \coll{Pb}{Pb} collisions as a function of $p_T$, as measured by ATLAS \cite{ATLAS:2022kqu} (left) and CMS \cite{CMS:2016xef} (right). The model results from the global extraction of $\alpha_s^{\text{eff.}}$ (described in \cref{sec:global_extraction}) are shown in red, where the width of the band corresponds to the theoretical uncertainty and uncertainty on extracted $\alpha_s^{\text{eff.}}$, added in quadrature. Additionally, model results produced with the extracted $\alpha_s^{\text{eff.}}$ from separate fits in the five disjoint $p_T$ regions and for each experimental dataset are shown in blue. The vertical dashed gray lines indicate the $p_T$ regions used for these local best fits. We observe in \cref{fig:raa_CMS_ATLAS} that the ATLAS and CMS datasets lead to qualitatively different conclusions about the inclusion of running coupling effects. 
	The ATLAS data favors the coupling staying relatively constant in the first three $p_T$ ranges, before slightly increasing in the fourth $p_T$ range, $p_T \geq 55 ~\mathrm{GeV}$, and then significantly increasing in the final $p_T$ range, $p_T \geq 132 ~\mathrm{GeV}$.
	The CMS data shows a preference for the coupling decreasing monotonically as a function of $p_T$ over all $p_T$ ranges, qualitatively consistent with the coupling running with $p_T$.
	One sees that our global extracted best fit is in good qualitative agreement with the ATLAS data except for the highest $p_T$ bins, while our global extracted best fit is systematically oversuppressed compared to the CMS data. Since our model explicitly uses fixed coupling, and we expect there to be running coupling effects, our agreement with the ATLAS data is surprising, while our systematic over-suppression compared to CMS data is what we would qualitatively expect.

Let us now consider various simple parametric estimates for scales at which the coupling might run, and compare our results to the extracted $\alpha_s^{\text{eff.}}$ at these scales. While at this time there exists no rigorous calculation of the scale at which the coupling runs in energy loss processes, there are many phenomenological implementations in the literature \cite{Peshier:2006ah,Djordjevic:2013xoa,Xu:2014ica,Xu:2015bbz,JETSCAPE:2024cqe}. 
While each diagram that contributes to energy loss could involve couplings that each run at different scales, it is typical to consider only three scales for radiative energy loss and two scales for collisional energy loss.
Phenomenological implementations of running coupling in the literature \cite{Peshier:2006ah,Djordjevic:2013xoa,Xu:2014ica,Xu:2015bbz,JETSCAPE:2024cqe} typically allow the various couplings in the problem to run as 
\begin{equation}
	\alpha_s(Q) = 2 \pi / (9 \ln (Q / \Lambda_{\text{QCD}})
	\label{eqn:running_coupling}
\end{equation}
where the scale $Q$ differs in different models and depending on the vertex (radiative or elastic scatter) in the relevant process. 
Examples of the scales used in the literature include \cite{Xu:2014ica, Xu:2015bbz, Peshier:2006ah,JETSCAPE:2024cqe,Djordjevic:2013xoa} $Q^2 \in \{ \mathbf{q}^2, \mathbf{k}^2 / [x (1 - x)], 4 E T, 2 E T, (2 \pi T)^2, \mu_D^2, E^2\}$, where $\mathbf{q}$ is the transverse momentum exchanged with the scattering center, $\mathbf{k}$ is the transverse radiated gluon momentum, $\mu_D$ is the Debye mass, $T$ is the temperature, and $E$ is the incident parton energy. 
We will follow the literature in using \cref{eqn:running_coupling} for the parameterization of the running coupling, and we will take $\Lambda_{\text{QCD}} = 0.2 ~\mathrm{GeV}$. 
At this level of analysis, we cannot consider parameters such as $\mathbf{k}$ and $\mathbf{q}$ as they are already integrated out. Instead, we consider a single characteristic value of $\alpha_s$ that we compare to the extracted $\alpha_s^{\text{eff.}}$ from our model, defined as
\begin{equation}
	\alpha_s^{\text{single}}(Q_{\text{coll.}}, Q_{\text{rad.}}) \equiv \left \{ 
	\begin{array}{ll}
		\left[\alpha_s^2\left(Q_{\text{coll.}}\right) \alpha_s\left(Q_{\text{rad.}}\right) \right]^{1 / 3}\\[5pt]
		\left[\alpha_s^2\left(Q_{\text{coll.}}\right)\right]^{1 / 2},
	\end{array}
	\right.
	\label{eqn:running_coupling_single_effective}
\end{equation}
where $Q_{\text{coll.}}$ and $Q_{\text{rad.}}$ are the scales at which the collisional and radiative couplings run, shown in \cref{tab:running_coupling_forms}, and the top form is used when there is a radiative scale while the bottom form is used when there is no radiative scale.

\begin{table}[!b]
    \centering
    \begin{tabular}{ccl}
        \toprule
        Collisional scale $Q^2_{\text{coll.}}$ & Radiative scale $Q^2_{\text{rad.}}$ \\
        \midrule
        $ E T$  & None\\
        $ E T$  & $E^2$\\
        $ (2 \pi T)^2 $  & None \\
				$ (2 \pi T)^2 $  & $E^2$ \\
        \bottomrule
    \end{tabular}
		\caption{The four pairs of scales ($Q_{\text{coll.}}$, $Q_{\text{rad.}}$) used in \cref{eqn:running_coupling_single_effective} considered in our analysis.}
\label{tab:running_coupling_forms}
\end{table}

\begin{figure}[!t]
	\includegraphics[width=\linewidth]{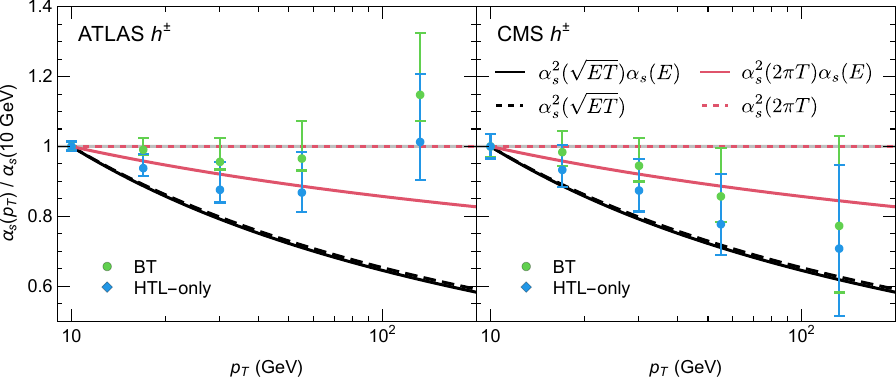}
	\caption{Ratio of the effective strong coupling $\alpha_s^{\text{eff.}}$ extracted in each $p_T$  range to the value of the strong coupling at  $p_T = 10 ~\mathrm{GeV}$  for ATLAS \cite{ATLAS:2022kqu} (left) and CMS \cite{CMS:2016xef} (right) charged hadron data from $0\text{--}5\%$ centrality $\sqrt{s_{NN}} = 5.02 ~\mathrm{TeV}$ \coll{Pb}{Pb} collisions. Points are placed at the lower bound of the corresponding  $p_T$ range. Error bars indicate uncertainties on the extracted value of $\alpha_s^{\text{eff.}}$ and theoretical uncertainties from the value of the $|\mathbf{k}|_{\text{max}}$ multiplier, added in quadrature. Lines represent different scales at which the coupling might run, as listed in \cref{tab:running_coupling_forms}. Extractions are shown separately for HTL-only and BT collisional energy loss implementations.}
	\label{fig:running_coupling_CMS_ATLAS}
\end{figure}

\Cref{fig:running_coupling_CMS_ATLAS} plots the ratio of the local best fit extracted $\alpha_s^{\text{eff.}}$ in each $p_T$ range to the local best fit extracted $\alpha_s^{\text{eff.}}$ in the smallest $p_T$ bin as a function of $p_T$ for the ATLAS charged hadron data \cite{ATLAS:2022kqu} (left) and the CMS charged hadron data \cite{CMS:2016xef} (right). The points are placed at the lower $p_T$ bound of the $p_T$ range used, since the lowest $p_T$ point has the smallest uncertainty and is therefore typically the most constraining on the value of $\alpha_s^{\text{eff.}}$ in the fit. Error bars represent the uncertainty associated with the extracted value of $\alpha_s^{\text{eff.}}$. 
Also shown in the figure are lines that correspond to different scales at which the coupling might run, as listed in \cref{tab:running_coupling_forms}. Specifically, these lines are evaluated as $\alpha_s^{\text{single}}(Q_{\text{coll.}}, Q_{\text{rad.}}) / \alpha_s^{\text{single}}(Q_{\text{coll.}}^{0}, Q_{\text{rad.}}^{0})$ where $Q_{\text{coll.}}$ and $Q_{\text{rad.}}$ are taken from \cref{tab:running_coupling_forms} and $Q_{\text{coll.}}^{0}$ and $Q_{\text{rad.}}^{0}$ are $Q_{\text{coll.}}$ and $Q_{\text{rad.}}$, respectively, evaluated at the lowest $p_T$ bin.

We observe from \cref{fig:running_coupling_CMS_ATLAS} that the extracted $\alpha_s^{\text{eff.}}$ from CMS data decreases monotonically in $p_T$, qualitatively consistent with the running coupling expectation, while the central value of the ATLAS extracted $\alpha_s^{\text{eff.}}$ decreases for $p_T \lesssim 50 ~\mathrm{GeV}$ before \emph{increasing} for $p_T \gtrsim 50 ~\mathrm{GeV}$. While the favored extracted $p_T$-dependence of $\alpha_s^{\text{eff.}}$ for the ATLAS data leads to $\alpha_s^{\text{eff.}}$ rising in $p_T$, the results are still consistent with no $p_T$ dependence in $\alpha_s^{\text{eff.}}$ within a 68\% confidence interval. In contrast, the CMS data favors the coupling running somewhere between $(\alpha_s^{\text{eff.}})^3(E) \propto \alpha_s^2(\sqrt{E T}) \alpha_s(E)$ and $(\alpha_s^{\text{eff.}})^3(E) \propto \alpha_s^2(2 \pi T) \alpha_s(E)$. 

We also see in \cref{fig:running_coupling_CMS_ATLAS} that there is a fairly large difference in the implied scales at which the couplings run when using BT vs.\ HTL-only elastic energy loss. Considering the CMS data, the BT collisional energy loss results favor the coupling running as $(\alpha_s^{\text{eff.}})^3(E) \propto \alpha_s^2(2 \pi T) \alpha_s(E)$, while the HTL-only collisional energy loss results favor a coupling running somewhere between $(\alpha_s^{\text{eff.}})^3(E) \propto \alpha_s^2(2 \pi T) \alpha_s (E)$ and $(\alpha_s^{\text{eff.}})^3(E) \propto \alpha_s^2(\sqrt{ET}) \alpha_s(E)$. 
We see that there is a negligible difference in the running of the coupling when the coupling runs as $(\alpha_s^{\text{eff.}})^3(E) \propto \alpha_s^2(\sqrt{ET}) \alpha_s(E)$ (solid black) compared to when the coupling runs as $(\alpha_s^{\text{eff.}})^2(E) \propto \alpha_s^2(\sqrt{ET})$ (dashed black); the similarity of these results is not a surprise since the coupling running at $\sqrt{E T}$ is parametrically the same as the coupling running at $E$ for $E \gg T$. Considering the ATLAS data, the BT collisional energy loss is compatible with the coupling running as $(\alpha_s^{\text{eff.}})^2 \propto \alpha_s^2(2 \pi T)$ while there is no consistent set of scales for the HTL-only to run at.

We conclude that the datasets from CMS and ATLAS lead to significantly different pictures of the scale at which the coupling might run. We note that while there is no physical mechanism that leads to the coupling increasing as a function of $p_T$, the ATLAS results are consistent with the coupling running only at the temperature scale. The CMS results, however, imply that at least one of the couplings present in the model run at the hard scale and are compatible with all three couplings running, at least partially, at the hard scale.
Focusing on the CMS data, which \cref{fig:running_coupling_CMS_ATLAS} shows leads to an extracted $\alpha_s^{\text{eff.}}$ running between $(\alpha_s^{\text{eff.}})^3(E) \propto \alpha_s^2(\sqrt{E T}) \alpha_s(E)$ and $(\alpha_s^{\text{eff.}})^3(E) \propto \alpha_s^2(2 \pi T) \alpha_s(E)$, it is likely that a more sophisticated derivation of energy loss that determines more carefully the running coupling on a diagram-by-diagram basis would yield a result in better agreement with CMS data.
LHC run 3 results for \coll{Pb}{Pb} collisions should yield better statistics, allowing for a more accurate determination of the $R_{AB}$ at large $p_T$ and potentially clarifying this discrepancy between ATLAS and CMS results. Future theoretical work might perform a rigorous calculation of the scales at which the strong coupling runs, allowing for a more rigorous comparison to high-$p_T$ experimental data \cite{Horowitz:2010yg}.

We now shift our attention to consider the potential effects of couplings running at the temperature scale. As shown in \cref{fig:phase_space_experiments}, different centrality classes of heavy-ion collisions probe mostly different path lengths, but the average temperature in these systems is largely the same, except in more peripheral centrality classes and in central small systems.
From \cref{fig:phase_space_experiments}, we see that central \coll{d}{Au} collisions have a temperature similar to central \coll{Pb}{Pb} collisions. While central \coll{p}{Pb} collisions have a higher temperature, the apparent large centrality bias in those systems prevents us from considering them in this analysis.
In order to change the temperature appreciably, one could consider going to more peripheral collisions, but we found that the large uncertainties on this data meant that the extracted couplings were not well constrained.
Therefore, we consider the two temperatures for central and semi-central \coll{A}{A} collisions available to us by the change in $\sqrt{s_{NN}}$ from $\sqrt{s_{NN}} = 0.2 ~\mathrm{TeV}$ at RHIC to $\sqrt{s_{NN}} = 5.02 ~\mathrm{TeV}$ at LHC.
We now perform our analysis separately for $0\text{--}50\%$ centrality light-flavor hadrons from \coll{Au}{Au} collisions at $\sqrt{s_{NN}} = 0.2 ~\mathrm{TeV}$ at RHIC and from \coll{Pb}{Pb} collisions at $\sqrt{s_{NN}} = 5.02 ~\mathrm{TeV}$ at LHC. 
We exclude the $D^0$ meson experimental data because we found in \cref{sec:flavor} that the extracted $\alpha_s^{\text{eff.}}$ for $D^0$ mesons was not in particularly good agreement with that extracted from charged hadrons.
These two data sets have respective average temperatures of $\langle T_{\text{RHIC}} \rangle = 0.23 ~\mathrm{GeV}$ and $\langle T_{\text{LHC}} \rangle = 0.31 ~\mathrm{GeV}$ in our model (see \cref{fig:phase_space_experiments}). %

\begin{figure}[!htbp]
	\centering
	\includegraphics[width=0.6\linewidth]{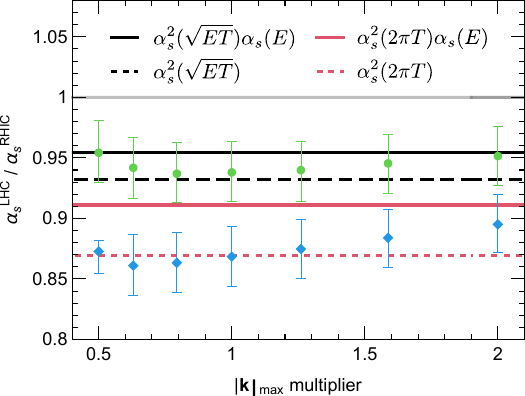}
	\caption{Ratio of the effective strong coupling $\alpha_s^{\text{eff.}}$ extracted from LHC data to that extracted from RHIC data as a function of the $|\mathbf{k}|_{\text{max}}$ multiplier. Results are shown for both BT (green circles) and HTL-only (blue diamonds) collisional energy loss implementations. 
		Error bars indicate the uncertainty on the extracted value of $\alpha_s^{\text{eff.}}$.
	Lines represent this ratio calculated under different assumptions about the scales at which the couplings might run, listed in \cref{tab:running_coupling_forms}.}
	\label{fig:running_coupling_rhic_lhc}
\end{figure}

\Cref{fig:running_coupling_rhic_lhc} shows the ratio of the extracted $\alpha_s^{\text{eff.}}$ from data at LHC to that extracted from data at RHIC as a function of the $|\mathbf{k}|_{\text{max}}$ multiplier. The lines represent the expected ratio if the coupling ran as various simple parametric forms (see \cref{eqn:running_coupling_single_effective} and \cref{tab:running_coupling_forms}). The bars represent the uncertainty associated with the extraction of $\alpha_s^{\text{eff.}}$. Results are shown for both BT (green circles) and HTL-only (blue diamonds) collisional energy loss implementations. 
We observe from the figure that the HTL-only collisional energy loss implementation favors the coupling running as $(\alpha_s^{\text{eff.}})^2 \propto \alpha_s^2(2 \pi T)$ or $(\alpha_s^{\text{eff.}})^3 \propto \alpha_s^2(2 \pi T) \alpha_s(E)$ while the BT collisional energy loss implementation is less constrained, but is incompatible with the coupling running as $(\alpha_s^{\text{eff.}})^2 \propto \alpha_s^2(2 \pi T)$. We additionally see that the ratio $\alpha_s^{\text{LHC}} / \alpha_s^{\text{RHIC}}$ is relatively insensitive to the $|\mathbf{k}|_{\text{max}}$ multiplier that is used.

	\begin{table}[!h]
    \centering
    \begin{tabular}{l||c|c}
        \toprule
				& HTL-only & BT \\
        \midrule
        \midrule
				Varying $p_T$ (ATLAS) & None & $\alpha_s^2(2 \pi T)$\\
        \midrule
				Varying $p_T$ (CMS) & \makecell{
				$\alpha_s^2(\sqrt{ET})$,
				$\alpha_s^2(\sqrt{ET}) \alpha_s(E)$,\\
			$\alpha_s^2(2 \pi T) \alpha_s(E)$
		} & $\alpha_s^2(2 \pi T) \alpha_s(E)$\\
        \midrule
				Varying $\sqrt{s_{NN}}$ & \makecell{
				$\alpha_s^2(2 \pi T)$,\\ 
			  $\alpha_s^2(2 \pi T) \alpha_s(E)$} & \makecell{$\alpha_s^2(\sqrt{ET})$, $\alpha_s^2(\sqrt{ET}) \alpha_s(E)$,\\$\alpha_s^2(2 \pi T) \alpha_s(E)$} \\
        \bottomrule
    \end{tabular}
	\caption{Parametric forms of the running coupling which are preferred by data from a comparison by varying $p_T$ with charged hadron data from $0\text{--}5\%$ centrality \coll{Pb}{Pb} collisions reported by ATLAS (first row) and CMS (second row), as well as by varying $\sqrt{s_{NN}}$ from $\sqrt{s_{NN}} = 0.2 ~\mathrm{TeV}$ to $\sqrt{s_{NN}} = 5.02 ~\mathrm{TeV}$ with data from STAR, PHENIX, ATLAS, CMS, and ALICE; consult \cref{tab:raa_experiments_all_data} for details.. Results are shown for HTL-only (first column) and BT (second column) collisional energy loss.}
		\label{tab:running_coupling_prefered}
\end{table}

	We now consider the joint implications of our results of the running of the strong coupling with temperature and with $p_T$. If the differences in the extracted $\alpha_s^{\text{eff.}}$ are actually a manifestation of running coupling effects, there should be a consistent set of scales at which the coupling runs. 
	\Cref{tab:running_coupling_prefered} shows the forms for the parametric running couplings that are approximately within one standard deviation of the extracted effective strong couplings $\alpha_s^{\text{eff.}}$. 
	From \cref{tab:running_coupling_prefered}, we see that the implications for the scales at which couplings run are sensitive to the tension in the ATLAS and CMS data for $p_T \gtrsim 50 ~\mathrm{GeV}$.
	The ATLAS data in conjunction with the $\sqrt{s_{NN}}$ dependence of the extracted $\alpha_s^{\text{eff.}}$ does not have a consistent set of scales that the coupling can run at within the set that we considered. Considering the coupling running with $p_T$ from the CMS data and the coupling running with temperature from the comparison of RHIC and LHC data, the coupling is likely to run as $(\alpha_s^{\text{eff.}})^3 \propto \alpha_s^2(2 \pi T) \alpha_s(E)$ and the preferred collisional energy loss implementation is likely somewhere between HTL-only and BT.
	While in this analysis we only considered four simple forms for the running coupling, future analysis could consider more realistic running coupling forms to gain further insight into the energy loss mechanisms and the running coupling. Additionally, the sensitivity of our results to the collisional energy loss implementation indicates that a more detailed study could place stronger constraints on the energy loss mechanism.

\section{Jet transport coefficient and comparison with other work}
\label{sec:jet_transport_coefficient_and_comparison_with_other_work}

While in our model it is most convenient to treat the effective strong coupling as the single fitted parameter, there is a significant literature \cite{JET:2013cls,Andres:2016iys,Xie:2020zdb,JETSCAPE:2021ehl,Xie:2022ght,Apolinario:2022vzg} that considers the extraction of the jet transport coefficient $\hat{q}$. In this section we will compute the jet transport coefficient $\hat{q}$ at RHIC and LHC temperatures using the effective strong couplings that were extracted in \cref{sec:global_extraction}. The jet transport coefficient $\hat{q}$ is defined as \cite{Armesto:2011ht}
\begin{equation}
	\hat{q} \equiv \rho \int d^2 \mathbf{q} \; \mathbf{q}^2 \frac{d \sigma}{d^2 \mathbf{q}} =\int_0^{q_{\max }} d^2 \mathbf{q} \frac{d \Gamma_{\text{el}}}{d^2 \mathbf{q}} \mathbf{q}^2 ,
	\label{eqn:qhat}
\end{equation}
where $\mathbf{q}$ is the transverse momentum exchanged with a scattering center, $\rho$ is the density of the medium, $\frac{d \sigma}{d^2 \mathbf{q}}$ is the elastic cross section, and $\frac{d \Gamma_{\text{el}}}{d^2 \mathbf{q}}$ is the elastic scattering rate. In principle, the elastic scattering rate is fully specified by the modeling of the collisional energy loss, discussed in \cref{sec:collisional_energy_loss}. In practice, the number of scatterings in the BT collisional energy loss is divergent in the infrared and so one cannot compute $\frac{d \Gamma_{\text{el}}}{d^2 \mathbf{q}}$, therefore, we use the following expression \cite{Armesto:2011ht}
\begin{equation}
	\hat{q} = 4 \pi \alpha_s^2 C_R \ln \left(\frac{q_{\max }^2(E,T)}{\mu_D^2(T)}+1\right) \times \frac{\zeta(3)}{\zeta(2)}\left(1+\frac{1}{4} n_f\right) T^3,
	\label{eqn:qhat_brick}
\end{equation}
where $C_R$ is the Casimir of the incident parton, $n_f$ is the number of active quark flavors in the plasma, $\mu_D$ is the Debye mass, and $\zeta$ is the Riemann zeta function. In \cref{eqn:qhat_brick} we will use $q_{\text{max}}^2(E, T) = 3 E \mu_D(T)$, consistent with what was used to compute the energy loss. 

For HTL-only collisional energy loss, we may compute $d \Gamma_{\text{el}} / d^2 \mathbf{q} \equiv dN / dz d^2\mathbf{q}$ from \cref{eqn:scattering_rate_final} by writing \cref{eqn:integral_q} as
\begin{equation}
	\int_q = \int_0^{q_{\text{max}}} \frac{d^2 \mathbf{q}}{2} \int_{-E+\sqrt{M^2+\mathbf{q}^2}}^{\infty} d \omega \quad \text{with} \quad q_z=\frac{\omega}{v},
	\label{eqn:}
\end{equation}
which allows for a straightforward computation of $d \Gamma_{\text{el}}/d^2 \mathbf{q}$ from \cref{eqn:scattering_rate_final},
\begin{align}
	\frac{dN}{d z d^2 q} =& \frac{2 C_R \alpha_s^2}{\pi} \int_{-E+\sqrt{M^2+\mathbf{q}^2}}^{\infty} d \omega \quad \frac{n_B(\omega)}{q} \left( 1 - \frac{\omega^2}{q^2} \right)^2\\
	&\sum_m\left[C_{L L}^{m}\left|\Delta_L\right|^2+2 C_{L T}^{m} \operatorname{Re}\left(\Delta_L \Delta_T\right)+C_{T T}^m\left|\Delta_T\right|^2\right],
	\label{eqn:wicks_rate}
\end{align}
where $q_z = \omega / v$.

In all cases of the computation of $\hat{q}$, we will compute $\hat{q}$ for light quarks with $E = 100 ~\mathrm{GeV}$ and $n_f = 2$. The Debye mass $\mu_D$ is computed from \cref{eqn:debye_mass}, consistent with all the other results in this work. These choices are consistent with those made by JETSCAPE \cite{JETSCAPE:2024cqe}; however, other models have made different choices, which may impact the extracted $\hat{q}$. Ensuring that different groups use compatible parameters and computation methods for extracting $\hat{q}$ is important, but is beyond the scope of this work.

We treat the extracted $\alpha_s^{\text{eff.}}$ for the various theoretical models as forming a one-sigma distribution, which was discussed in \cref{sec:global_extraction}. For models incorporating BT collisional energy loss, we use \cref{eqn:qhat_brick}, while for models employing HTL-only collisional energy loss, we use \cref{eqn:qhat} together with \cref{eqn:wicks_rate}.

\begin{figure}[!t]
	\centering
	\includegraphics[width=0.85\linewidth]{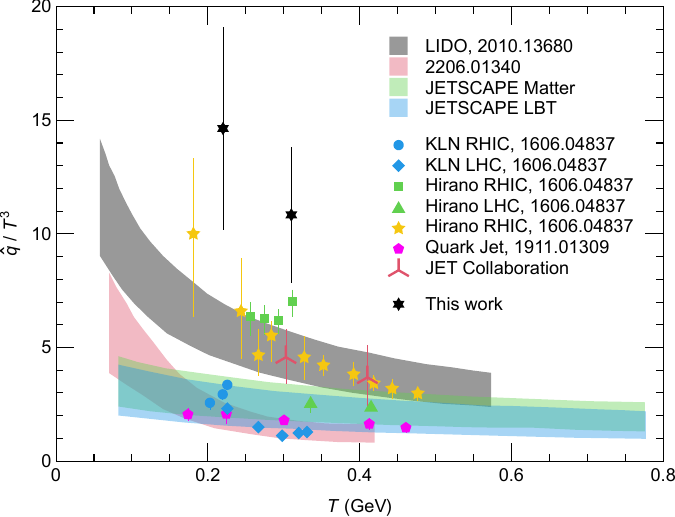}
	\caption{The jet transport coefficient over the temperature cubed $\hat{q} / T^3$ as a function of the temperature $T$  of the formed plasma. Extracted values are shown for analysis used in this work (black six-pointed stars) as well the values extracted from a variety of other works \cite{Andres:2016iys, Ke:2020clc, Xie:2020zdb, Xie:2022ght} (points and bands). Figure adapted from \cite{Apolinario:2022vzg}.}
	\label{fig:qhatOverTcubed}
\end{figure}

\Cref{fig:qhatOverTcubed} plots the extracted $\hat{q} / T^3$ as a function of the temperature of the plasma $T$ as extracted from our global analysis (black points) as well as from various other analyses of high-$p_T$ data \cite{JET:2013cls,Andres:2016iys,Xie:2020zdb,JETSCAPE:2021ehl,Xie:2022ght,Apolinario:2022vzg}. Our extraction yields a larger $\hat{q}/T^3$ than these other analyses, although with substantial theoretical uncertainties. 
We emphasize that the bands shown in other works reflect only uncertainties in the extracted parameters, whereas in our analysis theoretical uncertainties---which dominate the $\hat{q}$ uncertainty---are also included, highlighting their importance when determining physically relevant quantities. 
We further note that our larger $\hat{q}$ may reflect the omission of next-to-leading-order (NLO) corrections to both $\hat{q}$ \cite{Blaizot:2014bha,Arnold:2021pin,Muller:2021wri,Ghiglieri:2022gyv} and the energy loss calculation, which could likely be absorbed into a renormalization of $\alpha_s^{\text{eff.}}$ and, consequently, $\hat{q}$. Additionally, missing initial-state physics, such as nPDF effects \cite{Helenius:2012wd} or initial-state energy loss \cite{Vitev:2007ve}, could already modify the yield before the hard probe traverses the medium, thereby changing the amount of final-state suppression---and thus the $\hat{q}$---required to describe experimental data.
These considerations underscore the need for caution when interpreting the extracted $\hat{q}$ values.

\section{Conclusions}
\label{sec:conclusion}

In this work we used light- and heavy-flavor high-$p_T$ suppression data from central and semi-central \coll{Pb}{Pb} and \coll{Au}{Au} collisions in a $\chi^2$ minimization procedure to extract the best fit effective strong coupling $\alpha_s^{\text{eff.}}$ for our model and then generate zero-parameter extrapolations to peripheral \coll{A}{A} collisions and central \collThree{p}{d}{A} collisions.
Our model implements both collisional \cite{Braaten:1991jj,Braaten:1991we,Wicks:2008zz} and radiative energy loss \cite{Gyulassy:2001nm,Djordjevic:2003zk}, multiple gluon emission \cite{Gyulassy:2001nm}, realistic hadronization, realistic production spectra, and fluctuating IP-Glasma initial conditions \cite{Schenke:2020mbo}. Our pQCD-based energy loss model receives short path length corrections to both the collisional \cite{Wicks:2008zz,Faraday:2024gzx} and radiative energy loss \cite{Kolbe:2015rvk,Kolbe:2015suq,Faraday:2023mmx}, allowing us to make justifiable predictions for small and peripheral large systems. 
We quantitatively estimate the size of two large theoretical uncertainties in our energy loss model: first from the crossover between HTL and vacuum propagators \cite{Romatschke:2004au,Gossiaux:2008jv,Wicks:2008zz} and second from the kinematic restriction on the phase space for radiative gluon emission \cite{Faraday:2023mmx,Faraday:2023uay}.
The extracted value of $\alpha_s^{\text{eff.}}$ that we found was 
$\alpha_s^{\text{eff.}} = \num{0.41(14:10)}$ at RHIC and $\alpha_s^{\text{eff.}} = \num{0.37(11:8)}$ at LHC,
where the uncertainties are primarily due to theoretical uncertainties and not from the extraction procedure.
We converted our extracted value of $\alpha_s^{\text{eff.}}$ into an extraction of $\hat{q} / T^3$ in order to make contact with previous studies \cite{JET:2013cls,Andres:2016iys,Xie:2020zdb,JETSCAPE:2021ehl,Xie:2022ght,Apolinario:2022vzg}; we found $\hat{q} / T^3 = 15 \pm 4$ at RHIC and $\hat{q} / T^3 = 11\pm 3$ at LHC.
Since the extracted values of $\alpha_s^{\text{eff.}}$ were consistent across centrality cuts in central and semi-central collisions and across hadronic flavor, 
we extrapolated our model to peripheral \coll{A}{A} collisions and central \collThree{p}{d}{A} collisions.
Our model described very well the suppression in peripheral \coll{A}{A} collisions at RHIC and LHC and the suppression in central \coll{d}{Au} collisions at RHIC.
We applied our statistical analysis to various subsets of the data and found that the data implied that our model misses important low-$p_T$ physics at LHC (likely due to medium modifications to hadronization for $p_T \lesssim 10 ~\mathrm{GeV}$) and high-$p_T$ physics at LHC (likely due to running coupling effects that become important for $p_T \gtrsim 50 ~\mathrm{GeV}$). We further found that the data implied that, roughly, two strong couplings run with a scale of $2 \pi T$ and one with $E$, and that such an analysis has the potential to constrain the particular energy loss mechanisms.%

We previously found that a particular approximation in the DGLV radiative energy loss---the large formation time (LFT) approximation---is violated at large momenta by both the DGLV result and the DGLV result which receives a short path length correction (DGLV + SPLC) \cite{Faraday:2023mmx}. To mitigate the effects of the breakdown in the LFT approximation, we implemented for the first time a kinematic cutoff $|\mathbf{k}|_{\text{max}} = \operatorname{Min}[2 x E(1-x), \sqrt{2 x E} (\mu^2+\mathbf{q}^2)^{1 /4}]$ which ensures that no contributions to the energy loss are included from regions where either the collinear or large formation time approximations are violated \cite{Faraday:2023mmx}. We showed that using this cutoff dramatically reduces the size of the SPLC from $\mathcal{O}(200\%)$ to $\mathcal{O}(5\%)$ for $p_T \gtrsim 100 ~\mathrm{GeV}$ gluons, %
 and also reduces the standard DGLV energy loss by $\sim \! 20\%$. One may understand the difference in energy loss between the standard collinear cutoff $|\mathbf{k}|_{\text{max}} = 2 x E(1-x)$ and our new large formation time + collinear cutoff as a result of the $\sim E$ scaling of the collinear cutoff compared to the $\sim E^{1 / 2}$ scaling of the large formation time cutoff, which results in the phase space being significantly reduced at high-$p_T$ in the latter case.
Self consistently, we showed that the large formation time approximation is no longer violated by either DGLV or DGLV + SPLC when we use our new cutoff.

We quantitatively estimated the size of two large theoretical uncertainties in our model. The first uncertainty was related to the exact value chosen for the large formation time + collinear kinematic cutoff. Similarly to previous work \cite{Horowitz:2009eb}, we estimated this uncertainty by varying the value of the kinematic cutoff up and down by factors of two. The second uncertainty was related to the transition between HTL and vacuum propagators in the collisional energy loss \cite{Romatschke:2004au,Gossiaux:2008jv,Wicks:2008zz}. The propagator transition uncertainty was estimated by performing model calculations with two limiting implementations of the collisional energy loss: HTL-only \cite{Wicks:2008zz}, which uses only HTL propagators, and Braaten \& Thoma (BT) \cite{Braaten:1991jj,Braaten:1991we}, which transitions between the vacuum and HTL propagators such that one is minimally sensitive to the transition value chosen. We presented 14 different model extractions of $\alpha_s^{\text{eff.}}$, which included all combinations of the 2 collisional energy loss implementations and 7 different values in the range $[0.5, 2]$ multiplying the kinematic cutoff on the transverse radiated gluon momentum. While it is difficult to associate these different model choices with a distribution and hence a ``one sigma" confidence interval, we found that these different model choices resulted in an extracted value of $\alpha_s^{\text{eff.}} \in [0.3, 0.5]$. 
In our above presentation of the extracted value of $\alpha_s^{\text{eff.}}$ in the Conclusions, we naively treated each of these 14 model variations as representing a one-sigma variation of the theoretical parameter space for the extracted value of $\alpha_s^{\text{eff.}}$. We note that the uncertainties associated with the extraction procedure of $\alpha_s^{\text{eff.}}$ are negligible compared to the theoretical uncertainties.

We then performed our analysis on different subsets of the data used in the full analysis in order to understand the consistency of our model. 
We first considered applying our statistical analysis separately to different centrality classes. We found that at both LHC and RHIC, the extracted value of $\alpha_s^{\text{eff.}}$ was consistent between $0\text{--}10\%$, $10\text{--}30\%$, and $30\text{--}50\%$ centrality classes within a 68\% confidence interval. 
The results when using HTL-only collisional energy loss described the centrality dependence better than the results with BT collisional energy loss; however, neither model can be ruled out from this analysis.
We additionally considered the flavor dependence of our results by applying the statistical analysis separately to charged hadrons, $D^0$ mesons, and $B^{\pm}$ mesons produced in \coll{Pb}{Pb} collisions. We found that the $D^{0}$ mesons $\alpha_s^{\text{eff.}}$ was $5\text{--}15\%$ larger and within $1\text{--}2$ standard deviation of the value extracted from light-flavor data. The $\alpha_s^{\text{eff.}}$ extracted from $B^{\pm}$ meson data was $15\text{--}50\%$ larger and within $1\text{--}4$ standard deviations of the $\alpha_s^{\text{eff.}}$ extracted from charged hadrons.
Our results are clearly consistent across centrality classes and in moderate but not inconsistent tension with the heavy-flavor data. We concluded that it was reasonable to extrapolate our model to peripheral \coll{A}{A} and central \coll{p}{A} collisions.

We thus presented predictions for both peripheral large and central small collisions systems. 
We found that our model predictions constrained on central and semi-central heavy-ion collisions were in good agreement with the $p_T$ dependence of light-flavor hadron $R_{AB}$ data from both $60\text{--}80\%$ centrality \coll{Au}{Au} collisions and $60\text{--}80\%$ centrality \coll{Pb}{Pb} collisions. We then compared the $p_T$-integrated centrality dependence of our model predictions to \coll{Pb}{Pb} data from LHC \cite{ALICE:2018ekf}. Our predictions were in good agreement with the data for $0\text{--}80\%$ centrality, but then showed dramatic deviations for the centrality dependence from $80\text{--}100\%$, which is due to centrality bias \cite{ALICE:2018ekf}.
We further used our global model constrained on central and semi-central heavy-ion collisions to produce predictions for charged hadrons produced in central \coll{d}{Au} collisions and charged hadrons and $D^0$ mesons produced in central $\sqrt{s_{NN}} = 5.02 ~\mathrm{TeV}$ \coll{p}{Pb} collisions. 
Similar to our previous work \cite{Faraday:2024qtl}, we saw here that while the central small systems are smaller than the peripheral large systems they are hotter, and so we anticipated similar suppression in central small systems as in peripheral large systems.
Consistent with this expectation, we found that our large-system-constrained model predictions yielded a result that is in excellent agreement with $0\text{--}5\%$ centrality \coll{d}{Au} data, which is normalized by prompt photons \cite{PHENIX:2023dxl}; however, our large-system-constrained predictions are in stark disagreement with data from charged hadrons \cite{ALICE:2014xsp,ATLAS:2022kqu} and $D^0$ mesons \cite{ALICE:2019fhe} produced in $0\text{--}5\%$ centrality \coll{p}{Pb} collisions, which are normalized using the Glauber model. We also presented predictions for $\pi^0$ hadrons produced in minimum bias and central \coll{O}{O} and \coll{Ne}{Ne} collisions, finding $R_{AB} = 0.68 \pm 0.08$ at $p_T = 10 ~\mathrm{GeV}$. Our predicted suppression is significant enough that it should be distinguishable from the no-quenching baseline \cite{Gebhard:2024flv} within anticipated experimental uncertainties \cite{Huss:2020whe,Huss:2020dwe,Brewer:2021kiv}.

We showed in \cite{Faraday:2024qtl} that the suppression in $0\text{--}5\%$ central \collThree{p}{d}{A} collisions and $60\text{--}70\%$ \coll{A}{A} collisions is theoretically expected to be similar at both RHIC and LHC, across various simple energy loss model implementations---including in the limits of weak and strong coupling, multiple soft and single hard scattering, and collisional and radiative only energy loss. 
Further, at RHIC and LHC peripheral large system collisions should be less sensitive to centrality bias than central small systems: peripheral large systems have a larger number of binary collisions---$\langle N_{\text{coll}} \rangle \sim \! 50$ in $60\text{--}70\%$ centrality \coll{A}{A} collisions vs.\ $\langle N_{\text{coll}} \rangle \sim \! 10$ in $0\text{--}5\%$ centrality \collThree{p}{d}{A} collisions. 
Our conclusion is that our agreement with central \coll{d}{Au} suppression at RHIC and our disagreement with central \coll{p}{Pb} suppression at LHC is likely due to a larger centrality bias \cite{ALICE:2014xsp, PHENIX:2023dxl, ATLAS:2023zfx,ALICE:2018ekf} in the Glauber-normalized LHC analyses. LHC measurements of $R_{pA}$ normalized by weakly-coupled particles such as photons or $Z$ bosons could shed valuable light on the possibility of suppression in central small system collisions by reducing centrality bias and model dependencies in the measurement. Further, a $Z$ boson is produced at significantly larger Bjorken $x$ values of the proton than a corresponding photon with the same $p_T$ due to its large mass. Thus, while energy loss models would predict the same $R_{pA}$ normalized by either photons or $Z$ bosons, the color fluctuation model \cite{Alvioli:2014eda,Alvioli:2017wou} would likely predict significantly more suppression for the $Z$-normalized $R_{pA}$ than for the photon-normalized $R_{pA}$.%

We then considered various subsets of data in order to see if we could observe from data that there was missing physics in our energy loss model. We first examined the sensitivity to the lowest momentum data considered in the statistical analysis, motivated by the potential influence of, for example, non-perturbative energy loss processes \cite{Gubser:2006bz,Casalderrey-Solana:2011dxg}, the medium modification of hadronization \cite{Sato:1981ez,Biro:1994mp,Biro:1998dm,Braun-Munzinger:2003pwq}, or the effect of
 temperature gradients and medium flow on energy loss \cite{Lekaveckas:2013lha,Casalderrey-Solana:2016jvj,Reiten:2019fta,He:2020iow,Antiporda:2021hpk,Bahder:2024jpa}.
We considered high-$p_T$ suppression data from $0\text{--}50\%$ centrality heavy-ion collisions with a $p_T$ range of $p_T^{\text{min}} \leq p_T \leq 20 ~\mathrm{GeV}$, separately for $\pi^0$ mesons at RHIC, charged hadrons at LHC, and $D^0$ mesons at LHC. By calculating the $p$-value of our best model fit compared to these different datasets as a function of $p_T^{\text{min}}$, we found that our model well described $\pi^0$ mesons at RHIC for $p_T \gtrsim 5 ~\mathrm{GeV}$, charged hadrons at LHC for $p_T^{\text{min}} \gtrsim 9 ~\mathrm{GeV}$, and heavy-flavor hadrons at LHC for $p_T \gtrsim 5 ~\mathrm{GeV}$. Our findings are consistent with non-trivial modifications to vacuum fragmentation, for instance coalescence \cite{Sato:1981ez,Biro:1994mp,Biro:1998dm}, being an important contribution to the final light-flavor hadron spectra from heavy-ion collisions at LHC out to $p_T \simeq 9 ~\mathrm{GeV}$. In contrast, the light-flavor RHIC data did not require any such medium modifications to hadronization past $p_T \simeq 5 ~\mathrm{GeV}$, consistent with the lower temperatures of the produced medium at RHIC.
The good agreement between our model and the heavy-flavor LHC data for $p_T \gtrsim 5 ~\mathrm{GeV}$ suggests that non-trivial modifications to vacuum fragmentation may be less important for heavy-flavor hadrons. However, this good agreement could also stem from the larger uncertainties in heavy- compared to light-flavor measurements, which impose weaker constraints on the model.

To gain further insight into missing physics in our model, we examined the $p_T$ dependence of the extracted value of $\alpha_s^{\text{eff.}}$ by applying our statistical analysis over various $p_T$ ranges. We considered ATLAS \cite{ATLAS:2022kqu} and CMS \cite{CMS:2016xef} data for charged hadrons produced in $0\text{--}50\%$ centrality \coll{Pb}{Pb} collisions partitioned into five $p_T$ ranges covering $10 \leq p_T \leq 450 ~\mathrm{GeV}$.
From ATLAS data, the extracted $\alpha_s^{\text{eff.}}$ decreased as a function of $p_T$ for $p_T \lesssim 50 ~\mathrm{GeV}$ before \emph{increasing} for $p_T \gtrsim 50 ~\mathrm{GeV}$, while the extracted $\alpha_s^{\text{eff.}}$ from the CMS data decreased monotonically as a function of $p_T$ for all $p_T$. 
To probe the temperature dependence of the running coupling, we separately fit $R_{AB}$ data from light-flavor hadrons produced in $0\text{--}50\%$ centrality collisions from RHIC and LHC separately, with average temperatures of $\langle T_{\text{RHIC}} \rangle = 0.23 ~\mathrm{GeV}$ and $\langle T_{\text{LHC}} \rangle = 0.31 ~\mathrm{GeV}$, respectively. We argued that if the different values of the extracted $\alpha_s^{\text{eff.}}$ as a function of $p_T$ and $\sqrt{s_{NN}}$ are truly due to running coupling effects, then one should expect that there is a consistent prescription for the running coupling which describes both effects. 
We found that by comparing the data to a simple model of running couplings, the data prefers that one of the couplings runs with $E$ and two of the couplings run with $2 \pi T$. We further found that this analysis can discern the transition scale between vacuum and HTL propagators; however, neither of the two limiting cases that we considered---HTL-only and BT---provided a consistent description for the data as a function of both $\sqrt{s_{NN}}$ and $p_T$.
Our results indicate that such a comparison over a large $p_T$ and $\sqrt{s_{NN}}$ range may be used in the future to place strong constraints on both the scales at which the coupling runs and the transition scale between vacuum and HTL propagators.

Is there partonic energy loss induced by the presence of a quark-gluon plasma in small collision systems?  Despite the overwhelming evidence for the formation of quark-gluon plasma in small collision systems from low-$p_T$ multi-particle correlations \cite{ATLAS:2015hzw, ALICE:2023ulm,ATLAS:2013jmi, ALICE:2014dwt, CMS:2015yux} strangeness enhancement \cite{ALICE:2015mpp, ALICE:2013wgn}, and quarkonia suppression \cite{ALICE:2016sdt}, naively, the measured $R_{pA} > 1$ from LHC \cite{ATLAS:2022kqu,ALICE:2014xsp,ALICE:2019hno} indicates that the answer is no.  Our work here demonstrates that, absent centrality biases, one expects measurably significant hadronic suppression from perturbative QCD-based partonic energy loss in central small collision systems such as \coll{d}{A} collisions at RHIC and \coll{p}{A} collisions at LHC.  

We found that our suppression predictions for \coll{d}{A} collisions at RHIC, fully constrained by central \coll{A}{A} collisions, describe well recent PHENIX measurements that normalize the nuclear modification factor with direct photons \cite{PHENIX:2023dxl}, thus minimizing centrality bias.  An alternative explanation for the measured suppression at RHIC, which also describes the enhancement of $R_{pA}$ at LHC \cite{Alvioli:2017wou}, is the color fluctuation model \cite{Perepelitsa:2024eik,Alvioli:2014eda,Alvioli:2017wou}, which posits that the deviations from 1 in $R_{AB}$ are due to initial-state effects.  Experimentally, energy loss and the color fluctuation model predict opposite behavior as a function of collision system size \cite{Perepelitsa:2024eik}; a measurement of $R_{AB}$ for \collFour{p}{d}{He3}{Au} collisions at RHIC should show increased suppression as a function of system size for energy loss while the color fluctuation model predicts decreased suppression as a function of system size. We presented predictions for the double ratio of the nuclear modification factor in \coll{He3}{Au} collisions to that in \coll{p}{Au} collisions as a function of $p_T$ and centrality, yielding a $10\%$ relative suppression. The systematic uncertainties in the PHENIX measurement are dominated by the \coll{p}{p} baseline \cite{PHENIX:2023dxl} and so such a double ratio, where the systematic uncertainties are canceled, should significantly improve the experimental constraints on initial- vs.\ final state effects in small systems.

Further theoretical/experimental insight may be gained by considering self-normalizing observables such as high-$p_T$ $v_2$ \cite{ATLAS:2019vcm,CMS:2025kzg}, dijet \cite{ATLAS:2022iyq,CMS:2025jbv} and dihadron correlations, jet substructure \cite{Kolbe:2023rsq}, energy-energy correlators \cite{Andres:2022ovj}, and minimum bias smaller collision systems such as \coll{O}{O} collisions \cite{Huss:2020whe,Huss:2020dwe}, which should also reduce the influence of centrality bias.  Predictions of $v_2$ and dihadron correlations require a careful treatment of event-by-event fluctuations \cite{Noronha-Hostler:2016eow} and are likely influenced by the hydrodynamical evolution of the medium; since we do not include either of these effects in this current work, we leave simultaneous predictions for $R_{AB}$, $v_2$, and dihadron correlations for future work \cite{Bert:2024}.

\section*{Acknowledgments}
We thank Raymond Ehlers, Joseph Bahder, Jan Fiete Grosse-Oetringhaus, Peter Jacobs, Florian Jonas, Jaime Norman, Yen-Jie Lee, Chun Shen, and Liliana Apolinàrio for productive discussions. Computations were performed using facilities provided by the University of Cape Town’s ICTS High Performance Computing team: \href{http://hpc.uct.ac.za}{hpc.uct.ac.za}.
CF and WAH thank the National Research Foundation, the National Institute for Theoretical and Computational Sciences (NITheCS), and the SA-CERN collaboration for their generous financial support during the course of this work.
The authors gratefully acknowledge the CERN Theory Group for their hospitality during the course of this work.

\newpage

\appendix

\section{Full comparison of global fitted model results to data used in the global extraction of \texorpdfstring{$\alpha_s^{\text{eff.}}$}{alpha s}}
\label{sec:app_global}

\begin{figure*}[!htbp]
	\centering
	\includegraphics[width=0.95\linewidth]{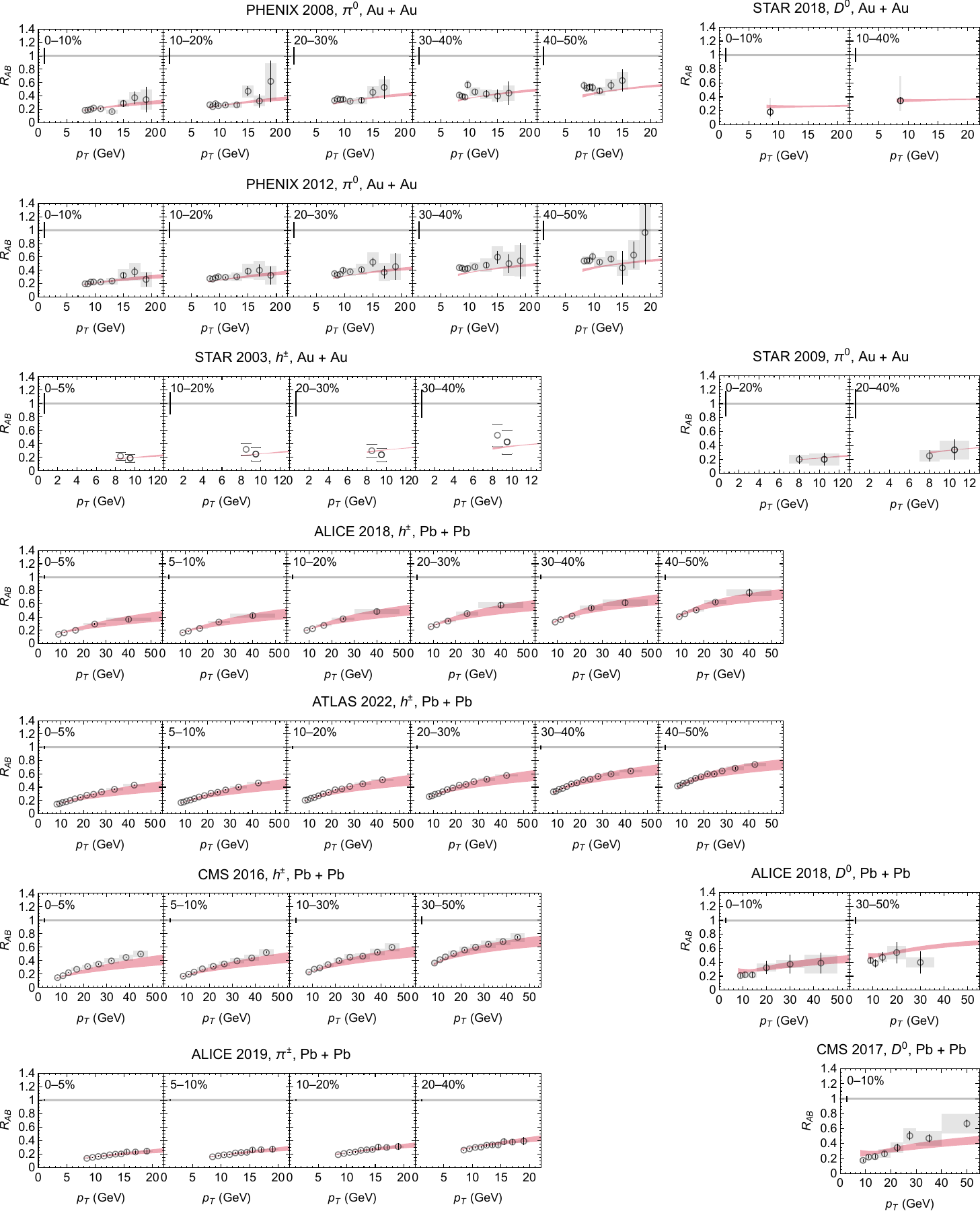}
	\caption{All measurements of leading hadron $R_{AB}$ (open circles) used in the global extraction of $\alpha_s^{\text{eff.}}$ in this work and th corresponding fitted model results after fitting (red bands). Statistical uncertainties are represented by error bars, systematic uncertainties by shaded boxes, and global normalization uncertainties by solid lines at $R_{AB} = 1$. The red band width corresponds to the uncertainties from the extraction and the theoretical uncertainties, added in quadrature.}
	\label{fig:all_data_vs_theory_global}
\end{figure*}

\Cref{fig:all_data_vs_theory_global} plots our model results evaluated with the globally extracted $\alpha_s^{\text{eff.}}$ compared to all experimental data which was used in the extraction. The full list of experimental data compared to in this figure is provided in \cref{tab:raa_experiments}.

\section{Full comparison of global fitted model results to all experimental data shown in this work}
\label{sec:app_all}

\begin{figure*}[!htbp]
	\centering
	\includegraphics[width=\linewidth]{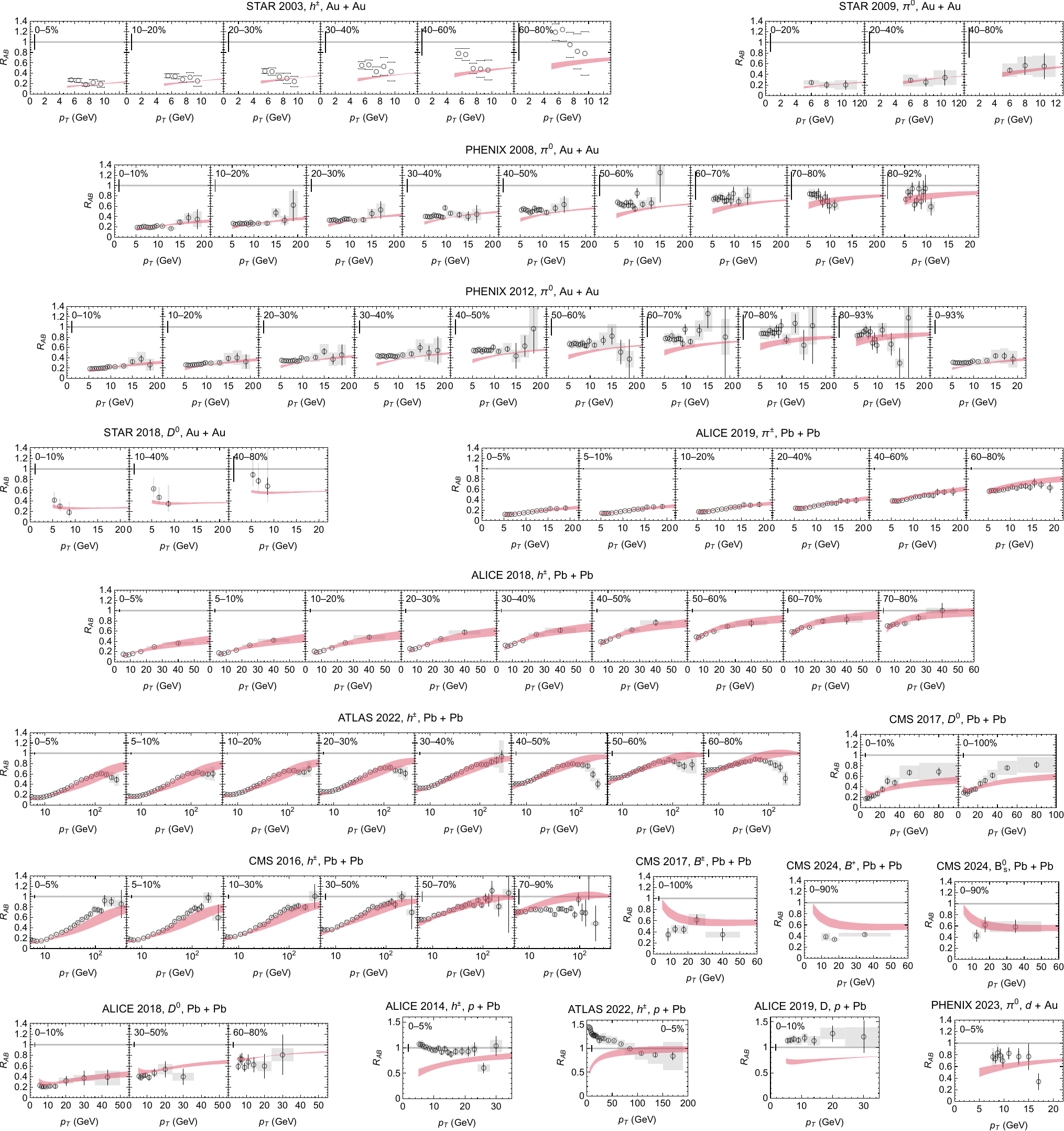}
	\caption{All measurements of leading hadron $R_{AB}$ (open circles) used in this analysis and corresponding fitted model results from our global extraction of $\alpha_s^{\text{eff.}}$ (red bands). Statistical uncertainties correspond to error bars, systematic uncertainties to shaded boxes, and global normalization uncertainties to the solid lines at $R_{AB} = 1$. The red band width corresponds to the uncertainties from the extraction and the theoretical uncertainties, added in quadrature.}
	\label{fig:all_data_vs_theory}
\end{figure*}

\Cref{fig:all_data_vs_theory} plots our model results evaluated with the globally extracted $\alpha_s^{\text{eff.}}$ compared to all experimental data which was used in this work. 
The set of experimental data shown in the figure includes an extended set of data in comparison to \cref{fig:all_data_vs_theory_global}, including data from a larger $p_T$ range, from all reported centrality ranges, and from small collisions. The full list of experimental data compared to in this figure is provided in \cref{tab:raa_experiments_all_data}. 

\newpage

\bibliographystyle{JHEP} %
\bibliography{manual,small_system_elastic_fitted}

\end{document}